# Statistical Detection of Potentially Fabricated Numerical Data: A Case Study

By

Joel H. Pitt[1] and Helene Z. Hill[2]

[1] Renaissance Associates, Princeton, NJ drjhpitt@yahoo.com

[2] Rutgers University, NJ Medical School, Newark, NJ 07101-1709 hill@njms.rutgers.edu

**Abstract**

Scientific fraud is an increasingly vexing problem. Many current programs for fraud detection focus on image manipulation, while techniques for detection based on anomalous patterns that may be discoverable in the underlying numerical data get much less attention, even though these techniques are often easy to apply. We employed three such techniques in a case study in which we considered data sets from several hundred experiments. We compared patterns in the data sets from one research teaching specialist (RTS), to those of 9 other members of the same laboratory and from 3 outside laboratories. Application of two conventional statistical tests and a newly developed test for anomalous patterns in the triplicate data commonly produced in such research to various data sets reported by the RTS resulted in repeated rejection of the hypotheses (often at p-levels well below 0.001) that anomalous patterns in his data may have occurred by chance. This analysis emphasizes the importance of access to raw data that form the bases of publications, reports and grant applications in order to evaluate the correctness of the conclusions, as well as the utility of methods for detecting anomalous, especially fabricated, numerical results.

## 1. INTRODUCTION

During the past decade, retractions of scientific articles have increased more than 10-fold (Van Noorden 2011). At least two-thirds of these retractions are attributable to scientific misconduct: fraud (data fabrication and falsification), suspected fraud, duplicate publication, and plagiarism (Fang, Steen et al. 2012). Techniques for early identification of fraudulent research are clearly needed. Much current attention has been focused on sophisticated methods for detecting image manipulation (Rossner and Yamada 2004) and their use is encouraged on the website of the Office of Research Integrity (ORI) of the United States Department of Health and Human Services. But statistical methods which can readily be used to identify potential data fabrication (Mosimann, Wiseman et al. 1995; Mosimann, Dahlberg et al. 2002; Al-Marzouki, Evans et al. 2005; Baggerly and Coombes 2009; Hudes, McCann et al. 2009; Carlisle 2012; Simonsohn 2012) are all but ignored by the ORI and the larger world. We believe that routine application of statistical tools to identify potential fabrication could help to avoid the pitfalls of undetected fabricated data just as tools such as, for example, CrossCheck and TurnItIn are currently used to detect plagiarism.

The first step in using statistical techniques to identify fabricated data is to look for anomalous patterns of data values in a given data set (or among statistical summaries presented for separate data sets), patterns that are inconsistent with those that might ordinarily appear in genuine empirical data. That such patterns are, indeed, anomalous may potentially be confirmed by using genuine data sets as controls, and by using simulations or probabilistic calculations based on appropriate models for the data to show that they would not ordinarily occur in genuine data.

The existence of these anomalous patterns in given suspect data sets may be indicative of serious issues of data integrity including data fabrication (Al-Marzouki, Evans et al. 2005), but they may also arise as a result of chance. Hence it is of considerable importance to have statistical methods available to test the hypothesis that a given anomalous pattern in a data set may have occurred as the result of chance.

For example, Mosimann *et al.* (Mosimann, Dahlberg et al. 2002) identified instances of fabricated data based on the observation that in experimental data sets containing count data in which the terminal (insignificant) digits are immaterial and inconsequential (hence not under the control of the investigator) it is reasonable to expect and generally the case that these inconsequential digits will appear to have been drawn at random from a uniform population. When terminal digits of the count values in a data set of this type do not appear to have been drawn from a uniform population (as may be tested using the Chi-square goodness of fit test) this may indicate that they have been fabricated.

A test like this is not entirely foolproof. Before applying it, one must ask whether there really is any evidence, beyond mere supposition, that terminal digits of data of the given kind should be random in the sense of uniform. Ideally one would like to have a probability model for the underlying randomness in the experimental data and use it to show that the distribution of terminal digits of counts values in data sets consistent with that model will be uniform (Hill and Schürger 2005). Alternately one might be able to run simulations based on an appropriate probability model and demonstrate that the terminal digits of the counts in the simulated data sets do generally appear to have been drawn uniformly. Finally, one could try to validate the assumption that terminal digits of counts in legitimate data sets are uniform, empirically, by testing the uniformity of terminal digits in indisputably legitimate experimental data sets of exactly the same type, constructed using the same protocols, as that of the suspect data.

Simonsohn (Simonsohn 2012) uncovered fabrications in several psychological research papers based entirely on the summary data available in published reports. He noted that despite the fact that the means of various variables measured in the study varied considerably, their standard deviations were remarkably similar, and hypothesized that this would not be the case were the results derived from genuine experimental data. He confirmed his hypothesis with simulation and empirical observation of the distribution of standard deviations in comparable studies.

When we have an appropriate probability model available for the underlying experiment that purportedly produced the suspect data, we can often apply our knowledge of probability theory to determine the probability that an anomalous pattern in question may have occurred by chance in the data set under consideration. Where that probability is less than some reasonable level, we term our tests significant, and, in the absence of any alternative explanation, may find any such significant results convincing evidence that the data in question has been fabricated.

**2. The Case Study:** Concerns about the legitimacy of raw data generated by one Research-Teaching Specialist (RTS) in the laboratory in which one of us was a member led us investigate data sets of his which had been used in several publications, a grant application and its renewal. We also had access to data sets generated by nine other researchers in the same laboratory who followed the same or similar protocols, as well as data from three outside laboratories that employed similar techniques. By applying the same investigating techniques to their data sets, we were able to use them as controls. Copies of the laboratory notebooks containing the raw data that we analyzed were in the form of PDF files which we transferred into Excel spreadsheets (cf Supplementary Material).

We believe that this was a unique situation, as we were able to review and compare essentially all the data from a single laboratory, data produced by a number of independent investigators using the same or similar research techniques, over such a long period of time. In particular it allowed us to determine whether suspect patterns that we had already noted in a limited number of data sets from the RTS whose data had raised the initial concerns appeared in other data sets of his and whether the same patterns might be found in the data sets from the other investigators.

These other than expected patterns in the RTS's data included: (1) a non-uniform distribution of insignificant terminal digits; (2) an unusually large frequency of equal terminal digit pairs (i.e. equal right-most and second right-most digits); and (3) a surprisingly large number of triplicate colony and cell counts in which a value near the average value of the triple or even that average value appeared as one of the constituent counts of the triple.

None of these patterns were evident in any of the data sets reported by the nine other investigators in the same laboratory, or in data sets that we obtained from three other independent, outside researchers. We believe this is a matter of significant concern.

We can use the well-known chi-square goodness of fit test to determine whether non-uniformity of terminal digits can be considered significant. Additionally, a straightforward test of significance based on the binomial distribution can be used to test the significance of an unusually high frequency of equal terminal digit pairs, but there is no such standard test to determine the significance of unusually large numbers of triplicate counts containing values near their average. Random variation in these triplicate data that are common components of pharmacological, cell biological and radiobiological experimentation, can be analyzed by modeling the triples as sets of three independent, identical Poisson random variables. A major focus of this study is on developing a method to calculate bounds and estimates for the probability that a given set of $n$ such triplicates contains $k$ or more triples which contain their own mean. We use these bounds and estimates in tests of the hypothesis that the observed unusually high incidence of mean containing triples in certain data sets may have occurred by chance.

Our methods should be useful to laboratory investigators in therapeutic, toxicological, cell and radiation biological studies involving evaluation of cell survival after various treatments. Much of our analyses pertain to triple replicates such as are commonly used in cell survival protocols (Bonifacino 1998; Munshi, Hobbs et al. 2005; Katz, Ito et al. 2008).

**3. Experimental Protocols:** The experiments we analyzed followed the same or very similar protocols and employed, with few exceptions, the same Chinese hamster cell line. The cells, harvested from mass culture, were counted, apportioned equally into culture tubes and incubated overnight with radioisotopes. They were washed free of radioactivity and transferred to new tubes for a 3-day incubation at low temperature ($10.5°$ C) to allow for the given isotope to decay. They were then harvested, triplicate aliquots were suspended for cell counts using a Coulter ZM particle counter and aliquots were diluted and plated onto tissue culture dishes in triplicate in order that single cells could grow into colonies which were stained and counted (manually) after about a week.

**4. Data sets and Probability Model:** The primary data sets with which we are concerned are collection of triples of integer Coulter ZM counts and triples of colony counts. The former are copied by hand into a notebook from an LED digital readout of the Coulter ZM counter that counts single cells as they pass randomly through a narrow orifice, the latter are counted by hand. The colony triples are counts of the number of colonies formed by the surviving cells. The counts in each Coulter triple and each colony triple are modeled probabilistically as independent, identical Poisson random variables. The Poisson parameter of these triples will, of course, vary from triple to triple.

Throughout this report, the accumulated data from the RTS's experiments are independently paralleled to the accumulated data of other investigators including nine members of the laboratory other than the RTS who utilized the same Coulter counter and/or counted colonies in the same manner, two professors from out-of-state universities who contributed triplicate data from their Coulter ZM counters, and triplicate colony counts from an additional independent laboratory.

**5. Analysis of Triplicate Data:** Many radiobiological experiments result in data sets consisting of triplicate counts where the means of the triples are the key values that are associated with the corresponding treatments in subsequent analyses. An investigator wishing to guide the results of such experiments would have to arrange the data values in each of the triples so that their means are consistent with the desired results. The quickest and easiest way to construct such triples would be to choose the desired mean (or a close approximation) as one of the three count values and then, using two roughly equal constants, calculate the other two values as this initial value plus or minus the selected constants.

Data sets constructed in this manner might then be expected to include either (1) an unusually high concentration of triples whose **mid-ratio** (the ratio of the difference between the middle value and the smallest value to the difference between the largest value and the smallest value (the **gap**) was close to 0.5; or (2) an unusually large number of triples that actually include their own (rounded) mean as one of their values.

**5.1 Initial mid-ratio review:** Having observed what appeared to us to be an unusual frequency of triples in RTS's data containing a value close to their mean, we used R to calculate the mid-ratios for all of the colony data triples that were available to us. We then constructed histograms of the resulting data sets. The results are shown in **Figures 1A** and 1**B.** The histogram of mid-ratios for RTS's colony triples exhibits a distinct predominance of mid-ratios in the range 0.4 to 0.6, while the histogram of mid-ratios of the data triples recorded by the nine other members of the laboratory is fairly uniform over the ten sub-intervals. The dramatic contrast between the two histograms seems a clear indication that RTS' data may have been manipulated to guide the mean values of its triples.

**Fig. 1: Distributions of the mid-ratios (middle – low)/(high – low) for colony triples  A.** RTS, 1343 triples, 128 experiments; **B.** Other investigators, 572 triples, 59 experiments.

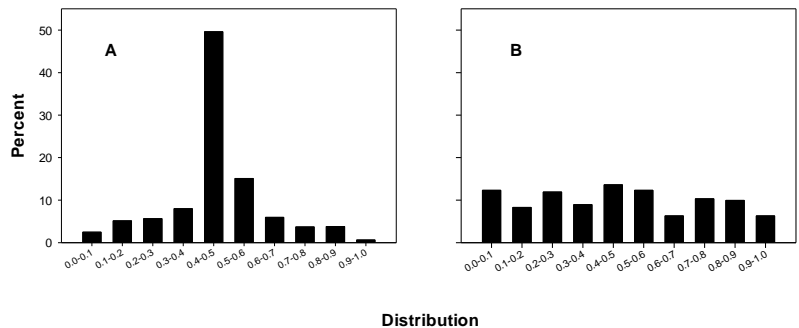

**5.2 Appearance of the Mean in Triplicate Samples:** We extended our investigation by writing an R program to identify and count triples that contained their rounded average. (**Figure 2** is a scan of a page from one of the RTS's notebooks. Triples that contain their rounded average are highlighted in blue. In this instance six of the ten triples are highlighted.) Of the 1343 complete colony triples in RTS's data, 690 (more than 50%) contained their rounded average, whereas only 109 (19%) of the 572 such triples from other investigators did.

**Figure. 2: PDF Image of Colony Counts from an experiment performed by RTS.** The rounded average (highlighted in blue) appears as one of the triplicate counts in 6 of the 10 samples (Ppoibin Prob = 0.00169, See Section 5.7, below.).

TABLE-4

Expt # : 2       Date: 12/11/98

| Tube.dilution | Colony 1 | Colony 2 | Colony 3 | Avg Colony | SD |
|---|---|---|---|---|---|
| 1·1 | 120 | 111 | 99 | 116.5 | |
| 2·2 | 131 | 124 | 117 | | |
| 3·2 | 104 | 114 | 96 | 104.6 | 0.5984 |
| 4·2 | 93 | 100 | 86 | 93 | 0.7982 |
| 5·1 | 70 | 78 | 63 | 70.3 | 0.6037 |
| 6·2 | 58 | 68 | 49 | 58.3 | 0.5007 |
| 7·1 | 29 | 22 | 19 | 23.3 | 0.2002 |
| 8·3 | 104 | 115 | 94 | 104 | 0.0895 |
| 9·4 | 116 | 126 | 107 | 1.16 | 0.0099 |
| 10·4 | 23 | 29 | 18 | 0.23 | 0.002 |

Given the marked difference between the percentage of the RTS's triples that contain their mean and the corresponding percentage of other investigators' triples that do so, and the similar disparity between the histograms of mid-ratios of the RTS's triples and those of other investigators, it is reasonable to ask whether the apparently excessive numbers of mean/near mean containing triples in the RTS's data sets might plausibly have occurred by mere chance. In order to answer that question we used a probability model for such triplicate data to calculate bounds and estimates of the probability that a given set of $n$ such triplicates contains $k$ or more triples. Using these estimates we are able to test the chance hypothesis.

**5.3 The Model for Triplicate Data:** The differences between the three actual count values in each colony count triple arise from random differences in the number of cells actually drawn and transferred to the three dishes and the randomness in the numbers of cells that survive the treatment applied to the cells in that triple. As noted above the random variables that correspond to the number of cells that are originally in each of the three dishes can be modeled probabilistically as the values of three independent, identical Poisson random variables. The common Poisson parameter $\lambda_0$ of those three variables will be the (unknown) expected value of those cell counts.

Since the cells in the three dishes have all been exposed to the same level of radiation, the probabilities that a given cell survives to generate colonies should be the same in each of the three dishes. Accordingly, the actual number of survivor colonies in the three dishes will have a binomial distribution with the same p parameter (the common individual cell survival rate) and differing n values corresponding to the numbers of cells on each dish. It is easy to show that these resulting counts have Poisson distributions with parameter $\lambda=\lambda_0 p$.

Thus the three values in each set can be modeled as the values of three independent Poisson random variables sharing a common parameter $\lambda$. The actual value of $\lambda$ varies from triple to triple as it depends both on the specific $\lambda_0$ associated with the initial cell count Poisson distribution and the specific p associated with the treatment which gave rise to the given triple. The likelihood that one of the counts in the triple is equal to or near the triple mean value depends on the value of this parameter.

Given the value of their common Poisson parameter $\lambda$ a relatively straightforward calculation can be used to find the probability that a triple generated by independent, identical Poisson random variables includes its mean (see Appendix). The values of the various $\lambda$ parameters of the Poisson random variables that gave rise to the triples in our data set are, of course, unknown to us, but, in as much as actual colony count values are all less than 400 we can safely assume that the λ parameters of the underlying Poisson random variables are certainly less than 1000.

We wrote an R program to calculate the probability that a triple generated by independent, identical Poisson variables with known parameter λ includes its own (rounded) mean value and used it to calculate and create a table (referred to below as the **MidProb** table) of this probability for all integer values of λ from 1 to 2000 and, as the variation of these probabilities between successive integer values of λ greater than 2000 was negligible we extended the table by calculating the value of the probability for values of λ that were multiples of 100 between 2100 and 10000, and multiples of 1000 between 11000 and 59000 (see **Table 1** for the first 25 entries). Our calculations showed that as λ increased from 1 to 3, the probability that a randomly generated triple contains its own mean increases from about 0.27 to slightly more than 0.40 and then decreases as λ continued to increase. We were thus assured that no matter what the value of λ for the Poisson variables that generated a given triple, the probability that the triple would have included its mean as one of its three elements would not exceed 0.42.

**5.5 Hypothesis testing I -- A non-parametric test:** The observation that the probability that a triple generated by independent, identical Poisson variables with known parameter λ includes its own (rounded) mean value never exceeds 0.42 gives us the ability to construct a crude test of the hypothesis that an observed, suspect high number of mean containing triples in a given collection of triples may have occurred by chance. Using the number k of triples with gap two or more that contain their means and the number n of triples in the collection we need only find the binomial probability p of k or more successes in n independent Bernoulli trials where the probability of success is 0.42. If the probability p is less than the chosen α level of the test we reject the null hypothesis at that significance level. The test is crude in the sense that the calculated p is not the p-level of the test, it is simply a (possibly gross) over estimate of the p-level.

**Table 1. Partial MidProb Table. Probability that a triple generated by 3 independent Poisson random variables with parameter $\lambda$ contains its mean for $\lambda = 1$ to 25.** It is clear that $\lambda$ continues to decrease after $\lambda = 4$.

| $\lambda$ | P | $\lambda$ | P | $\lambda$ | P | $\lambda$ | P | $\lambda$ | P |
|---|---|---|---|---|---|---|---|---|---|
| 1 | 0.267 | 6 | 0.372 | 11 | 0.317 | 16 | 0.281 | 21 | 0.254 |
| 2 | 0.387 | 7 | 0.359 | 12 | 0.309 | 17 | 0.275 | 22 | 0.250 |
| 3 | 0.403 | 8 | 0.348 | 13 | 0.301 | 18 | 0.269 | 23 | 0.246 |
| 4 | 0.397 | 9 | 0.337 | 14 | 0.294 | 19 | 0.264 | 24 | 0.242 |
| 5 | 0.385 | 10 | 0.327 | 15 | 0.287 | 20 | 0.259 | 25 | 0.238 |

When we apply this test to determine how likely it is that 690 or more of the 1343 colony triples in RTS's data might have contained their rounded average by chance, we find that it is less than $2.85 \times 10^{-12}$, an extremely significant result.

Since there are only 109 mean containing triples among the 572 from other investigators, and 109 is considerably less than the expected number of successes in 572 Bernoulli trials with a success probability of 0.42 it is immediately clear that the probability of having 109 or more mean containing triples is reasonably large -- indeed it is essentially one.

**5.7 Hypothesis testing II -- Using $\lambda$ to obtain p-values**: It is important to have a more sensitive test, as we can use it to confirm the validity of our model by applying it to what we believe to be legitimate experimental data. To do so we use a heuristic method to estimate the actual probability that a given collection of n triples includes k mean containing triples. This allows us to provide an actual p-value for the one-tailed test we apply for seemingly high numbers of mean containing triples and thereby allows us to determine whether the numbers of mean containing triples in our controls are consistent with our model or whether they are also significantly different from what our model indicates.

We start with the observation that the results of our calculations in the **MidProb** table show that the probability that a triple of independent, identical Poisson random variables includes its own mean decreases rapidly as $\lambda$ increases. For example the probability that a triple contains its mean if it is generated by Poisson random variables with $\lambda = 20$ is about 0.26, but with $\lambda = 50$ it is less than 0.18, and with $\lambda = 100$ it is less than 0.14 (it even falls to 0.032 when $\lambda = 2000$).

We applied a heuristic approach to use our table of calculated values of this probability to estimate (rather than merely bound) the probability that a given collection of n triples that are hypothesized to have been drawn as triples of independent, identical Poisson variables has as many or more than the actual number of mean containing triples than it was observed to contain. We do not know the actual $\lambda$ values of the Poisson random variables that (hypothetically) generated the triples in the data sets, but the mean of any actual triple is a reasonable estimate of the $\lambda$ parameter of the variables that gave rise to it. (The mean is the maximum likelihood estimator in this case.). We can then look up these (rounded) $\lambda$ values in the **MidProb** table to obtain an estimate of the probability that had the triple been randomly drawn it would contain its own mean.

We are thus able to consider the events that the various individual triples of the collection contain their own means as successes in individual, independent, Bernoulli trials each with a known probability of success. The random variable (statistic) which takes as its value the number of triples in the given data set that contain their own means is the sum of the Bernoulli random variables that indicate success in the various trials. These Bernoulli trials have the known (actually estimated) probabilities of taking the value 1 that we obtain from the **MidProb** table as described above.

Sums of such independent, not necessarily identically, distributed Bernoulli random variables are said to be Poisson binomial random variables and to have a Poisson binomial distribution. A Poisson binomial random variable that is the sum of n Bernoulli random variables can potentially take any of the values 0,1,..., n, and the probability that it takes, or is greater than, or equal to any of these potential values is completely determined by the probabilities $p_1, p_2, ..., p_n$ that the constituent Bernoulli random variables take the value 1.

Few (if any) standard statistical packages include functions for calculating Poisson binomial distributions. Although there is a straightforward algorithm which can, in principle, be used to calculate probabilities for the distribution function of a Poisson binomial random variable given the success probabilities of the individual Bernoulli variables $p_1, p_2, ..., p_n$, issues of numerical stability in these calculations can arise for even moderately large values of n, and processing times increase exponentially as n increases. Nonetheless, we were able to take advantage of an efficient algorithm that has recently been developed and implemented as a package for R (Hong 2011) to find exact values for the tail p-values that we wish to have in testing our null hypothesis.

The function *ppoibin* in the R package *poibin* accepts as input two parameters, an integer j and a vector of probabilities $p_1, p_2, ..., p_n$ and returns the probability that the Poisson binomial random variable that corresponds to that vector of probabilities takes a value less than or equal to j. To use it to find the probability that there are k or more mean containing triples in a collection of triples generated by groups of three Poisson random variables with common probabilities $p_1, p_2, ..., p_n$, we execute *ppoibin* with the value j=k-1 as the first parameter and the given probabilities as the second and subtract the result from 1.

We applied this more refined test to the RTS collection of 1343 complete colony triples and found that, given the likely λ values that had given rise to the individual triples in the collection, the probability of the observed 690 or more mean containing triples is approximately 6.26 x $10^{-13}$ (not surprisingly an extremely significant result). Applying the same test to find the probability of finding 109 or more mean containing triples among the 572 complete colony triples that had been recorded by the other investigators in the same laboratory, we found that the probability was 0.47, and the probability of 109 or fewer such triples is 0.58; results that are entirely consistent with our hypothesis.

**5.8 Hypothesis testing III -- Normal estimation of p-values**: Given the success probabilities of the individual Bernoulli variables $p_1, p_2, ..., p_n$ the expectation of their Poisson binomial sum is μ = $\sum_{i=1}^{n} p_i$ and its variance $\sigma^2 = \sum_{i=1}^{n} p_i (1 - p_i)$. Both are easy to calculate. When the values of the $p_i's$ are bounded below, the (Lindeberg-Feller) Central Limit Theorem applies and we can obtain reasonable approximations of the (upper) tail probabilities of a Poisson binomial random variable using normal probabilities.

Where an efficient implementation of an algorithm for calculating exact Poisson binomial probabilities is not available, we can use a normal approximation which with a second order correction (Volkova 1995) provides a quite precise estimate. Hong (2011) reports the results of multiple simulations that indicate that by including the second order correction the normal approximations to upper tail probabilities will usually -- but certainly not always -- return probability values marginally higher than the true tail probabilities. The normal distribution we use to approximate a Poisson binomial is the normal with the same mean and standard deviation as the Poisson binomial.

Using the normal approximation has a second advantage, in as much as the z-values we calculate in order to look up normal probabilities are informative without recourse to an actual table of normal probabilities. Virtually all students of statistics learn that in normal populations upper tails corresponding to z-scores of 2 or more or 3 or more are quite unlikely -- with the first having a probability of less the 0.025 and the second having a probability of less than .0015.

To use this approach to approximate the probability of the 690 or more mean containing triples among the RTS' 1343 complete triples, we first obtain (to two decimal places) µ=220.31 and σ=13.42. Using a standard correction for continuity, the z-value we use to find the probability of 690 or more mean containing triples is $\frac{689.5-220.31}{13.42} = 34.97$ so large that the upper tail probability is effectively indistinguishable from 0, hence significant at virtually any level.

It is important to keep in mind that the normal distribution probabilities are approximations, not exact values, of the Poisson binomial probabilities. Unfortunately the normal approximations of upper tail Poisson binomial probabilities are generally less than the true values. In this instance, however, the aforementioned Volkova correction provides the same estimate.

**5.9 Application to Coulter Counts:** While the means of colony triples are the key values of interest to investigators, means of Coulter triples are not as significant. Thus there is less reason to believe that an investigator wishing to guide results might be inclined to construct Coulter triples that include their own means as one of their values. Nonetheless we extended our investigation and counted the number of mean contain triples in both the RTS' Coulter triples and those from other investigators. The results are interesting and illustrate the power and importance of the more sensitive tests we discussed in 5.7 and 5.8 above.

Coulter data from the RTS included 1717 complete triples, 173 of which included their rounded mean, while we had 929 complete Coulter triples from other investigators in the same lab, 36 of which included their rounded means. Application of the crude test described in 5.6 gives no reason for concern as in both cases the numbers of mean containing triples are consistent with our belief that the probability that any given triple includes its mean will be less than 0.42.

When, however, we apply the more refined analysis introduced in sections 5.7 and 5.8, we find reason once again to question RTS' data. Coulter count values are in a much higher range than colony count values, thus the Poisson random variables that give rise to them have $\lambda$ values in a higher range and probabilities that Coulter triples include their means tend to be lower. Using our table of probabilities, triples of independent Poisson random variables with given $\lambda$ parameters that contain their own mean, we found that were we to randomly generate 1717 Poisson triples with respective $\lambda$ parameters set equal to the means of the RTS' actual triples the expected number of mean containing triples would be 97.74 and the standard deviation 9.58. Given this (and using the normal approximation to the Poisson binomial) the 7.80 z-score that corresponds to the actual number of 173 mean containing triples in the RTS' data makes it immediately clear that it is exceedingly unlikely we might have encountered such a large number of mean contain triples by chance. The actual Poisson binomial tail probability is $6.26 \times 10^{-13}$.

When we apply the same analysis to the Coulter triples we obtained from other investigators in the same lab the results are well within the expected range. According to our calculations the expected number of mean containing triples would be 39.85 and the standard deviation is 6.11.

Hence the z-value corresponding to the actual number of 36 mean containing triples is -0.71 and the actual p-value is 0.76, entirely consistent with our model.

We applied the same analysis to the triplicate Coulter count data sets we had from two investigators in other labs and triplicate colony counts from an investigator in another lab and the results for all of these sets are summarized in **Table 2** below.

**Table 2: Summary results for analysis of mean containing triples for colony and Coulter count triples from RTS, 9 other investigators from the same lab, and investigators in outside labs**

| TYPE | INVESTIGATOR | NO. EXPS | NO. COMPL/TOTAL | NO. W MEAN | NO. EXPECTED | STD | Z-VALUE | P≥K |
|---|---|---|---|---|---|---|---|---|
| COLONIES | RTS | 128 | 1343/1361 | 690 | 220.3 | 13.42 | 34.97 | 0 |
| COLONIES | Others | 59 | 572/591 | 109 | 107.8 | 9.23 | 0.08 | 0.466 |
| COLONIES | Outside lab 1 | 1 | 49/50 | 3 | 7.9 | 2.58 | -2.11 | 0.991 |
| COULTER | RTS | 174 | 1716/1717 | 173 | 97.7 | 9.58 | 7.80 | $6.26 \times 10^{-13}$ |
| COULTER | Others | 103 | 929/929 | 36 | 39.9 | 6.11 | -0.71 | 0.758 |
| COULTER | Outside lab 2 | 11 | 97/97 | 0 | 4.4 | 2.03 | -2.42 | 1.00 |
| COULTER | Outside lab 3 | 17 | 120/120 | 1 | 3.75 | 1.90 | -1.71 | 0.990 |

**5.10 Probability model for Mid-Ratios:** We took a similar approach to evaluating the significance of the occurrence of high percentages of triples having mid-ratios close to 0.5 to that which we used when dealing with triples that contain their mean. In like manner, we wrote an R function to calculate the probability that the mid-ratio of a triple with a given parameter λ falls within the interval [0.40,0.60]. Using this function we calculated these probabilities for each of the integer values of λ from 1 to 2000 and stored them in a table. The results of these calculations showed that as λ increases from 1 to 10 the probability that a triple has a mid-ratio in the interval [0.40, 0.60] increases from about 0.184 to slightly more than 0.251 and decreases thereafter. Thus our calculated results tell us that for every value of λ, the probability that the mid-ratio is in the interval [0.40, 0.60] is less than 0.26. Hence, given a collection of n triples the probability that k or more of those triples have mid-ratios in the interval [0.40, 0.60] cannot be greater than the probability of k or more successes in n independent Bernoulli trials in which the probability of success is 0.26. As was the case when we considered triples which contain their mean, these Binomial probabilities can be used to provide a crude but potentially useful test of significance.

We used the same heuristic approach that we had used to develop a more refined significance test for the occurrence of triples that contain their own means to develop a more refined significance test for the incidence of mid-ratios in the [0.40, 0.60] interval. This test could be of use in detecting instances in which an investigator wishing to guide the mean values of triplicates employs a reasonably subtle technique.

**6. Terminal Digit Analysis:** J. E. Mosimann and colleagues (Mosimann, Wiseman et al. 1995; Mosimann, Dahlberg et al. 2002) recommend a technique for identifying aberrant data sets based on the observation that under many ordinary circumstances the least significant (rightmost) digits of genuine experimental count data can be expected to be uniformly distributed and the further observation that when people invent numbers they are generally not uniform.

As per our introductory remarks it is important to confirm the applicability of this expected uniformity in any context in which we hope to use it. The fact that, in as much as the cells counted in a single batch by the Coulter counter typically number in the several hundreds up to the many thousands, control in selecting the batches of cells to be counted is far from precise enough to extend to the last digit, lends some *a priori* support to the expectation that terminal digits will be uniform. But we also ran simulations generating data sets of triples of independent identical Poisson random variables with comparable means, and the distributions of terminal digits in these sets were consistent with the hypothesis of uniformity.

Based on these considerations we believe it is reasonable to suppose that the Mossiman technique applies to the various Coulter count data sets under consideration. The fact that we are able to apply our tests of uniformity to what we believe to be uncontested experimental data in the course of our test provides further of empirical confirmation of the applicability of the Mossiman test.

**6.1 Application of terminal digit analysis to the data sets:** We counted the number of times each of the digits 0,1,...,9 occurred as the rightmost digit of counts copied from the Coulter ZM counter screen and from colony counts. (Note that these analyses do not require that the data be arranged in triplicate sets.) If these least significant digits were indeed uniform -- as they should be if the data was truly generated experimentally -- then our counts for each of these ten digits should be roughly the same.

We obtain a more precise measure of the degree to which these distributions diverge from the expected uniform by applying the Chi-square test for goodness of fit. We show the actual distribution of terminal digits for the various full data sets we considered in **Table 3**, along with the computed Chi-square statistics and the associated p-values. The p-values for RTS's terminal digit sets result in our rejecting the null hypothesis of uniformity at any reasonable level (and even unreasonable levels) of significance; results for all other investigators' data sets are consistent with our null hypothesis.

**Table 3. Terminal digit analysis of Coulter and colony counts.** "Others" refers to other investigators in the laboratory. Outside labs contributed two sets of Coulter data and one set of colony data. Probabilities of 0 were too small to estimate.

| Type | Investigator | 0 | 1 | 2 | 3 | 4 | 5 | 6 | 7 | 8 | 9 | Total | Chi-sq | P-value |
|---|---|---|---|---|---|---|---|---|---|---|---|---|---|---|
| Coulter | RTS 174 exps | 472 | 612 | 730 | 416 | 335 | 725 | 362 | 422 | 370 | 711 | 5155 | 456.4 | 0 |
| Coulter | Others 103 exps | 261 | 311 | 295 | 259 | 318 | 290 | 298 | 283 | 331 | 296 | 2942 | 16.0 | 0.067 |
| Coulter | Outside lab 11 exps | 28 | 34 | 29 | 24 | 27 | 36 | 44 | 33 | 26 | 33 | 314 | 9.9 | 0.36 |
| Coulter | Outside lab 17 exps | 34 | 38 | 45 | 35 | 32 | 42 | 31 | 35 | 35 | 33 | 360 | 4.9 | 0.84 |
| Colonies | RTS 128 exps | 564 | 324 | 463 | 313 | 290 | 478 | 336 | 408 | 383 | 526 | 3501 | 200.7 | 0 |
| Colonies | Others 59 exps | 187 | 180 | 193 | 178 | 183 | 173 | 176 | 183 | 183 | 178 | 1814 | 1.65 | 0.996 |
| Colonies | Outside lab 1 exp | 21 | 9 | 15 | 16 | 19 | 19 | 9 | 19 | 11 | 12 | 150 | 12.1 | 0.21 |

**7. Equal Digit Analysis:** Just as it is reasonable to expect that insignificant terminal digits in experimental data would be approximately uniform, it is also seems reasonable to expect that the last two digits of three plus digit experimental data (in which the terminal digits are relatively immaterial) will be equal approximately 10% of the time. We used R to count the number of terminal digit pairs in the RTS' and other investigators' Coulter count data and found that there were 291 (9.9%) equal pairs of rightmost digit pairs among the 2942 Coulter count values produced by investigators in the laboratory other than the RTS, while there were 636 (12.3%) such pairs in the RTS's 5155 recorded Coulter counts. Assuming that these right-most pairs were generated uniformly, the probability of 636 or more equal pairs in 5155 Coulter values is less than 3.3 x $10^{-8}$, which significantly contraindicates their expected randomness. In contrast, the probability of 291 or more equal pairs among 2942 Coulter values for the other researchers is 0.587 which is consistent with our randomness hypothesis.

## 8. Summary

1. In the RTS's experiments, the averages of triplicate colony counts appear as one of those counts at improbably high levels based on our model. The rates at which triplicate colony counts reported by other investigators include their averages is consistent with our model.

2. In the RTS's experiments, the mid-ratio values of triplicate colony counts fall in the interval [0.4,0.6] at improbably high levels based on our model. The rates at which mid-ratios of triplicate colony counts reported by other investigators fall in that interval is consistent with our model.

3. Distributions of terminal digits of values in the RTS 's Coulter counts and colonies differ significantly from expected uniformity. This does not hold for the colony and Coulter terminal digits of other workers.

5. Significantly more than the expected one tenth of the data values the RTS recorded in his Coulter counts have equal terminal digits. This does not hold for the occurrences and distributions of terminal doubles in the Coulter counts of other workers.

## 9. Discussion

**9.1 Limitations** In most case studies, the number of controls is either equal to or greater than the number of test values. Since this is a *post hoc* study, we had no control over the numbers of data we analyzed. To address our concern about smaller control sample sizes in one such instance, we randomly selected 314 terminal digits from the RTS's Coulter results and ran chi-square analyses 100,000 times to test for uniformity. All of the runs would have rejected the null hypothesis for uniformity at the 0.00001 level; one run rejected the hypothesis at the 0.000000001 level. The value of 314 was selected because it is the total number of digits supplied by one of the two outside contributors and was the smallest of the Coulter sample sets with which we worked (cf **Table 3**).

During the time that the RTS was working in the laboratory, few experiments were being performed simultaneously by others, which resulted in some temporal disparity. However, the protocols that we analyzed were followed almost identically by all of the members of the laboratory. There is no *a priori* evidence that the cells, instrumentation, equipment and consumable supplies used by the other researchers were any different from those utilized by the RTS. There is also no evidence that different operators could influence the terminal digits seen on the display of the Coulter counter. All of the investigators used similar techniques to stain and count the colonies.

**9.2 Power of statistics:** In a recent editorial in *Science*, Davidian and Louis emphasize the increasing importance of statistics in science and in world affairs as a "route to a data-informed future" (Davidian and Louis 2012). Statistical analysis of numerical data can be used to identify aberrant results (Tomkins, Penrose et al. 2010; Postma 2011; Tomkins, Penrose et al. 2011), even in esoteric studies (Brown, Cronk et al. 2005) (Trivers, Palestris et al. 2009). Recently, a rigorous statistical analysis of data that purported to predict the responses to chemotherapeutic agents of human lung, breast and ovarian cancers demonstrated the erroneous nature of the results (Baggerly and Coombes 2009; Baggerly and Coombes 2011) and led to several retractions (Baggerly and Coombes 2010; Goldberg 2010; 2011; 2011) and a resignation. In this case, patients were potentially directly affected by the use of the wrong drug and/or the withholding of the right drug.

Statistics were used to uncover fraudulent behavior on the part of Japanese anesthesiologist Y. Fujii who is believed to have fabricated data in as many as 168 publications (Carlisle 2012). In like manner, Al-Marzouki, et al. (Al-Marzouki, Evans et al. 2005) used statistics to implicate R.B. Singh for fabricating data in a clinical trial involving dietary habits. Their control, like our controls, was a similar trial performed using comparable methods by an outside group. Of interest is the fact that Singh was unable to produce his original data for re-examination because it had been, he alleged, consumed by termites. Hudes, et al. and McCann, et al. (Hudes, McCann et al. 2009) used statistics to detect unusual clustering of coefficients of variation in a number of articles produced by members from the same biochemistry department in India. The controls for these studies were obtained by searching for similar studies in PubMed. Once data manipulation is suspected, it is up to the statistician to find the proper test(s) to reveal discrepancies – to "let the punishment fit the crime", so to speak.

**9.3 Are** the RTS **'s data real:** The consistent and highly significant improbability that any of the multiple anomalies observed in the RTS's data sets are likely to have occurred by chance, and the fact that none of these anomalies appear in either the many data sets we examined from the nine other investigators in the same laboratory, working under the same conditions with the same equipment or in the comparable data sets we obtained from investigators outside the laboratory, leaves us with no alternative than to believe that the RTS's data is simply not genuine experimental data.

## 10. Remedies

**10.1 Automated analysis can deter tampering with results:** Automatic colony counters are commercially available, and their use in colony survival and other such studies should be encouraged. The counts from particle counters such as the Coulter ZM should be recorded on a printer.

**10.2 Journals should require the availability and archiving of raw data.** Many now do. This will permit verification, help to avoid unnecessary duplication of experimental results and facilitate interactions and interchanges among researchers.

**10.3 An Excel spreadsheet, available on request** to perform the calculations that we have proposed in this article, understanding that most researchers performing these types of survival and related experiments are not versed in the use of the statistical program R. The spreadsheet is available from Dr Pitt on request.

## Appendix

**Calculating the probability that a Poisson triple contains its rounded mean:**

As a preliminary to determining the probability that a triple contains its rounded mean, we first calculated the probability that a triple randomly generated by three independent Poisson random variables with a given $\lambda$ has a gap of two or more and contains its own mean. This event is the union of the infinite collection of mutually exclusive events:

$A_j =$ the event that the gap is j and the triple contains its own (rounded) mean, for j = 2, 3, 4, 5, ...

Hence its probability is, the sum of the separate probabilities of the $A'_j s$, $\sum_{j=2}^{\infty} P(A_j)$

For each j the event $A_j$ is itself the union of the infinite collection of mutually exclusive events:

$A_{j,k}$ = the event that the largest value in the triple is k (hence the smallest is k-j) and the triple includes as one of its elements its own (rounded) mean where, for any given j, the admissible values of k are j, j+1, j+2, j+3, ... Hence $P(A_{j,k}) = \sum_{k=j}^{\infty} P(A_{jk})$

To calculate $P(A_{j,k})$ we observe that in order for the event $A_{j,k}$ to occur, the smallest of the three elements of the triple must be k-j, and, of course, the largest must be k, but depending on the parity of j there may be one or two different possible values completing the triple. When j is even the third must be k-j/2 as it is easy to see that this is the only integer value that can complete a triple {k-j,n,k} that has mean n. However, when j is odd, there are two distinct integer values that can complete the triple {k-j,n,k} so that its mean is n, these are: k-[j/2] (where [x] is the greatest integer function, i.e. [x] = greatest integer less than or equal to x) and k-[j/2]-1.

Since the elements of our triples are assumed to be independently generated Poisson random variables with common parameter $\lambda$ we can obtain formulas in terms of Poisson probabilities for $P(A_{j,k})$. We first consider the case j even. Writing $p(n, \lambda)$ for the Poisson probability ($e^{-\lambda} \frac{\lambda^n}{n!}$) of obtaining the value n from a Poisson random variable with parameter $\lambda$, the probability that a triple consists of the values {k-j, k-[j/2],k} in any one of the six different orders in which these numbers can be permuted is $p(k - j, \lambda)p(k - [j/2], \lambda)p(k, \lambda)$ and hence the the probability of obtaining the triple for j even is

$$P(A_{j,k}) = 6p(k - j, \lambda)p(k - [j/2], \lambda)p(k, \lambda)$$

Applying a similar analysis with the two distinct triple types that could result in the event $A_{jk}$ when j is odd we get for odd j

$$P(A_{j,k}) = 6p(k-j,\lambda)(p(k-[j/2],\lambda) + p(k-[j/2]-1,\lambda))p(k,\lambda)$$

We combine the preceding observations to obtain a formula for the probability $P(A)$ that a triplet of numbers chosen independently from the same Poisson distribution contains its (rounded) mean. We get

$$P(A) = 6\left(\sum_{odd\, j=3}^{\infty} \sum_{k=j}^{\infty} p(k-j,\lambda)(p(k-[j/2],\lambda) + p(k-[j/2]-1,\lambda))p(k,\lambda)\right.$$
$$\left. + \sum_{even\, j=2}^{\infty} \sum_{k=j}^{\infty} p(k-j,\lambda)p(k-[j/2],\lambda)p(k,\lambda)\right)$$

And writing odd(x) for the function that is 1 when x is odd and 0 when x is even we can rewrite this as the single double sum:

$$P(A) = 6\left(\sum_{j=2}^{\infty} \sum_{k=j}^{\infty} p(k-j,\lambda)(p(k-[j/2],\lambda) + odd(j)p(k-[j/2]-1,\lambda))p(k,\lambda)\right)$$

Since we wish to obtain decimal values for these probabilities for various values of $\lambda$ we note that if, for a given $\lambda$ we choose N such that $\sum_{j=N+1}^{\infty} p(j,\lambda) < 10^{-9}$ or, equivalently, $\sum_{j=0}^{N} p(j,\lambda) \geq 1 - 10^{-9}$, then we can obtain a value of P(A) accurate to 5 decimal places using the formula:

$$P(A) = 6(\sum_{j=2}^{N}\sum_{k=j}^{N} p(k-j,\lambda)(p(k-[j/2],\lambda) + odd(j)p(k-[j/2]-1,\lambda))p(k,\lambda))$$

Using this formula, we wrote an R program to calculate the probability that a triple of independent Poisson random variables with a common parameter λ includes its mean as one of its three elements. We ran this program to create a table of the values of this probability for each of the integer values of λ from 1 to 2000. As a double check on the applicability of our calculation, we performed bootstrap calculations of selected probabilities using R to perform sets of 200,000 trials. The results were consistent with our calculations.

PAGE 31: Statistical Detection of Potentially Fabricated Data: A Case StudyPostma, E. (2011). "Comment on "Additive genetic breeding values correlate with the load of partially deleterious mutations"." Science **333**(6047): 1221.

Rossner, M. and K. M. Yamada (2004). "What's in a picture? The temptation of image manipulation." J Cell Biol **166**(1): 11-15.

Simonsohn, U. (2012). "Just post it: The lesson from two cases of fabricated data detected by statistics alone." Available at SSRN: http://ssrn.com/abstract=2114571 or http://dx.doi.org/10.2139/ssrn.2114571

Tomkins, J. L., M. A. Penrose, et al. (2010). "Additive genetic breeding values correlate with the load of partially deleterious mutations." Science **328**(5980): 892-894.

Tomkins, J. L., M. A. Penrose, et al. (2011). "Retraction." Science **333**(6047): 1220.

Trivers, R., B. G. Palestris, et al. (2009). The Anatomy of a Fraud: Symmetry and Dance. Antioch, CA 94509, TPZ Publishers.

Van Noorden, R. (2011). "Science publishing: The trouble with retractions." Nature **478**(7367): 26-28.

Volkova, A. Y. (1995). "A refinement of the Central Limit Theorem for Sums of Independent Random Indicators." Theory Probab. Appl. **40**(4): 791-794.

**Raw Data to accompany Statistical Detection of Potentially Fabricated Data**

The numbers were copied from PDF files obtained from the laboratory in question and from 3 outside laboratories and span the period from April, 1992 to April, 2005

**RTS Colonies**

| Date | col1 | col2 | col3 |
| --- | --- | --- | --- |
| 10/27/1997 | 78 | 91 | 93 |
| 10/27/1997 | 90 | 88 | 90 |
| 10/27/1997 | 80 | 66 | 69 |
| 10/27/1997 | 63 | 67 | 71 |
| 10/27/1997 | 44 | 58 | 64 |
| 10/27/1997 | 38 | 53 | 51 |
| 10/27/1997 | 247 | 264 | 258 |
| 10/27/1997 | 46 | 24 | 27 |
| 10/27/1997 | 64 | 63 | 61 |
| 10/27/1997 | 77 | 82 | 98 |
| 11/24/1997 | 115 | 98 | 109 |
| 11/24/1997 | 87 | 95 | 98 |
| 11/24/1997 | 41 | 31 | 38 |
| 11/24/1997 | 146 | 155 | 178 |
| 11/24/1997 | 112 | 105 | 104 |
| 11/24/1997 | 117 | 143 | 136 |
| 11/24/1997 | 117 | 133 | 114 |
| 11/24/1997 | 38 | 57 | 53 |
| 11/24/1997 | 170 | 171 | 176 |
| 11/24/1997 | 102 | 108 | |
| 12/1/1997 | 74 | 100 | 79 |
| 12/1/1997 | 85 | 90 | 70 |

| Date | | | |
|---|---|---|---|
| 12/1/1997 | 38 | 32 | 44 |
| 12/1/1997 | 28 | 41 | 26 |
| 12/1/1997 | 28 | 29 | 27 |
| 12/1/1997 | 103 | 91 | 123 |
| 12/1/1997 | 114 | 120 | 103 |
| 12/1/1997 | 26 | 25 | 24 |
| 12/1/1997 | 160 | 162 | 170 |
| 12/1/1997 | 104 | 103 | 100 |
| 12/15/1997 | 68 | 55 | 61 |
| 12/15/1997 | 66 | 61 | 65 |
| 12/15/1997 | 39 | 36 | 38 |
| 12/15/1997 | 53 | 50 | 47 |
| 12/15/1997 | 100 | 96 | 98 |
| 12/15/1997 | 62 | 68 | 77 |
| 12/15/1997 | 58 | 58 | 59 |
| 12/15/1997 | 30 | 35 | 37 |
| 12/15/1997 | 46 | 48 | 44 |
| 12/15/1997 | 83 | 95 | 87 |
| 12/19/1997 | 68 | 68 | 67 |
| 12/19/1997 | 57 | 62 | 64 |
| 12/19/1997 | 40 | 32 | 38 |
| 12/19/1997 | 50 | 48 | 52 |
| 12/19/1997 | 112 | 100 | 93 |
| 12/19/1997 | 53 | 64 | 65 |
| 12/19/1997 | 58 | 49 | 57 |
| 12/19/1997 | 27 | 28 | 30 |
| 12/19/1997 | 40 | 38 | 36 |
| 12/19/1997 | 82 | 78 | 83 |

| Date | | | |
|---|---|---|---|
| 12/22/1997 | 182 | 159 | 169 |
| 12/22/1997 | 155 | 150 | 168 |
| 12/22/1997 | 150 | 139 | 145 |
| 12/22/1997 | 130 | 127 | 122 |
| 12/22/1997 | 111 | 112 | 122 |
| 12/22/1997 | 174 | 177 | 150 |
| 12/22/1997 | 164 | 165 | 168 |
| 12/22/1997 | 151 | 134 | 130 |
| 12/22/1997 | 128 | 126 | 123 |
| 2/9/1998 | 44 | 37 | 44 |
| 2/9/1998 | 118 | 107 | 113 |
| 2/9/1998 | 73 | 91 | 93 |
| 2/9/1998 | 71 | 78 | 66 |
| 2/9/1998 | 69 | 71 | 68 |
| 2/9/1998 | 60 | 61 | 62 |
| 2/9/1998 | 55 | 45 | 60 |
| 2/9/1998 | 44 | 54 | 53 |
| 2/9/1998 | 28 | 25 | 31 |
| 2/20/1998 | 40 | 41 | 39 |
| 2/20/1998 | 55 | 34 | 39 |
| 2/20/1998 | 25 | 29 | 40 |
| 2/20/1998 | 95 | 98 | 105 |
| 2/20/1998 | 80 | 73 | 75 |
| 2/20/1998 | 75 | 100 | 184 |
| 2/20/1998 | 115 | 136 | 210 |
| 2/20/1998 | 121 | 91 | 64 |
| 2/20/1998 | 89 | 89 | 85 |
| 2/20/1998 | 51 | 54 | 56 |

| Date | | | |
|---|---|---|---|
| 2/23/1998 | 50 | 55 | 51 |
| 2/23/1998 | 72 | 70 | 55 |
| 2/23/1998 | 47 | 35 | 28 |
| 2/23/1998 | 94 | 95 | 98 |
| 2/23/1998 | 67 | 68 | 65 |
| 2/23/1998 | 57 | 52 | 50 |
| 2/23/1998 | 74 | 55 | 54 |
| 2/23/1998 | 28 | 30 | 25 |
| 2/23/1998 | 50 | 51 | 48 |
| 2/23/1998 | 29 | 27 | 23 |
| 2/27/1998 | 80 | 89 | 90 |
| 2/27/1998 | 90 | 102 | 81 |
| 2/27/1998 | 65 | 68 | 67 |
| 2/27/1998 | 29 | 25 | 26 |
| 2/27/1998 | 65 | 70 | 59 |
| 2/27/1998 | 113 | 129 | 138 |
| 2/27/1998 | 138 | 139 | 150 |
| 2/27/1998 | 50 | 47 | 47 |
| 2/27/1998 | 81 | 76 | 80 |
| 2/27/1998 | 134 | 130 | 128 |
| 3/9/1998 | 59 | 55 | 65 |
| 3/9/1998 | 75 | 55 | 62 |
| 3/9/1998 | 77 | 70 | 69 |
| 3/9/1998 | 120 | 125 | 129 |
| 3/9/1998 | 42 | 38 | 45 |
| 3/9/1998 | 44 | 46 | 49 |
| 3/9/1998 | 59 | 56 | 59 |
| 3/9/1998 | 36 | 41 | 38 |

| Date | | | |
|---|---|---|---|
| 3/9/1998 | 42 | 46 | 40 |
| 3/9/1998 | 8 | 6 | 5 |
| 3/13/1998 | 65 | 75 | 69 |
| 3/13/1998 | 60 | 72 | 74 |
| 3/13/1998 | 57 | 46 | 43 |
| 3/13/1998 | 160 | 179 | 163 |
| 3/13/1998 | 48 | 52 | 44 |
| 3/13/1998 | 87 | 89 | 106 |
| 3/13/1998 | 96 | 93 | 112 |
| 3/13/1998 | 25 | 27 | 29 |
| 3/13/1998 | 36 | 30 | 33 |
| 3/13/1998 | 79 | 70 | 72 |
| 3/16/1998 | 100 | 76 | 86 |
| 3/16/1998 | 96 | 92 | 94 |
| 3/16/1998 | 72 | 69 | 70 |
| 3/16/1998 | 36 | 34 | 32 |
| 3/16/1998 | 82 | 89 | 76 |
| 3/16/1998 | 140 | 132 | 152 |
| 3/16/1998 | 127 | 133 | 133 |
| 3/16/1998 | 42 | 34 | 50 |
| 3/16/1998 | 64 | 60 | 58 |
| 3/16/1998 | 21 | 22 | 20 |
| 3/20/1998 | 119 | 125 | 117 |
| 3/20/1998 | 135 | 139 | 130 |
| 3/20/1998 | 84 | 85 | 96 |
| 3/20/1998 | 79 | 78 | 78 |
| 3/20/1998 | 71 | 63 | 65 |
| 3/20/1998 | 59 | 54 | 47 |

PAGE 37: Statistical Detection of Potentially Fabricated Data: A Case Study| Date | | | |
|---|---|---|---|
| 3/20/1998 | 69 | 75 | 71 |
| 3/20/1998 | 130 | 127 | 133 |
| 3/20/1998 | 99 | 111 | 101 |
| 3/20/1998 | 33 | 38 | 37 |
| 3/30/1998 | 79 | 85 | 82 |
| 3/30/1998 | 69 | 68 | 62 |
| 3/30/1998 | 30 | 27 | 26 |
| 3/30/1998 | 99 | 92 | 82 |
| 3/30/1998 | 36 | 30 | 29 |
| 3/30/1998 | 130 | 117 | 121 |
| 3/30/1998 | 43 | 37 | 36 |
| 3/30/1998 | 65 | 62 | 57 |
| 3/30/1998 | 26 | 24 | 21 |
| 3/30/1998 | 40 | 32 | 35 |
| 4/3/1998 | 95 | 105 | 119 |
| 4/3/1998 | 121 | 116 | 125 |
| 4/3/1998 | 90 | 88 | 82 |
| 4/3/1998 | 80 | 72 | 76 |
| 4/3/1998 | 72 | 68 | 65 |
| 4/3/1998 | 110 | 106 | 112 |
| 4/3/1998 | 183 | 178 | 172 |
| 4/3/1998 | 220 | 216 | 230 |
| 4/3/1998 | 195 | 199 | 211 |
| 4/3/1998 | 127 | 121 | 129 |
| 4/20/1998 | 80 | 68 | 60 |
| 4/20/1998 | 11 | 14 | 14 |
| 4/20/1998 | 47 | 49 | 35 |
| 4/20/1998 | 82 | 90 | 78 |

| Date | | | |
|---|---|---|---|
| 4/20/1998 | 61 | 40 | 52 |
| 4/20/1998 | 72 | 82 | 95 |
| 4/20/1998 | 20 | 28 | 32 |
| 4/20/1998 | 59 | 53 | 54 |
| 4/20/1998 | 20 | 27 | 29 |
| 5/1/1998 | 149 | 152 | 165 |
| 5/1/1998 | 150 | 150 | 152 |
| 5/1/1998 | 130 | 132 | 122 |
| 5/1/1998 | 106 | 113 | 105 |
| 5/1/1998 | 95 | 100 | 105 |
| 5/1/1998 | 82 | 81 | 78 |
| 5/1/1998 | 95 | 93 | 91 |
| 5/1/1998 | 140 | 142 | 138 |
| 5/1/1998 | 160 | 164 | 158 |
| 5/1/1998 | 95 | 93 | 98 |
| 5/4/1998 | 77 | 82 | 88 |
| 5/4/1998 | 68 | 73 | 79 |
| 5/4/1998 | 65 | 69 | 68 |
| 5/4/1998 | 45 | 42 | 47 |
| 5/4/1998 | 39 | 41 | 38 |
| 5/4/1998 | 29 | 36 | 34 |
| 5/4/1998 | 32 | 30 | 29 |
| 5/4/1998 | 26 | 26 | 25 |
| 5/4/1998 | 154 | 162 | 149 |
| 5/4/1998 | 140 | 136 | 130 |
| 5/8/1998 | 58 | 54 | 56 |
| 5/8/1998 | 69 | 61 | 65 |
| 5/8/1998 | 39 | 33 | 37 |

| Date | | | |
|---|---|---|---|
| 5/8/1998 | 82 | 97 | 91 |
| 5/8/1998 | 58 | 45 | 46 |
| 5/8/1998 | 70 | 75 | 78 |
| 5/8/1998 | 69 | 68 | 66 |
| 5/8/1998 | 23 | 29 | 22 |
| 5/8/1998 | 38 | 40 | 30 |
| 5/8/1998 | 27 | 25 | 36 |
| 5/15/1998 | 125 | 129 | 128 |
| 5/15/1998 | 110 | 122 | 130 |
| 5/15/1998 | 105 | 100 | 113 |
| 5/15/1998 | 90 | 89 | 99 |
| 5/15/1998 | 74 | 78 | 82 |
| 5/15/1998 | 90 | 92 | 82 |
| 5/15/1998 | 98 | 90 | 89 |
| 5/15/1998 | 120 | 130 | 140 |
| 5/15/1998 | 189 | 179 | 195 |
| 5/15/1998 | 129 | 120 | 118 |
| 5/18/1998 | 65 | 60 | 62 |
| 5/18/1998 | 59 | 67 | 72 |
| 5/18/1998 | 64 | 58 | 61 |
| 5/18/1998 | 52 | 47 | 62 |
| 5/18/1998 | 42 | 40 | 38 |
| 5/18/1998 | 41 | 31 | 32 |
| 5/18/1998 | 28 | 30 | 27 |
| 5/18/1998 | 25 | 29 | 24 |
| 5/18/1998 | 88 | 84 | 84 |
| 5/18/1998 | 65 | 62 | 60 |
| 5/22/1998 | 59 | 55 | 57 |

| Date | | | |
|---|---|---|---|
| 5/22/1998 | 56 | 62 | 64 |
| 5/22/1998 | 33 | 42 | 36 |
| 5/22/1998 | 120 | 135 | 140 |
| 5/22/1998 | 38 | 45 | 50 |
| 5/22/1998 | 77 | 70 | 62 |
| 5/22/1998 | 65 | 63 | 71 |
| 5/22/1998 | 26 | 23 | 28 |
| 5/22/1998 | 35 | 42 | 49 |
| 5/22/1998 | 48 | 52 | 42 |
| 5/25/1998 | 85 | 82 | 87 |
| 5/25/1998 | 74 | 76 | 78 |
| 5/25/1998 | 69 | 75 | 72 |
| 5/25/1998 | 58 | 46 | 52 |
| 5/25/1998 | 48 | 52 | 40 |
| 5/25/1998 | 56 | 44 | 50 |
| 5/25/1998 | 58 | 64 | 60 |
| 5/25/1998 | 194 | 160 | 184 |
| 5/25/1998 | 120 | 140 | 116 |
| 5/25/1998 | 65 | 62 | 63 |
| 5/29/1998 | 80 | 72 | 70 |
| 5/29/1998 | 70 | 60 | 69 |
| 5/29/1998 | 58 | 78 | 68 |
| 5/29/1998 | 67 | 57 | 63 |
| 5/29/1998 | 55 | 49 | 43 |
| 5/29/1998 | 45 | 47 | 42 |
| 5/29/1998 | 37 | 29 | 34 |
| 5/29/1998 | 30 | 33 | 25 |
| 5/29/1998 | 26 | 24 | 25 |

| Date | | | |
|---|---|---|---|
| 5/29/1998 | 100 | 91 | 82 |
| 6/1/1998 | 130 | 137 | 128 |
| 6/1/1998 | 125 | 116 | 108 |
| 6/1/1998 | 82 | 89 | 70 |
| 6/1/1998 | 48 | 40 | 32 |
| 6/1/1998 | 200 | 217 | 214 |
| 6/1/1998 | 144 | 134 | 150 |
| 6/1/1998 | 132 | 149 | 134 |
| 6/1/1998 | 27 | 26 | 30 |
| 6/1/1998 | 61 | 55 | 62 |
| 6/1/1998 | 76 | 70 | 62 |
| 6/5/1998 | 125 | 129 | 130 |
| 6/5/1998 | 120 | 119 | 135 |
| 6/5/1998 | 68 | 78 | 74 |
| 6/5/1998 | 40 | 31 | 35 |
| 6/5/1998 | 218 | 199 | 184 |
| 6/5/1998 | 138 | 142 | 136 |
| 6/5/1998 | 126 | 139 | 132 |
| 6/5/1998 | 176 | 182 | 170 |
| 6/5/1998 | 34 | 30 | 38 |
| 6/5/1998 | 68 | 76 | 60 |
| 6/26/1998 | 110 | 119 | 105 |
| 6/26/1998 | 108 | 99 | 105 |
| 6/26/1998 | 90 | 97 | 106 |
| 6/26/1998 | 80 | 66 | 72 |
| 6/26/1998 | 70 | 60 | 64 |
| 6/26/1998 | 59 | 62 | 57 |
| 6/26/1998 | 70 | 75 | 65 |

| Date | | | |
|---|---|---|---|
| 6/26/1998 | 225 | 220 | 230 |
| 6/26/1998 | 204 | 209 | 199 |
| 6/26/1998 | 71 | 76 | 66 |
| 6/29/1998 | 228 | 215 | 214 |
| 6/29/1998 | 212 | 190 | 209 |
| 6/29/1998 | 156 | 163 | 148 |
| 6/29/1998 | 95 | 91 | 80 |
| 6/29/1998 | 246 | 237 | 231 |
| 6/29/1998 | 234 | 226 | 210 |
| 6/29/1998 | 223 | 216 | 225 |
| 6/29/1998 | 113 | 120 | 106 |
| 6/29/1998 | 150 | 162 | 150 |
| 6/29/1998 | 200 | 207 | 193 |
| 7/3/1998 | 180 | 189 | 176 |
| 7/3/1998 | 190 | 182 | 193 |
| 7/3/1998 | 144 | 150 | 138 |
| 7/3/1998 | 82 | 89 | 73 |
| 7/3/1998 | 39 | 42 | 35 |
| 7/3/1998 | 210 | 204 | 196 |
| 7/3/1998 | 215 | 210 | 199 |
| 7/3/1998 | 51 | 61 | 42 |
| 7/3/1998 | 120 | 127 | 113 |
| 7/3/1998 | 168 | 161 | 152 |
| 7/10/1998 | 185 | 170 | 178 |
| 7/10/1998 | 190 | 214 | 203 |
| 7/10/1998 | 170 | 182 | 161 |
| 7/10/1998 | 152 | 150 | 149 |
| 7/10/1998 | 124 | 142 | 132 |

| Date | | | |
|---|---|---|---|
| 7/10/1998 | 97 | 86 | 94 |
| 7/10/1998 | 25 | 28 | 28 |
| 7/10/1998 | 58 | 62 | 54 |
| 7/10/1998 | 72 | 79 | 64 |
| 7/10/1998 | 25 | 26 | 24 |
| 7/13/1998 | 170 | 162 | 181 |
| 7/13/1998 | 156 | 168 | 169 |
| 7/13/1998 | 159 | 145 | 154 |
| 7/13/1998 | 136 | 131 | 142 |
| 7/13/1998 | 128 | 123 | 118 |
| 7/13/1998 | 109 | 106 | 118 |
| 7/13/1998 | 38 | 32 | 37 |
| 7/13/1998 | 49 | 56 | 44 |
| 7/13/1998 | 67 | 60 | 53 |
| 7/13/1998 | 25 | 24 | 25 |
| 7/24/1998 | 161 | 169 | 153 |
| 7/24/1998 | 150 | 149 | 156 |
| 7/24/1998 | 117 | 120 | 113 |
| 7/24/1998 | 78 | 70 | 82 |
| 7/24/1998 | 40 | 45 | 36 |
| 7/24/1998 | 140 | 141 | 152 |
| 7/24/1998 | 139 | 141 | 143 |
| 7/24/1998 | 57 | 50 | 64 |
| 7/24/1998 | 180 | 188 | 196 |
| 7/24/1998 | 244 | 240 | 248 |
| 7/27/1998 | 118 | 107 | 105 |
| 7/27/1998 | 117 | 121 | 118 |
| 7/27/1998 | 102 | 100 | 89 |

| Date | | | |
|---|---|---|---|
| 7/27/1998 | 101 | 113 | 113 |
| 7/27/1998 | 90 | 122 | 119 |
| 7/27/1998 | 108 | 97 | 98 |
| 7/27/1998 | 89 | 101 | 112 |
| 7/27/1998 | 82 | 97 | 83 |
| 7/27/1998 | 50 | 39 | 47 |
| 7/31/1998 | 212 | 198 | 225 |
| 7/31/1998 | 196 | 177 | 168 |
| 7/31/1998 | 170 | 178 | 163 |
| 7/31/1998 | 125 | 131 | 120 |
| 7/31/1998 | 110 | 117 | 114 |
| 7/31/1998 | 116 | 122 | 110 |
| 7/31/1998 | 218 | 219 | 214 |
| 7/31/1998 | 52 | 60 | 44 |
| 7/31/1998 | 43 | 50 | 36 |
| 7/31/1998 | 27 | 25 | 23 |
| 8/3/1998 | 150 | 165 | 149 |
| 8/3/1998 | 132 | 147 | 140 |
| 8/3/1998 | 126 | 135 | 130 |
| 8/3/1998 | 100 | 109 | 92 |
| 8/3/1998 | 90 | 93 | 86 |
| 8/3/1998 | 90 | 95 | 86 |
| 8/3/1998 | 188 | 195 | 180 |
| 8/3/1998 | 48 | 58 | 39 |
| 8/3/1998 | 30 | 39 | 22 |
| 8/3/1998 | 170 | 177 | 169 |
| 8/7/1998 | 136 | 142 | 151 |
| 8/7/1998 | 129 | 136 | 139 |

| Date | | | |
|---|---|---|---|
| 8/7/1998 | 109 | 119 | 100 |
| 8/7/1998 | 40 | 38 | 43 |
| 8/7/1998 | 130 | 136 | 143 |
| 8/7/1998 | 120 | 122 | 124 |
| 8/7/1998 | 129 | 120 | 119 |
| 8/7/1998 | 28 | 26 | 24 |
| 8/7/1998 | 68 | 64 | 61 |
| 8/7/1998 | 70 | 64 | 67 |
| 8/10/1998 | 150 | 142 | 139 |
| 8/10/1998 | 131 | 129 | 126 |
| 8/10/1998 | 116 | 101 | 108 |
| 8/10/1998 | 36 | 43 | 31 |
| 8/10/1998 | 185 | 190 | 201 |
| 8/10/1998 | 106 | 105 | 112 |
| 8/10/1998 | 92 | 119 | 99 |
| 8/10/1998 | 40 | 30 | 26 |
| 8/10/1998 | 32 | 27 | 37 |
| 8/10/1998 | 40 | 45 | 51 |
| 8/24/1998 | 134 | 131 | 142 |
| 8/24/1998 | 153 | 142 | 139 |
| 8/24/1998 | 125 | 130 | 127 |
| 8/24/1998 | 105 | 120 | 108 |
| 8/24/1998 | 108 | 101 | 93 |
| 8/24/1998 | 96 | 86 | 100 |
| 8/24/1998 | 75 | 74 | 79 |
| 8/24/1998 | 44 | 49 | 52 |
| 8/24/1998 | 227 | 240 | 238 |
| 8/24/1998 | 150 | 159 | 131 |

| Date | | | |
|---|---|---|---|
| 8/28/1998 | 101 | 92 | 93 |
| 8/28/1998 | 89 | 95 | 99 |
| 8/28/1998 | 60 | 66 | 73 |
| 8/28/1998 | 24 | 28 | 25 |
| 8/28/1998 | 157 | 148 | 165 |
| 8/28/1998 | 110 | 108 | 102 |
| 8/28/1998 | 120 | 112 | 107 |
| 8/28/1998 | 40 | 51 | 29 |
| 8/28/1998 | 50 | 58 | 63 |
| 8/28/1998 | 100 | 114 | 82 |
| 8/31/1998 | 109 | 118 | 108 |
| 8/31/1998 | 111 | 106 | 127 |
| 8/31/1998 | 85 | 95 | 76 |
| 8/31/1998 | 48 | 38 | 29 |
| 8/31/1998 | 226 | 206 | 240 |
| 8/31/1998 | 134 | 145 | 142 |
| 8/31/1998 | 115 | 129 | 131 |
| 8/31/1998 | 48 | 44 | 41 |
| 8/31/1998 | 59 | 54 | 47 |
| 8/31/1998 | 109 | 99 | 88 |
| 9/4/1998 | 123 | 113 | 136 |
| 9/4/1998 | 116 | 128 | 108 |
| 9/4/1998 | 109 | 103 | 106 |
| 9/4/1998 | 92 | 76 | 86 |
| 9/4/1998 | 75 | 85 | 66 |
| 9/4/1998 | 99 | 80 | 88 |
| 9/4/1998 | 194 | 201 | 187 |
| 9/4/1998 | 29 | 35 | 24 |
| 9/4/1998 | 26 | 25 | 23 |
| 9/4/1998 | 120 | 129 | 136 |
| 10/1/1998 | 125 | 115 | 132 |
| 10/1/1998 | 136 | 121 | 129 |
| 10/1/1998 | 129 | 119 | 115 |
| 10/1/1998 | 113 | 107 | 117 |
| 10/1/1998 | 112 | 102 | 91 |
| 10/1/1998 | 65 | 76 | 84 |
| 10/1/1998 | 47 | 41 | 35 |
| 10/1/1998 | 127 | 134 | 119 |
| 10/1/1998 | 39 | 32 | 24 |
| 10/1/1998 | 68 | 62 | 68 |
| 10/13/1998 | 187 | 182 | 190 |
| 10/13/1998 | 228 | 199 | 213 |
| 10/13/1998 | 66 | 68 | 61 |
| 10/13/1998 | 39 | 37 | 44 |
| 10/13/1998 | 43 | 37 | 33 |
| 10/13/1998 | 160 | 153 | 175 |
| 10/13/1998 | 250 | 150 | 170 |
| 10/13/1998 | 122 | 133 | 125 |
| 10/13/1998 | 137 | 132 | 131 |
| 10/13/1998 | 58 | 60 | 50 |
| 10/19/1998 | 115 | 114 | 112 |
| 10/19/1998 | 113 | 140 | 146 |
| 10/19/1998 | 87 | 85 | 88 |
| 10/19/1998 | 61 | 67 | 64 |
| 10/19/1998 | 42 | 38 | 44 |
| 10/19/1998 | 97 | 110 | 113 |

| Date | | | |
|---|---|---|---|
| 10/19/1998 | 115 | 112 | 105 |
| 10/19/1998 | 66 | 78 | 70 |
| 10/19/1998 | 60 | 56 | 59 |
| 10/19/1998 | 35 | 31 | 36 |
| 10/23/1998 | 73 | 83 | 75 |
| 10/23/1998 | 24 | 26 | 29 |
| 10/23/1998 | 39 | 37 | 43 |
| 10/23/1998 | 58 | 49 | 67 |
| 10/23/1998 | 61 | 60 | 68 |
| 10/23/1998 | 68 | 69 | 78 |
| 10/26/1998 | 175 | 157 | 153 |
| 10/26/1998 | 188 | 179 | 192 |
| 10/26/1998 | 56 | 44 | 63 |
| 10/26/1998 | 70 | 67 | 69 |
| 10/26/1998 | 20 | 25 | 21 |
| 10/26/1998 | 184 | 186 | 198 |
| 10/26/1998 | 157 | 189 | 180 |
| 10/26/1998 | 50 | 49 | 52 |
| 10/26/1998 | 80 | 89 | 79 |
| 10/26/1998 | 15 | 16 | 19 |
| 10/30/1998 | 160 | 175 | 150 |
| 10/30/1998 | 152 | 148 | 162 |
| 10/30/1998 | 190 | 182 | 199 |
| 10/30/1998 | 158 | 165 | 172 |
| 10/30/1998 | 37 | 47 | 29 |
| 10/30/1998 | 99 | 105 | 111 |
| 10/30/1998 | 56 | 63 | 70 |
| 10/30/1998 | 25 | 30 | 36 |

| Date | | | |
|---|---|---|---|
| 10/30/1998 | 21 | 23 | 26 |
| 10/30/1998 | 63 | 70 | 57 |
| 11/2/1998 | 110 | 102 | 101 |
| 11/2/1998 | 103 | 90 | 95 |
| 11/2/1998 | 121 | 117 | 107 |
| 11/2/1998 | 20 | 23 | 27 |
| 11/2/1998 | 107 | 123 | 118 |
| 11/2/1998 | 125 | 120 | 129 |
| 11/2/1998 | 100 | 115 | 92 |
| 11/2/1998 | 131 | 126 | 116 |
| 11/2/1998 | 30 | 27 | 24 |
| 11/2/1998 | 68 | 64 | 59 |
| 11/6/1998 | 124 | 128 | 123 |
| 11/6/1998 | 164 | 153 | 174 |
| 11/6/1998 | 74 | 85 | 72 |
| 11/6/1998 | 55 | 57 | 70 |
| 11/6/1998 | 33 | 34 | 42 |
| 11/6/1998 | 108 | 115 | 111 |
| 11/6/1998 | 34 | 42 | 40 |
| 11/6/1998 | 30 | 39 | 43 |
| 11/6/1998 | 14 | 18 | 19 |
| 11/6/1998 | 2 | 3 | 4 |
| 11/16/1998 | 98 | 118 | 110 |
| 11/16/1998 | 142 | 154 | 129 |
| 11/16/1998 | 96 | 91 | 90 |
| 11/16/1998 | 67 | 71 | 69 |
| 11/16/1998 | 44 | 46 | 42 |
| 11/16/1998 | 23 | 30 | 28 |
| 11/16/1998 | 70 | 63 | 68 |
| 11/16/1998 | 36 | 33 | 30 |
| 11/16/1998 | 91 | 96 | 94 |
| 11/16/1998 | 51 | 46 | 42 |
| 11/20/1998 | 145 | 159 | 152 |
| 11/20/1998 | 135 | 147 | 140 |
| 11/20/1998 | 58 | 50 | 66 |
| 11/20/1998 | 23 | 24 | 22 |
| 11/20/1998 | 11 | 12 | 14 |
| 11/20/1998 | 162 | 169 | 170 |
| 11/20/1998 | 159 | 149 | 161 |
| 11/20/1998 | 38 | 44 | 40 |
| 11/20/1998 | 56 | 66 | 52 |
| 11/20/1998 | 39 | 38 | 34 |
| 11/23/1998 | 112 | 95 | 102 |
| 11/23/1998 | 89 | 79 | 90 |
| 11/23/1998 | 69 | 61 | 55 |
| 11/23/1998 | 36 | 41 | 39 |
| 11/23/1998 | 43 | 40 | 31 |
| 11/23/1998 | 33 | 28 | 29 |
| 11/23/1998 | 25 | 23 | 21 |
| 11/23/1998 | 22 | 24 | 16 |
| 11/23/1998 | 146 | 149 | 139 |
| 11/23/1998 | 90 | 74 | 78 |
| 11/30/1998 | 129 | 121 | 135 |
| 11/30/1998 | 112 | 139 | 109 |
| 11/30/1998 | 107 | 99 | 92 |
| 11/30/1998 | 50 | 57 | 42 |
| 11/30/1998 | 123 | 130 | 116 |
| 11/30/1998 | 145 | 155 | 139 |
| 11/30/1998 | 121 | 132 | 135 |
| 11/30/1998 | 57 | 48 | 46 |
| 11/30/1998 | 120 | 109 | 99 |
| 11/30/1998 | 15 | 19 | 13 |
| 12/11/1998 | 120 | 111 | 99 |
| 12/11/1998 | 131 | 121 | 117 |
| 12/11/1998 | 104 | 114 | 96 |
| 12/11/1998 | 93 | 100 | 86 |
| 12/11/1998 | 70 | 78 | 63 |
| 12/11/1998 | 58 | 68 | 49 |
| 12/11/1998 | 29 | 22 | 19 |
| 12/11/1998 | 104 | 115 | 94 |
| 12/11/1998 | 116 | 126 | 107 |
| 12/11/1998 | 23 | 29 | 18 |
| 12/14/1998 | 183 | 194 | 201 |
| 12/14/1998 | 142 | 160 | 162 |
| 12/14/1998 | 114 | 116 | 110 |
| 12/14/1998 | 75 | 84 | 91 |
| 12/14/1998 | 54 | 50 | 45 |
| 12/14/1998 | 99 | 120 | 119 |
| 12/14/1998 | 32 | 20 | 26 |
| 12/14/1998 | 52 | 44 | 49 |
| 12/14/1998 | 21 | 20 | 17 |
| 12/14/1998 | 20 | 15 | 14 |
| 12/17/1998 | 135 | 145 | 147 |
| 12/17/1998 | 122 | 130 | 119 |

| Date | | | |
|---|---|---|---|
| 12/17/1998 | 31 | 27 | 35 |
| 12/17/1998 | 20 | 24 | 18 |
| 12/17/1998 | 123 | 134 | 114 |
| 12/17/1998 | 165 | 152 | 158 |
| 12/17/1998 | 137 | 129 | 141 |
| 12/17/1998 | 58 | 62 | 54 |
| 12/17/1998 | 25 | 27 | 29 |
| 12/17/1998 | 19 | 22 | 17 |
| 12/21/1998 | 143 | 159 | 138 |
| 12/21/1998 | 126 | 129 | 141 |
| 12/21/1998 | 140 | 135 | 129 |
| 12/21/1998 | 132 | 129 | 124 |
| 12/21/1998 | 106 | 113 | 119 |
| 12/21/1998 | 100 | 92 | 85 |
| 12/21/1998 | 50 | 43 | 36 |
| 12/21/1998 | 20 | 17 | 19 |
| 12/21/1998 | 79 | 70 | 62 |
| 12/21/1998 | 49 | 38 | 30 |
| 12/21/1998 | 156 | 166 | 142 |
| 12/21/1998 | 139 | 149 | 129 |
| 12/21/1998 | 137 | 130 | 123 |
| 12/21/1998 | 132 | 140 | 151 |
| 12/21/1998 | 121 | 125 | 129 |
| 12/21/1998 | 108 | 100 | 117 |
| 12/21/1998 | 83 | 91 | 99 |
| 12/21/1998 | 56 | 64 | 52 |
| 12/21/1998 | 76 | 70 | 82 |
| 12/21/1998 | 79 | 72 | 65 |

| Date | | | |
|---|---|---|---|
| 1/8/1999 | 99 | 89 | 91 |
| 1/8/1999 | 80 | 74 | 68 |
| 1/8/1999 | 65 | 71 | 59 |
| 1/8/1999 | 40 | 48 | 57 |
| 1/8/1999 | 39 | 34 | 37 |
| 1/8/1999 | 28 | 26 | 23 |
| 1/8/1999 | 102 | 110 | 92 |
| 1/8/1999 | 33 | 47 | 40 |
| 1/8/1999 | 90 | 84 | 79 |
| 1/15/1999 | 172 | 161 | 167 |
| 1/15/1999 | 156 | 149 | 161 |
| 1/15/1999 | 133 | 140 | 147 |
| 1/15/1999 | 101 | 112 | 90 |
| 1/15/1999 | 80 | 86 | 92 |
| 1/15/1999 | 60 | 69 | 65 |
| 1/15/1999 | 25 | 32 | 40 |
| 1/15/1999 | 26 | 29 | 23 |
| 1/15/1999 | 19 | 22 | 15 |
| 1/15/1999 | 158 | 162 | 140 |
| 1/22/1999 | 145 | 139 | 121 |
| 1/22/1999 | 129 | 119 | 137 |
| 1/22/1999 | 130 | 135 | 124 |
| 1/22/1999 | 125 | 115 | 109 |
| 1/22/1999 | 95 | 101 | 107 |
| 1/22/1999 | 85 | 92 | 78 |
| 1/22/1999 | 59 | 69 | 78 |
| 1/22/1999 | 49 | 58 | 67 |
| 1/22/1999 | 38 | 44 | 33 |

| Date | | | |
|---|---|---|---|
| 1/22/1999 | 100 | 95 | 106 |
| 1/25/1999 | 176 | 162 | 155 |
| 1/25/1999 | 135 | 145 | 155 |
| 1/25/1999 | 128 | 135 | 142 |
| 1/25/1999 | 110 | 127 | 97 |
| 1/25/1999 | 96 | 108 | 85 |
| 1/25/1999 | 44 | 46 | 39 |
| 1/25/1999 | 155 | 162 | 148 |
| 1/25/1999 | 63 | 70 | 56 |
| 1/25/1999 | 120 | 112 | 104 |
| 1/25/1999 | 77 | 68 | 60 |
| 1/29/1999 | 152 | 169 | 159 |
| 1/29/1999 | 141 | 146 | 150 |
| 1/29/1999 | 133 | 138 | 143 |
| 1/29/1999 | 110 | 117 | 124 |
| 1/29/1999 | 92 | 100 | 107 |
| 1/29/1999 | 43 | 48 | 54 |
| 1/29/1999 | 36 | 42 | 50 |
| 1/29/1999 | 63 | 70 | 77 |
| 1/29/1999 | 116 | 126 | 107 |
| 1/29/1999 | 73 | 80 | 88 |
| 2/5/1999 | 150 | 135 | 139 |
| 2/5/1999 | 129 | 140 | 131 |
| 2/5/1999 | 107 | 100 | 98 |
| 2/5/1999 | 111 | 109 | 107 |
| 2/5/1999 | 35 | 40 | 29 |
| 2/5/1999 | 18 | 20 | 16 |
| 2/5/1999 | 13 | 13 | 14 |

| Date | | | |
|---|---|---|---|
| 2/5/1999 | 40 | 47 | 32 |
| 2/5/1999 | 146 | 157 | 167 |
| 2/5/1999 | 14 | 16 | 15 |
| 2/12/1999 | 120 | 135 | 118 |
| 2/12/1999 | 125 | 137 | 113 |
| 2/12/1999 | 115 | 120 | 129 |
| 2/12/1999 | 114 | 104 | 124 |
| 2/12/1999 | 112 | 118 | 123 |
| 2/12/1999 | 141 | 150 | 160 |
| 2/12/1999 | 219 | 225 | 232 |
| 2/12/1999 | 37 | 42 | 33 |
| 2/12/1999 | 42 | 46 | 50 |
| 2/12/1999 | 50 | 57 | 65 |
| 2/22/1999 | 130 | 149 | 142 |
| 2/22/1999 | 131 | 137 | 143 |
| 2/22/1999 | 123 | 131 | 138 |
| 2/22/1999 | 128 | 134 | 140 |
| 2/22/1999 | 125 | 130 | 136 |
| 2/22/1999 | 115 | 126 | 137 |
| 2/22/1999 | 17 | 20 | 24 |
| 2/22/1999 | 29 | 35 | 41 |
| 2/22/1999 | 62 | 70 | 54 |
| 2/22/1999 | 70 | 79 | 62 |
| 2/26/1999 | 129 | 131 | 120 |
| 2/26/1999 | 115 | 109 | 117 |
| 2/26/1999 | 111 | 115 | 120 |
| 2/26/1999 | 103 | 113 | 94 |
| 2/26/1999 | 89 | 94 | 100 |

| Date | | | |
|---|---|---|---|
| 2/26/1999 | 66 | 75 | 85 |
| 2/26/1999 | 56 | 63 | 70 |
| 2/26/1999 | 53 | 58 | 64 |
| 2/26/1999 | 54 | 56 | 53 |
| 2/26/1999 | 38 | 46 | 55 |
| 3/1/1999 | 145 | 137 | 129 |
| 3/1/1999 | 121 | 115 | 135 |
| 3/1/1999 | 116 | 125 | 134 |
| 3/1/1999 | 108 | 114 | 119 |
| 3/1/1999 | 98 | 107 | 117 |
| 3/1/1999 | 80 | 90 | 99 |
| 3/1/1999 | 78 | 84 | 90 |
| 3/1/1999 | 61 | 70 | 78 |
| 3/1/1999 | 65 | 75 | 56 |
| 3/1/1999 | 37 | 27 | 48 |
| 3/5/1999 | 145 | 165 | 153 |
| 3/5/1999 | 140 | 147 | 153 |
| 3/5/1999 | 139 | 148 | 157 |
| 3/5/1999 | 123 | 130 | 137 |
| 3/5/1999 | 109 | 118 | 128 |
| 3/5/1999 | 145 | 154 | 164 |
| 3/5/1999 | 67 | 75 | 84 |
| 3/5/1999 | 72 | 88 | 56 |
| 3/5/1999 | 58 | 66 | 50 |
| 3/5/1999 | 66 | 75 | 56 |
| 3/8/1999 | 155 | 160 | 149 |
| 3/8/1999 | 145 | 139 | 152 |
| 3/8/1999 | 21 | 20 | 17 |

| Date | | | |
|---|---|---|---|
| 3/8/1999 | 23 | 26 | 21 |
| 3/8/1999 | 2 | 2 | 1 |
| 3/8/1999 | 174 | 185 | 170 |
| 3/8/1999 | 169 | 172 | 177 |
| 3/8/1999 | 25 | 27 | 29 |
| 3/8/1999 | 26 | 30 | 35 |
| 3/8/1999 | 2 | 3 | 3 |
| 3/8/1999 | 182 | 169 | 178 |
| 3/8/1999 | 177 | 161 | 163 |
| 3/8/1999 | 26 | 31 | 37 |
| 3/8/1999 | 22 | 26 | 29 |
| 3/8/1999 | 2 | 2 | 2 |
| 3/12/1999 | 160 | 155 | 168 |
| 3/12/1999 | 152 | 157 | 161 |
| 3/12/1999 | 17 | 22 | 27 |
| 3/12/1999 | 23 | 30 | 38 |
| 3/12/1999 | 3 | 2 | 4 |
| 3/12/1999 | 156 | 142 | 150 |
| 3/12/1999 | 143 | 149 | 145 |
| 3/12/1999 | 18 | 25 | 30 |
| 3/12/1999 | 14 | 19 | 25 |
| 3/12/1999 | 3 | 4 | 5 |
| 3/12/1999 | 160 | 167 | 172 |
| 3/12/1999 | 151 | 149 | 165 |
| 3/12/1999 | 20 | 24 | 28 |
| 3/12/1999 | 23 | 27 | 35 |
| 3/12/1999 | 3 | 3 | 2 |
| 3/19/1999 | 150 | 142 | 149 |

| Date | | | |
|---|---|---|---|
| 3/19/1999 | 135 | 149 | 120 |
| 3/19/1999 | 135 | 145 | 126 |
| 3/19/1999 | 105 | 115 | 127 |
| 3/19/1999 | 95 | 104 | 114 |
| 3/19/1999 | 77 | 84 | 94 |
| 3/19/1999 | 58 | 67 | 77 |
| 3/19/1999 | 45 | 53 | 61 |
| 3/19/1999 | 47 | 35 | 23 |
| 3/19/1999 | 85 | 92 | 77 |
| 3/29/1999 | 165 | 151 | 149 |
| 3/29/1999 | 145 | 157 | 150 |
| 3/29/1999 | 155 | 136 | 146 |
| 3/29/1999 | 128 | 135 | 142 |
| 3/29/1999 | 104 | 116 | 127 |
| 3/29/1999 | 91 | 98 | 107 |
| 3/29/1999 | 63 | 69 | 77 |
| 3/29/1999 | 53 | 61 | 70 |
| 3/29/1999 | 26 | 32 | 40 |
| 3/29/1999 | 114 | 122 | 106 |
| 4/9/1999 | 140 | 155 | 159 |
| 4/9/1999 | 132 | 139 | 149 |
| 4/9/1999 | 131 | 129 | 125 |
| 4/9/1999 | 152 | 147 | 130 |
| 4/9/1999 | 116 | 126 | 107 |
| 4/9/1999 | 104 | 112 | 96 |
| 4/9/1999 | 75 | 86 | 97 |
| 4/9/1999 | 73 | 80 | 66 |
| 4/9/1999 | 47 | 56 | 65 |

| Date | | | |
|---|---|---|---|
| 4/9/1999 | 32 | 37 | 42 |
| 5/14/1999 | 122 | 135 | 119 |
| 5/14/1999 | 140 | 129 | 137 |
| 5/14/1999 | 120 | 125 | 131 |
| 5/14/1999 | 104 | 111 | 97 |
| 5/14/1999 | 93 | 100 | 88 |
| 5/14/1999 | 82 | 88 | 93 |
| 5/14/1999 | 70 | 79 | 89 |
| 5/14/1999 | 48 | 61 | 54 |
| 5/14/1999 | 30 | 35 | 41 |
| 5/14/1999 | 16 | 20 | 25 |
| 5/21/1999 | 140 | 135 | 129 |
| 5/21/1999 | 119 | 137 | 121 |
| 5/21/1999 | 122 | 127 | 135 |
| 5/21/1999 | 95 | 110 | 103 |
| 5/21/1999 | 86 | 96 | 106 |
| 5/21/1999 | 74 | 96 | 85 |
| 5/21/1999 | 43 | 50 | 57 |
| 5/21/1999 | 43 | 32 | 22 |
| 5/21/1999 | 18 | 26 | 21 |
| 5/21/1999 | 103 | 123 | 114 |
| 6/7/1999 | 111 | 107 | 119 |
| 6/7/1999 | 132 | 125 | 115 |
| 6/7/1999 | 106 | 120 | 113 |
| 6/7/1999 | 96 | 105 | 115 |
| 6/7/1999 | 100 | 117 | 84 |
| 6/7/1999 | 104 | 93 | 82 |
| 6/7/1999 | 86 | 63 | 75 |
| 6/7/1999 | 62 | 48 | 50 |
| 6/7/1999 | 55 | 62 | 48 |
| 6/7/1999 | 47 | 23 | 34 |
| 6/18/1999 | 135 | 149 | 152 |
| 6/18/1999 | 125 | 141 | 121 |
| 6/18/1999 | 88 | 95 | 102 |
| 6/18/1999 | 85 | 96 | 73 |
| 6/18/1999 | 32 | 42 | 49 |
| 6/18/1999 | 15 | 20 | 26 |
| 6/18/1999 | 123 | 130 | 117 |
| 6/18/1999 | 89 | 95 | 101 |
| 6/18/1999 | 48 | 57 | 67 |
| 6/18/1999 | 82 | 92 | 87 |
| 6/28/1999 | 149 | 162 | 151 |
| 6/28/1999 | 140 | 135 | 122 |
| 6/28/1999 | 131 | 123 | 117 |
| 6/28/1999 | 103 | 109 | 115 |
| 6/28/1999 | 105 | 115 | 96 |
| 6/28/1999 | 96 | 103 | 110 |
| 6/28/1999 | 111 | 93 | 85 |
| 6/28/1999 | 69 | 75 | 82 |
| 6/28/1999 | 45 | 38 | 31 |
| 6/28/1999 | 23 | 27 | 32 |
| 7/5/1999 | 121 | 132 | 145 |
| 7/5/1999 | 119 | 117 | 111 |
| 7/5/1999 | 118 | 108 | 99 |
| 7/5/1999 | 96 | 89 | 83 |
| 7/5/1999 | 82 | 87 | 77 |

PAGE 61: Statistical Detection of Potentially Fabricated Data: A Case Study| Date | | | |
|---|---|---|---|
| 7/5/1999 | 73 | 80 | 67 |
| 7/5/1999 | 44 | 53 | 35 |
| 7/5/1999 | 23 | 19 | 15 |
| 7/5/1999 | 110 | 131 | 120 |
| 7/5/1999 | 73 | 77 | 82 |
| 7/8/1999 | 120 | 115 | 130 |
| 7/8/1999 | 115 | 107 | 119 |
| 7/8/1999 | 117 | 107 | 98 |
| 7/8/1999 | 98 | 77 | 87 |
| 7/8/1999 | 73 | 80 | 67 |
| 7/8/1999 | 57 | 73 | 65 |
| 7/8/1999 | 62 | 55 | 49 |
| 7/8/1999 | 48 | 54 | 60 |
| 7/8/1999 | 29 | 38 | 48 |
| 7/8/1999 | 16 | 27 | 22 |
| 7/19/1999 | 130 | 149 | 122 |
| 7/19/1999 | 142 | 129 | 115 |
| 7/19/1999 | 114 | 122 | 130 |
| 7/19/1999 | 98 | 83 | 90 |
| 7/19/1999 | 76 | 86 | 67 |
| 7/19/1999 | 63 | 67 | 72 |
| 7/19/1999 | 47 | 56 | 66 |
| 7/19/1999 | 47 | 53 | 41 |
| 7/19/1999 | 32 | 41 | 36 |
| 7/19/1999 | 128 | 152 | 140 |
| 7/23/1999 | 125 | 132 | 119 |
| 7/23/1999 | 145 | 130 | 125 |
| 7/23/1999 | 116 | 125 | 135 |

| Date | | | |
|---|---|---|---|
| 7/23/1999 | 99 | 106 | 113 |
| 7/23/1999 | 89 | 100 | 112 |
| 7/23/1999 | 99 | 84 | 91 |
| 7/23/1999 | 72 | 90 | 81 |
| 7/23/1999 | 67 | 73 | 80 |
| 7/23/1999 | 75 | 82 | 68 |
| 7/23/1999 | 50 | 41 | 32 |
| 8/2/1999 | 135 | 122 | 119 |
| 8/2/1999 | 118 | 107 | 129 |
| 8/2/1999 | 113 | 103 | 92 |
| 8/2/1999 | 101 | 93 | 86 |
| 8/2/1999 | 76 | 82 | 70 |
| 8/2/1999 | 38 | 29 | 47 |
| 8/2/1999 | 16 | 21 | 26 |
| 8/2/1999 | 98 | 107 | 88 |
| 8/2/1999 | 18 | 21 | 25 |
| 8/2/1999 | 31 | 25 | 20 |
| 8/6/1999 | 110 | 107 | 99 |
| 8/6/1999 | 120 | 115 | 109 |
| 8/6/1999 | 111 | 103 | 96 |
| 8/6/1999 | 93 | 100 | 87 |
| 8/6/1999 | 82 | 90 | 99 |
| 8/6/1999 | 69 | 88 | 78 |
| 8/6/1999 | 53 | 59 | 65 |
| 8/6/1999 | 60 | 54 | 49 |
| 8/6/1999 | 41 | 48 | 55 |
| 8/6/1999 | 23 | 24 | 27 |
| 8/9/1999 | 130 | 145 | 155 |

| Date | | | |
|---|---|---|---|
| 8/9/1999 | 129 | 151 | 135 |
| 8/9/1999 | 125 | 134 | 143 |
| 8/9/1999 | 124 | 130 | 118 |
| 8/9/1999 | 109 | 115 | 122 |
| 8/9/1999 | 92 | 108 | 101 |
| 8/9/1999 | 51 | 60 | 69 |
| 8/9/1999 | 49 | 54 | 60 |
| 8/9/1999 | 39 | 45 | 52 |
| 8/9/1999 | 115 | 126 | 137 |
| 9/6/1999 | 160 | 151 | 142 |
| 9/6/1999 | 135 | 149 | 129 |
| 9/6/1999 | 137 | 127 | 117 |
| 9/6/1999 | 69 | 81 | 92 |
| 9/6/1999 | 46 | 35 | 55 |
| 9/6/1999 | 147 | 159 | 168 |
| 9/6/1999 | 22 | 29 | 36 |
| 9/6/1999 | 39 | 44 | 49 |
| 9/6/1999 | 11 | 9 | 7 |
| 9/10/1999 | 159 | 167 | 147 |
| 9/10/1999 | 129 | 140 | 137 |
| 9/10/1999 | 125 | 132 | 140 |
| 9/10/1999 | 95 | 102 | 110 |
| 9/10/1999 | 40 | 47 | 54 |
| 9/10/1999 | 23 | 28 | 33 |
| 9/10/1999 | 79 | 88 | 98 |
| 9/10/1999 | 37 | 52 | 33 |
| 9/10/1999 | 7 | 9 | 11 |
| 9/10/1999 | 3 | 4 | 5 |

| Date | | | |
|---|---|---|---|
| 9/20/1999 | 85 | 72 | 92 |
| 9/20/1999 | 15 | 21 | 12 |
| 9/20/1999 | 7 | 5 | 6 |
| 9/20/1999 | 3 | 1 | 4 |
| 10/1/1999 | 140 | 166 | 152 |
| 10/1/1999 | 135 | 149 | 157 |
| 10/1/1999 | 52 | 63 | 73 |
| 10/1/1999 | 36 | 41 | 31 |
| 10/1/1999 | 53 | 60 | 68 |
| 10/1/1999 | 121 | 115 | 139 |
| 10/1/1999 | 137 | 152 | 144 |
| 10/1/1999 | 69 | 75 | 82 |
| 10/1/1999 | 47 | 56 | 66 |
| 10/1/1999 | 100 | 111 | 121 |
| 10/4/1999 | 165 | 141 | 153 |
| 10/4/1999 | 135 | 152 | 149 |
| 10/4/1999 | 129 | 134 | 140 |
| 10/4/1999 | 109 | 119 | 128 |
| 10/4/1999 | 108 | 97 | 87 |
| 10/4/1999 | 67 | 74 | 61 |
| 10/4/1999 | 20 | 30 | 41 |
| 10/4/1999 | 60 | 68 | 76 |
| 10/4/1999 | 40 | 30 | 21 |
| 10/4/1999 | 6 | 9 | 11 |
| 10/11/1999 | 150 | 166 | 149 |
| 10/11/1999 | 137 | 129 | 152 |
| 10/11/1999 | 121 | 137 | 145 |
| 10/11/1999 | 152 | 130 | 119 |

| Date | | | |
|---|---|---|---|
| 10/11/1999 | 117 | 125 | 139 |
| 10/11/1999 | 165 | 147 | 155 |
| 10/11/1999 | 141 | 159 | 139 |
| 10/11/1999 | 129 | 125 | 138 |
| 10/11/1999 | 147 | 152 | 118 |
| 10/11/1999 | 167 | 145 | 149 |
| 10/22/1999 | 137 | 123 | 145 |
| 10/22/1999 | 119 | 127 | 131 |
| 10/22/1999 | 132 | 126 | 121 |
| 10/22/1999 | 106 | 115 | 124 |
| 10/22/1999 | 99 | 107 | 116 |
| 10/22/1999 | 91 | 95 | 100 |
| 10/22/1999 | 72 | 65 | 59 |
| 10/22/1999 | 32 | 49 | 60 |
| 10/22/1999 | 24 | 30 | 19 |
| 10/22/1999 | 91 | 99 | 84 |
| 11/9/1999 | 134 | 144 | 146 |
| 11/9/1999 | 112 | 100 | 116 |
| 11/9/1999 | 95 | 99 | 102 |
| 11/9/1999 | 70 | 72 | 76 |
| 11/9/1999 | 34 | 40 | 25 |
| 11/9/1999 | 91 | 90 | 82 |
| 11/9/1999 | 43 | 45 | 46 |
| 11/9/1999 | 16 | 15 | 23 |
| 11/15/1999 | 156 | 166 | 149 |
| 11/15/1999 | 139 | 149 | 152 |
| 11/15/1999 | 137 | 146 | 157 |
| 11/15/1999 | 145 | 131 | 138 |

| Date | | | |
|---|---|---|---|
| 11/15/1999 | 137 | 126 | 116 |
| 11/15/1999 | 115 | 112 | 130 |
| 11/15/1999 | 95 | 102 | 88 |
| 11/15/1999 | 68 | 87 | 76 |
| 11/15/1999 | 68 | 77 | 72 |
| 11/15/1999 | 40 | 32 | 24 |
| 11/22/1999 | 102 | 96 | 108 |
| 11/22/1999 | 90 | 88 | 94 |
| 11/22/1999 | 81 | 98 | 89 |
| 11/22/1999 | 79 | 81 | 85 |
| 11/22/1999 | 72 | 69 | 74 |
| 11/22/1999 | 59 | 53 | 51 |
| 11/22/1999 | 293 | 283 | 299 |
| 11/22/1999 | 103 | 111 | 106 |
| 11/22/1999 | 54 | 63 | 50 |
| 11/22/1999 | 55 | 57 | 44 |
| 12/13/1999 | 170 | 181 | 185 |
| 12/13/1999 | 169 | 179 | 178 |
| 12/13/1999 | 108 | 99 | 117 |
| 12/13/1999 | 45 | 54 | 64 |
| 12/13/1999 | 86 | 90 | 95 |
| 12/13/1999 | 31 | 36 | 42 |
| 12/13/1999 | 3 | 4 | 5 |
| 12/13/1999 | 1 | 1 | 0 |
| 12/17/1999 | 148 | 159 | 162 |
| 12/17/1999 | 165 | 172 | 157 |
| 12/17/1999 | 123 | 129 | 136 |
| 12/17/1999 | 58 | 68 | 79 |

| Date | | | |
|---|---|---|---|
| 12/17/1999 | 129 | 118 | 137 |
| 12/17/1999 | 81 | 90 | 72 |
| 12/17/1999 | 12 | 21 | 16 |
| 12/17/1999 | 135 | 149 | 151 |
| 12/17/1999 | 141 | 129 | 137 |
| 12/17/1999 | 123 | 117 | 112 |
| 12/17/1999 | 73 | 82 | 92 |
| 12/17/1999 | 23 | 26 | 20 |
| 12/17/1999 | 39 | 50 | 42 |
| 12/17/1999 | 99 | 112 | 88 |
| 12/28/1999 | 155 | 165 | 142 |
| 12/28/1999 | 135 | 129 | 145 |
| 12/28/1999 | 49 | 58 | 68 |
| 12/28/1999 | 89 | 101 | 110 |
| 12/28/1999 | 68 | 62 | 75 |
| 12/28/1999 | 12 | 15 | 9 |
| 12/28/1999 | 5 | 7 | 9 |
| 12/28/1999 | 175 | 185 | 181 |
| 12/28/1999 | 169 | 173 | 160 |
| 12/28/1999 | 51 | 56 | 62 |
| 12/28/1999 | 112 | 104 | 119 |
| 12/28/1999 | 75 | 87 | 64 |
| 12/28/1999 | 27 | 35 | 44 |
| 12/28/1999 | 9 | 14 | 20 |
| 1/7/2000 | 157 | 162 | 149 |
| 1/7/2000 | 168 | 172 | 159 |
| 1/7/2000 | 106 | 99 | 93 |
| 1/7/2000 | 98 | 87 | 77 |

PAGE 68: Statistical Detection of Potentially Fabricated Data: A Case Study| Date | | | |
|---|---|---|---|
| 1/7/2000 | 79 | 80 | 92 |
| 1/7/2000 | 37 | 32 | 28 |
| 1/7/2000 | 39 | 48 | 57 |
| 1/7/2000 | 142 | 138 | 132 |
| 1/7/2000 | 131 | 129 | 119 |
| 1/7/2000 | 36 | 44 | 51 |
| 1/7/2000 | 18 | 26 | 37 |
| 1/7/2000 | 40 | 47 | 34 |
| 1/7/2000 | 132 | 139 | 118 |
| 1/7/2000 | 20 | 16 | 13 |
| 1/13/2000 | 150 | 161 | 147 |
| 1/13/2000 | 139 | 151 | 159 |
| 1/13/2000 | 84 | 90 | 97 |
| 1/13/2000 | 40 | 45 | 51 |
| 1/13/2000 | 12 | 15 | 13 |
| 1/13/2000 | 17 | 15 | 14 |
| 1/13/2000 | 17 | 23 | 30 |
| 1/13/2000 | 165 | 160 | 179 |
| 1/13/2000 | 160 | 149 | 152 |
| 1/13/2000 | 48 | 56 | 62 |
| 1/13/2000 | 27 | 33 | 40 |
| 1/13/2000 | 15 | 17 | 19 |
| 1/13/2000 | 19 | 26 | 32 |
| 1/13/2000 | 1 | 2 | 1 |
| 1/20/2000 | 140 | 148 | 137 |
| 1/20/2000 | 161 | 152 | 157 |
| 1/20/2000 | 105 | 112 | 120 |
| 1/20/2000 | 67 | 72 | 63 |

| Date | | | |
|---|---|---|---|
| 1/20/2000 | 13 | 20 | 16 |
| 1/20/2000 | 24 | 37 | 30 |
| 1/20/2000 | 38 | 43 | 49 |
| 1/20/2000 | 169 | 152 | 160 |
| 1/20/2000 | 142 | 138 | 129 |
| 1/20/2000 | 95 | 104 | 114 |
| 1/20/2000 | 35 | 44 | 54 |
| 1/20/2000 | 60 | 68 | 53 |
| 1/20/2000 | 60 | 67 | 52 |
| 1/20/2000 | 11 | 7 | 5 |
| 1/21/2000 | 158 | 148 | 152 |
| 1/21/2000 | 141 | 139 | 156 |
| 1/21/2000 | 130 | 138 | 123 |
| 1/21/2000 | 49 | 60 | 54 |
| 1/21/2000 | 97 | 103 | 91 |
| 1/21/2000 | 67 | 58 | 76 |
| 1/21/2000 | 16 | 13 | 11 |
| 1/21/2000 | 132 | 142 | 129 |
| 1/21/2000 | 140 | 128 | 131 |
| 1/21/2000 | 113 | 120 | 128 |
| 1/21/2000 | 77 | 83 | 90 |
| 1/21/2000 | 13 | 16 | 20 |
| 1/21/2000 | 43 | 51 | 60 |
| 1/21/2000 | 80 | 90 | 71 |
| 1/31/2000 | 168 | 157 | 149 |
| 1/31/2000 | 151 | 142 | 138 |
| 1/31/2000 | 97 | 105 | 115 |
| 1/31/2000 | 75 | 64 | 84 |

| Date | | | |
|---|---|---|---|
| 1/31/2000 | 36 | 45 | 52 |
| 1/31/2000 | 32 | 25 | 40 |
| 1/31/2000 | 62 | 75 | 89 |
| 1/31/2000 | 170 | 158 | 162 |
| 1/31/2000 | 147 | 156 | 160 |
| 1/31/2000 | 92 | 109 | 100 |
| 1/31/2000 | 64 | 69 | 77 |
| 1/31/2000 | 26 | 35 | 45 |
| 1/31/2000 | 12 | 15 | 20 |
| 1/31/2000 | 24 | 32 | 41 |
| 2/14/2000 | 157 | 139 | 142 |
| 2/14/2000 | 88 | 96 | 105 |
| 2/14/2000 | 23 | 25 | 27 |
| 2/14/2000 | 95 | 103 | 112 |
| 2/14/2000 | 83 | 92 | 87 |
| 2/14/2000 | 132 | 141 | 129 |
| 2/14/2000 | 99 | 108 | 118 |
| 2/14/2000 | 71 | 78 | 96 |
| 2/14/2000 | 27 | 33 | 40 |
| 2/14/2000 | 130 | 120 | 109 |
| 3/6/2000 | 126 | 139 | 119 |
| 3/6/2000 | 109 | 120 | 115 |
| 3/6/2000 | 104 | 115 | 124 |
| 3/6/2000 | 98 | 107 | 117 |
| 3/6/2000 | 100 | 109 | 92 |
| 3/6/2000 | 95 | 88 | 102 |
| 3/6/2000 | 80 | 89 | 72 |
| 3/6/2000 | 67 | 87 | 76 |

| Date | | | |
|---|---|---|---|
| 3/6/2000 | 76 | 83 | 90 |
| 3/6/2000 | 68 | 79 | 59 |
| 3/6/2000 | 43 | 51 | 61 |
| 3/6/2000 | 37 | 44 | 52 |
| 3/31/2000 | 145 | 155 | 139 |
| 3/31/2000 | 150 | 161 | 172 |
| 3/31/2000 | 132 | 136 | 140 |
| 3/31/2000 | 130 | 142 | 123 |
| 3/31/2000 | 120 | 127 | 114 |
| 3/31/2000 | 106 | 116 | 97 |
| 3/31/2000 | 83 | 90 | 77 |
| 3/31/2000 | 59 | 65 | 72 |
| 3/31/2000 | 27 | 34 | 42 |
| 3/31/2000 | 13 | 17 | 25 |
| 4/7/2000 | 135 | 145 | 152 |
| 4/7/2000 | 129 | 121 | 119 |
| 4/7/2000 | 100 | 109 | 119 |
| 4/7/2000 | 69 | 75 | 82 |
| 4/7/2000 | 87 | 93 | 100 |
| 4/7/2000 | 58 | 67 | 77 |
| 4/7/2000 | 107 | 117 | 98 |
| 4/7/2000 | 33 | 55 | 45 |
| 4/7/2000 | 19 | 24 | 30 |
| 4/7/2000 | 5 | 7 | 9 |
| 6/23/2000 | 172 | 162 | 159 |
| 6/23/2000 | 153 | 160 | 149 |
| 6/23/2000 | 146 | 157 | 138 |
| 6/23/2000 | 125 | 110 | 97 |
| 6/23/2000 | 65 | 60 | 72 |
| 6/23/2000 | 30 | 52 | 41 |
| 6/23/2000 | 12 | 16 | 25 |
| 6/23/2000 | 21 | 27 | 18 |
| 6/23/2000 | 3 | 4 | 3 |
| 6/23/2000 | 2 | 2 | 1 |
| 6/30/2000 | 110 | 109 | 117 |
| 6/30/2000 | 97 | 115 | 121 |
| 6/30/2000 | 110 | 100 | 90 |
| 6/30/2000 | 82 | 87 | 78 |
| 6/30/2000 | 59 | 77 | 68 |
| 6/30/2000 | 67 | 58 | 49 |
| 6/30/2000 | 41 | 50 | 33 |
| 6/30/2000 | 67 | 78 | 54 |
| 6/30/2000 | 20 | 28 | 39 |
| 6/30/2000 | 17 | 26 | 12 |
| 7/3/2000 | 135 | 147 | 129 |
| 7/3/2000 | 121 | 111 | 118 |
| 7/3/2000 | 107 | 114 | 124 |
| 7/3/2000 | 110 | 119 | 99 |
| 7/3/2000 | 75 | 67 | 60 |
| 7/3/2000 | 38 | 45 | 56 |
| 7/3/2000 | 32 | 27 | 21 |
| 7/3/2000 | 83 | 101 | 117 |
| 7/3/2000 | 80 | 120 | 160 |
| 7/3/2000 | 11 | 7 | 4 |
| 7/24/2000 | 135 | 161 | 149 |
| 7/24/2000 | 122 | 149 | 151 |
| 7/24/2000 | 97 | 108 | 88 |
| 7/24/2000 | 65 | 75 | 87 |
| 7/24/2000 | 160 | 142 | 176 |
| 7/24/2000 | 35 | 44 | 54 |
| 7/24/2000 | 66 | 73 | 80 |
| 7/24/2000 | 156 | 166 | 169 |
| 7/24/2000 | 161 | 149 | 150 |
| 7/24/2000 | 63 | 82 | 71 |
| 7/24/2000 | 52 | 59 | 66 |
| 7/24/2000 | 91 | 80 | 72 |
| 7/24/2000 | 111 | 121 | 98 |
| 7/24/2000 | 9 | 14 | 20 |
| 7/28/2000 | 135 | 149 | 165 |
| 7/28/2000 | 152 | 119 | 139 |
| 7/28/2000 | 107 | 117 | 95 |
| 7/28/2000 | 64 | 82 | 73 |
| 7/28/2000 | 51 | 60 | 69 |
| 7/28/2000 | 36 | 45 | 27 |
| 7/28/2000 | 21 | 25 | 27 |
| 7/28/2000 | 16 | 19 | 13 |
| 7/28/2000 | 58 | 69 | 49 |
| 7/28/2000 | 10 | 19 | 15 |
| 7/31/2000 | 107 | 119 | 105 |
| 7/31/2000 | 99 | 101 | 89 |
| 7/31/2000 | 73 | 66 | 60 |
| 7/31/2000 | 41 | 49 | 59 |
| 7/31/2000 | 16 | 25 | 20 |
| 7/31/2000 | 39 | 47 | 58 |

| Date | | | |
|---|---|---|---|
| 7/31/2000 | 55 | 64 | 73 |
| 7/31/2000 | 120 | 117 | 110 |
| 7/31/2000 | 107 | 100 | 92 |
| 7/31/2000 | 54 | 63 | 73 |
| 7/31/2000 | 29 | 42 | 35 |
| 7/31/2000 | 24 | 18 | 14 |
| 7/31/2000 | 75 | 84 | 69 |
| 7/31/2000 | 8 | 6 | 4 |
| 8/7/2000 | 111 | 105 | 95 |
| 8/7/2000 | 125 | 135 | 119 |
| 8/7/2000 | 89 | 102 | 117 |
| 8/7/2000 | 82 | 70 | 89 |
| 8/7/2000 | 83 | 74 | 64 |
| 8/7/2000 | 58 | 67 | 49 |
| 8/7/2000 | 18 | 22 | 15 |
| 8/7/2000 | 22 | 19 | 17 |
| 8/7/2000 | 23 | 26 | 30 |
| 8/7/2000 | 12 | 15 | 17 |
| 8/11/2000 | 110 | 117 | 99 |
| 8/11/2000 | 99 | 101 | 102 |
| 8/11/2000 | 46 | 55 | 67 |
| 8/11/2000 | 21 | 31 | 15 |
| 8/11/2000 | 107 | 115 | 125 |
| 8/11/2000 | 73 | 82 | 64 |
| 8/11/2000 | 13 | 16 | 20 |
| 8/11/2000 | 132 | 125 | 118 |
| 8/11/2000 | 117 | 109 | 115 |
| 8/11/2000 | 67 | 77 | 58 |

| Date | | | |
|---|---|---|---|
| 8/11/2000 | 30 | 36 | 24 |
| 8/11/2000 | 18 | 21 | 23 |
| 8/11/2000 | 95 | 107 | 87 |
| 8/11/2000 | 13 | 15 | 18 |
| 8/14/2000 | 131 | 111 | 119 |
| 8/14/2000 | 143 | 129 | 107 |
| 8/14/2000 | 39 | 48 | 58 |
| 8/14/2000 | 22 | 30 | 37 |
| 8/14/2000 | 19 | 20 | 22 |
| 8/14/2000 | 47 | 55 | 65 |
| 8/14/2000 | 90 | 99 | 80 |
| 8/14/2000 | 151 | 149 | 161 |
| 8/14/2000 | 143 | 137 | 129 |
| 8/14/2000 | 38 | 48 | 29 |
| 8/14/2000 | 16 | 18 | 14 |
| 8/14/2000 | 65 | 72 | 80 |
| 8/14/2000 | 24 | 30 | 37 |
| 8/14/2000 | 9 | 7 | 6 |
| 8/14/2000 | 138 | 156 | 121 |
| 8/14/2000 | 129 | 119 | 109 |
| 8/14/2000 | 105 | 95 | 87 |
| 8/14/2000 | 70 | 81 | 68 |
| 8/14/2000 | 19 | 20 | 22 |
| 8/14/2000 | 78 | 68 | 99 |
| 8/14/2000 | 155 | 165 | 148 |
| 8/14/2000 | 151 | 142 | 138 |
| 8/14/2000 | 117 | 129 | 137 |
| 8/14/2000 | 63 | 90 | 76 |
| 8/14/2000 | 58 | 60 | 71 |
| 8/14/2000 | 95 | 107 | 87 |
| 8/14/2000 | 17 | 26 | 37 |
| 8/14/2000 | 48 | 54 | 60 |
| 8/18/2000 | 165 | 155 | 147 |
| 8/18/2000 | 139 | 149 | 141 |
| 8/18/2000 | 115 | 105 | 97 |
| 8/18/2000 | 95 | 84 | 75 |
| 8/18/2000 | 10 | 13 | 15 |
| 8/18/2000 | 43 | 52 | 34 |
| 8/18/2000 | 13 | 30 | 21 |
| 8/18/2000 | 123 | 137 | 141 |
| 8/18/2000 | 111 | 119 | 126 |
| 8/18/2000 | 92 | 82 | 74 |
| 8/18/2000 | 83 | 73 | 62 |
| 8/18/2000 | 69 | 76 | 82 |
| 8/18/2000 | 70 | 63 | 57 |
| 8/18/2000 | 24 | 30 | 19 |
| 9/25/2000 | 111 | 119 | 107 |
| 9/25/2000 | 98 | 89 | 72 |
| 9/25/2000 | 47 | 56 | 66 |
| 9/25/2000 | 43 | 32 | 24 |
| 9/25/2000 | 22 | 24 | 20 |
| 9/25/2000 | 88 | 78 | 97 |
| 9/25/2000 | 22 | 18 | 15 |
| 9/25/2000 | 109 | 113 | 110 |
| 9/25/2000 | 99 | 97 | 89 |
| 9/25/2000 | 55 | 45 | 36 |

| Date | | | |
|---|---|---|---|
| 9/25/2000 | 29 | 26 | 24 |
| 9/25/2000 | 14 | 16 | 12 |
| 9/25/2000 | 15 | 17 | 13 |
| 9/25/2000 | 70 | 63 | 56 |
| 10/2/2000 | 137 | 142 | 157 |
| 10/2/2000 | 119 | 109 | 121 |
| 10/2/2000 | 102 | 112 | 94 |
| 10/2/2000 | 80 | 90 | 72 |
| 10/2/2000 | 11 | 14 | 16 |
| 10/2/2000 | 34 | 40 | 47 |
| 10/2/2000 | 36 | 27 | 19 |
| 10/2/2000 | 117 | 121 | 132 |
| 10/2/2000 | 149 | 151 | 139 |
| 10/2/2000 | 78 | 88 | 99 |
| 10/2/2000 | 55 | 62 | 70 |
| 10/2/2000 | 38 | 44 | 31 |
| 10/2/2000 | 81 | 90 | 72 |
| 10/2/2000 | 29 | 34 | 40 |
| 10/13/2000 | 125 | 119 | 107 |
| 10/13/2000 | 110 | 99 | 129 |
| 10/13/2000 | 23 | 26 | 30 |
| 10/13/2000 | 119 | 129 | 139 |
| 10/13/2000 | 15 | 13 | 12 |
| 10/13/2000 | 9 | 11 | 8 |
| 10/13/2000 | 1 | 2 | 1 |
| 10/13/2000 | 131 | 112 | 109 |
| 10/13/2000 | 107 | 102 | 119 |
| 10/13/2000 | 27 | 33 | 40 |

| Date | | | |
|---|---|---|---|
| 10/13/2000 | 80 | 86 | 73 |
| 10/13/2000 | 23 | 30 | 19 |
| 10/13/2000 | 8 | 7 | 9 |
| 10/13/2000 | 1 | 2 | 3 |
| 10/16/2000 | 150 | 161 | 143 |
| 10/16/2000 | 132 | 141 | 121 |
| 10/16/2000 | 22 | 25 | 27 |
| 10/16/2000 | 115 | 127 | 135 |
| 10/16/2000 | 18 | 22 | 27 |
| 10/16/2000 | 15 | 19 | 13 |
| 10/16/2000 | 3 | 5 | 1 |
| 10/16/2000 | 116 | 132 | 109 |
| 10/16/2000 | 156 | 129 | 139 |
| 10/16/2000 | 31 | 38 | 35 |
| 10/16/2000 | 109 | 111 | 99 |
| 10/16/2000 | 19 | 28 | 22 |
| 10/16/2000 | 12 | 17 | 15 |
| 10/16/2000 | 1 | 4 | 3 |
| 10/23/2000 | 145 | 129 | 120 |
| 10/23/2000 | 111 | 132 | 109 |
| 10/23/2000 | 36 | 37 | 29 |
| 10/23/2000 | 111 | 122 | 99 |
| 10/23/2000 | 148 | 154 | 137 |
| 10/23/2000 | 19 | 22 | 17 |
| 10/23/2000 | 1 | 2 | 2 |
| 10/23/2000 | 155 | 142 | 139 |
| 10/23/2000 | 128 | 136 | 125 |
| 10/23/2000 | 20 | 15 | 19 |

| Date | | | |
|---|---|---|---|
| 10/23/2000 | 70 | 84 | 76 |
| 10/23/2000 | 88 | 74 | 80 |
| 10/23/2000 | 10 | 9 | 7 |
| 10/23/2000 | 1 | 1 | 0 |
| 10/23/2000 | 123 | 152 | 134 |
| 10/23/2000 | 122 | 115 | 129 |
| 10/23/2000 | 28 | 22 | 26 |
| 10/23/2000 | 80 | 77 | 78 |
| 10/23/2000 | 12 | 11 | 11 |
| 10/23/2000 | 10 | 12 | 10 |
| 10/23/2000 | 2 | 1 | 0 |
| 10/23/2000 | 114 | 109 | 128 |
| 10/23/2000 | 107 | 120 | 117 |
| 10/23/2000 | 30 | 36 | 27 |
| 10/23/2000 | 90 | 82 | 88 |
| 10/23/2000 | 16 | 19 | 21 |
| 10/23/2000 | 9 | 8 | 10 |
| 10/23/2000 | 1 | 2 | 3 |
| 3/8/2001 | 93 | 86 | 111 |
| 3/8/2001 | 103 | 82 | 91 |
| 3/8/2001 | 79 | 73 | 68 |
| 3/8/2001 | 51 | 60 | 43 |
| 3/8/2001 | 85 | 90 | 95 |
| 3/8/2001 | 20 | 25 | 30 |
| 3/8/2001 | 6 | 8 | 10 |
| 3/8/2001 | 112 | 109 | 89 |
| 3/8/2001 | 99 | 92 | 81 |
| 3/8/2001 | 92 | 99 | 89 |

| Date | | | |
|---|---|---|---|
| 3/8/2001 | 90 | 85 | 81 |
| 3/8/2001 | 59 | 66 | 78 |
| 3/8/2001 | 44 | 60 | 51 |
| 3/8/2001 | 23 | 40 | 33 |
| 10/8/1999 | 160 | 141 | 157 |
| 10/8/1999 | 137 | 149 | 158 |
| 10/8/1999 | 131 | 138 | 126 |
| 10/8/1999 | 106 | 125 | 115 |
| 10/8/1999 | 89 | 95 | 102 |
| 10/8/1999 | 48 | 53 | 42 |
| 10/8/1999 | 30 | 36 | 42 |
| 10/8/1999 | 38 | 47 | 29 |
| 10/8/1999 | 51 | 56 | 61 |
| 10/8/1999 | 2 | 3 | 4 |
| 10/13/1998 | 187 | 182 | 190 |
| 10/13/1998 | 228 | 199 | 213 |
| 10/13/1998 | 66 | 68 | 61 |
| 10/13/1998 | 39 | 37 | 44 |
| 10/13/1998 | 43 | 37 | 33 |
| 10/13/1998 | 160 | 153 | 175 |
| 10/13/1998 | 250 | 150 | 170 |
| 10/13/1998 | 122 | 133 | 125 |
| 10/13/1998 | 137 | 132 | 131 |
| 10/13/1998 | 58 | 60 | 50 |

**Other investigators Colonies**

| Date | col1 | col2 | col3 | investigator |
|---|---|---|---|---|
| 12/13/2001 | 140 | 141 | 160 | Inv1 |
| 12/13/2001 | 130 | 139 | 148 | Inv1 |
| 12/13/2001 | 54 | 56 | 75 | Inv1 |
| 12/13/2001 | 127 | 144 | 148 | Inv1 |
| 12/13/2001 | 59 | 55 | 49 | Inv1 |
| 12/13/2001 | 161 | 148 | 172 | Inv1 |
| 12/13/2001 | 81 | 83 | 72 | Inv1 |
| 12/13/2001 | 20 | 23 | 25 | Inv1 |
| 12/13/2001 | 7 | 7 | 13 | Inv1 |
| 12/13/2001 | 7 | 5 | 4 | Inv1 |
| 12/26/2001 | 124 | 99 | 109 | Inv1 |
| 12/26/2001 | 91 | 98 | 113 | Inv1 |
| 12/26/2001 | 75 | 92 | 84 | Inv1 |
| 12/26/2001 | 125 | 106 | 121 | Inv1 |
| 12/26/2001 | 97 | 96 | 101 | Inv1 |
| 12/26/2001 | 101 | 103 | 124 | Inv1 |
| 12/26/2001 | 91 | 93 | 84 | Inv1 |
| 12/26/2001 | 46 | | | Inv1 |
| 12/26/2001 | 210 | | | Inv1 |
| 12/26/2001 | 128 | | | Inv1 |
| 1/6/2002 | 27 | 22 | 22 | Inv1 |
| 1/6/2002 | 15 | 17 | 22 | Inv1 |
| 1/6/2002 | 37 | 46 | 47 | Inv1 |
| 1/6/2002 | 20 | 29 | 30 | Inv1 |
| 1/6/2002 | 30 | 34 | 37 | Inv1 |
| 1/6/2002 | 90 | 97 | 102 | Inv1 |

| Date | | | | |
|---|---|---|---|---|
| 1/6/2002 | 80 | 83 | 86 | Inv1 |
| 1/6/2002 | 168 | 174 | 194 | Inv1 |
| 1/6/2002 | 123 | 144 | 149 | Inv1 |
| 1/6/2002 | 46 | 60 | 67 | Inv1 |
| 1/6/2002 | 20 | 22 | 27 | Inv1 |
| 1/6/2002 | 18 | 23 | 23 | Inv1 |
| 10/11/1992 | 41 | 36 | 50 | Inv2 |
| 10/11/1992 | 49 | 45 | 31 | Inv2 |
| 10/11/1992 | 25 | 28 | 30 | Inv2 |
| 10/11/1992 | 14 | 19 | 20 | Inv2 |
| 10/11/1992 | 142 | 147 | 157 | Inv2 |
| 10/11/1992 | 51 | 60 | 67 | Inv2 |
| 10/11/1992 | 81 | 81 | 82 | Inv2 |
| 10/11/1992 | 89 | 93 | 97 | Inv2 |
| 10/11/1992 | 41 | 57 | 48 | Inv2 |
| 10/11/1992 | 74 | 71 | 70 | Inv2 |
| 10/11/1992 | 53 | 73 | 54 | Inv2 |
| 4/4/1995 | 19 | 12 | 12 | Inv2 |
| 4/4/1995 | 7 | 7 | 11 | Inv2 |
| 4/4/1995 | 19 | 36 | 20 | Inv2 |
| 4/4/1995 | 10 | 7 | 20 | Inv2 |
| 4/4/1995 | 73 | 65 | 59 | Inv2 |
| 4/4/1995 | 7 | 12 | 14 | Inv2 |
| 4/4/1995 | 1 | 7 | 8 | Inv2 |
| 4/4/1995 | 2 | 8 | 10 | Inv2 |
| 4/4/1995 | 3 | 2 | 0 | Inv2 |
| 4/25/1995 | 2 | 11 | 5 | Inv2 |
| 4/25/1995 | 2 | 6 | 7 | Inv2 |

| Date | | | | |
|---|---|---|---|---|
| 4/25/1995 | 14 | 33 | 37 | Inv2 |
| 4/25/1995 | 10 | 13 | 16 | Inv2 |
| 4/25/1995 | 52 | 61 | 51 | Inv2 |
| 4/25/1995 | 0 | 1 | 1 | Inv2 |
| 4/25/1995 | 0 | 0 | 1 | Inv2 |
| 4/25/1995 | 2 | 2 | 2 | Inv2 |
| 4/25/1995 | 0 | 0 | 2 | Inv2 |
| 4/25/1995 | 0 | 2 | 2 | Inv2 |
| 6/27/1995 | 72 | 75 | 61 | Inv2 |
| 6/27/1995 | 98 | 79 | 82 | Inv2 |
| 6/27/1995 | 30 | 20 | 42 | Inv2 |
| 6/27/1995 | 107 | 112 | 106 | Inv2 |
| 6/27/1995 | 37 | 35 | 44 | Inv2 |
| 6/27/1995 | 112 | 119 | 89 | Inv2 |
| 6/27/1995 | 112 | 109 | 96 | Inv2 |
| 6/27/1995 | 30 | 32 | 25 | Inv2 |
| 6/27/1995 | 24 | 26 | 30 | Inv2 |
| 6/27/1995 | 25 | 18 | 31 | Inv2 |
| 6/27/1995 | 192 | 192 | 186 | Inv2 |
| 6/27/1995 | 183 | 171 | 193 | Inv2 |
| 7/24/1995 | 50 | 56 | 84 | Inv2 |
| 7/24/1995 | 57 | 37 | 55 | Inv2 |
| 7/24/1995 | 32 | 29 | 34 | Inv2 |
| 7/24/1995 | 109 | 126 | 135 | Inv2 |
| 7/24/1995 | 63 | 58 | 62 | Inv2 |
| 7/24/1995 | 57 | 52 | 49 | Inv2 |
| 7/24/1995 | 55 | 45 | 54 | Inv2 |
| 7/24/1995 | 153 | 163 | 153 | Inv2 |

| Date | | | | |
|---|---|---|---|---|
| 7/24/1995 | 263 | 253 | 267 | Inv2 |
| 7/24/1995 | 38 | 23 | 35 | Inv2 |
| 9/25/1995 | 81 | 90 | 72 | Inv2 |
| 9/25/1995 | 109 | 93 | 111 | Inv2 |
| 9/25/1995 | 65 | 69 | 66 | Inv2 |
| 9/25/1995 | 52 | 61 | 30 | Inv2 |
| 9/25/1995 | 132 | 127 | 149 | Inv2 |
| 9/25/1995 | 128 | 123 | 116 | Inv2 |
| 9/25/1995 | 110 | 104 | 99 | Inv2 |
| 9/25/1995 | 78 | 78 | 82 | Inv2 |
| 9/25/1995 | 17 | 19 | 14 | Inv2 |
| 9/25/1995 | 153 | 145 | 152 | Inv2 |
| 9/25/1995 | 14 | 13 | 12 | Inv2 |
| 12/11/1995 | 125 | 123 | 126 | Inv2 |
| 12/11/1995 | 189 | 166 | 200 | Inv2 |
| 12/11/1995 | 131 | 120 | 108 | Inv2 |
| 12/11/1995 | 62 | 59 | 62 | Inv2 |
| 12/11/1995 | 53 | 54 | 49 | Inv2 |
| 12/11/1995 | 177 | 178 | 194 | Inv2 |
| 12/11/1995 | 108 | 108 | 93 | Inv2 |
| 12/11/1995 | 154 | 172 | 157 | Inv2 |
| 12/11/1995 | 50 | 49 | 44 | Inv2 |
| 12/11/1995 | 26 | 30 | 26 | Inv2 |
| 1/23/1996 | 94 | 84 | 79 | Inv2 |
| 1/23/1996 | 86 | 66 | 84 | Inv2 |
| 1/23/1996 | 26 | 27 | 42 | Inv2 |
| 1/23/1996 | 51 | 53 | 53 | Inv2 |
| 1/23/1996 | 20 | 18 | 12 | Inv2 |

| Date | | | | |
|---|---|---|---|---|
| 1/23/1996 | 37 | 57 | 56 | Inv2 |
| 1/23/1996 | 59 | 83 | 71 | Inv2 |
| 1/23/1996 | 87 | 91 | 108 | Inv2 |
| 1/23/1996 | 32 | 32 | 36 | Inv2 |
| 1/23/1996 | 45 | 48 | 37 | Inv2 |
| 1/23/1996 | 19 | 12 | 17 | Inv2 |
| 1/23/1996 | 50 | 42 | 43 | Inv2 |
| 2/1/1996 | 121 | 137 | 133 | Inv2 |
| 2/1/1996 | 121 | 125 | 117 | Inv2 |
| 2/1/1996 | 128 | 127 | 111 | Inv2 |
| 2/1/1996 | 45 | 54 | 60 | Inv2 |
| 2/1/1996 | 144 | 146 | 154 | Inv2 |
| 2/1/1996 | 60 | 46 | 43 | Inv2 |
| 2/12/1996 | 44 | 57 | 54 | Inv2 |
| 2/12/1996 | 48 | 49 | 50 | Inv2 |
| 2/12/1996 | 9 | 12 | 7 | Inv2 |
| 2/12/1996 | 17 | 25 | 20 | Inv2 |
| 2/12/1996 | 7 | 6 | 6 | Inv2 |
| 2/12/1996 | 65 | 61 | 67 | Inv2 |
| 2/12/1996 | 5 | 8 | 3 | Inv2 |
| 2/12/1996 | 13 | 8 | 9 | Inv2 |
| 2/12/1996 | 2 | 4 | 1 | Inv2 |
| 4/23/1996 | 164 | 171 | 163 | Inv2 |
| 4/23/1996 | 150 | 141 | 115 | Inv2 |
| 4/23/1996 | 42 | 61 | 55 | Inv2 |
| 4/23/1996 | 39 | 70 | 44 | Inv2 |
| 4/23/1996 | 36 | 29 | 31 | Inv2 |
| 4/23/1996 | 144 | 155 | 168 | Inv2 |

| Date | | | | |
|---|---|---|---|---|
| 4/23/1996 | 152 | 144 | 150 | Inv2 |
| 4/23/1996 | 53 | 48 | 61 | Inv2 |
| 4/23/1996 | 43 | 51 | 50 | Inv2 |
| 4/23/1996 | 34 | 47 | 36 | Inv2 |
| 5/7/1996 | 222 | 202 | 201 | Inv2 |
| 5/7/1996 | 210 | 183 | 180 | Inv2 |
| 5/7/1996 | 48 | 60 | 62 | Inv2 |
| 5/7/1996 | 66 | 88 | 73 | Inv2 |
| 5/7/1996 | 96 | 80 | 98 | Inv2 |
| 5/7/1996 | 170 | 168 | 189 | Inv2 |
| 5/7/1996 | 162 | 179 | 191 | Inv2 |
| 5/7/1996 | 58 | 41 | 68 | Inv2 |
| 5/7/1996 | 38 | 34 | 32 | Inv2 |
| 5/7/1996 | 42 | 50 | 51 | Inv2 |
| 8/20/1996 | 136 | 118 | 105 | Inv2 |
| 8/20/1996 | 78 | 96 | 95 | Inv2 |
| 8/20/1996 | 68 | 78 | 55 | Inv2 |
| 8/20/1996 | 73 | 57 | 53 | Inv2 |
| 8/20/1996 | 28 | 37 | 40 | Inv2 |
| 8/20/1996 | 182 | 166 | 203 | Inv2 |
| 8/20/1996 | 235 | 269 | 249 | Inv2 |
| 8/20/1996 | 88 | 95 | 81 | Inv2 |
| 8/20/1996 | 56 | 62 | 66 | Inv2 |
| 8/20/1996 | 120 | 98 |  | Inv2 |
| 1/23/1997 | 26 | 28 | 14 | Inv2 |
| 1/23/1997 | 34 | 40 | 39 | Inv2 |
| 1/23/1997 | 71 | 104 | 106 | Inv2 |
| 1/23/1997 | 165 | 161 | 169 | Inv2 |

| Date | | | | |
|---|---|---|---|---|
| 1/23/1997 | 26 | 23 | 19 | Inv2 |
| 1/23/1997 | 46 | 66 | 61 | Inv2 |
| 1/23/1997 | 72 | 43 | 46 | Inv2 |
| 1/23/1997 | 32 | 35 | 36 | Inv2 |
| 1/23/1997 | 65 | 47 | 68 | Inv2 |
| 1/23/1997 | 80 | 81 | 78 | Inv2 |
| 4/18/1997 | 28 | 28 | 25 | Inv2 |
| 4/18/1997 | 27 | 35 | 28 | Inv2 |
| 4/18/1997 | 18 | 20 | 21 | Inv2 |
| 4/18/1997 | 12 | 24 | 19 | Inv2 |
| 4/18/1997 | 17 | 12 | 14 | Inv2 |
| 4/18/1997 | 35 | 40 | 37 | Inv2 |
| 4/18/1997 | 39 | 34 | 31 | Inv2 |
| 4/18/1997 | 19 | 18 | 18 | Inv2 |
| 4/18/1997 | 10 | 14 | 10 | Inv2 |
| 4/18/1997 | 8 | 16 | 12 | Inv2 |
| 4/21/1997 | 11 | 3 | 8 | Inv2 |
| 4/21/1997 | 5 | 5 | 7 | Inv2 |
| 4/21/1997 | 70 | 66 | 72 | Inv2 |
| 4/21/1997 | 4 | 2 | 2 | Inv2 |
| 4/21/1997 | 66 | 66 | 49 | Inv2 |
| 4/21/1997 | 8 | 9 | 11 | Inv2 |
| 4/21/1997 | 3 | 10 | 17 | Inv2 |
| 4/21/1997 | 16 | 16 | 21 | Inv2 |
| 4/21/1997 | 4 | | | Inv2 |
| 4/21/1997 | 6 | 8 | 8 | Inv2 |
| 4/22/1997 | 22 | 19 | 25 | Inv2 |
| 4/22/1997 | 18 | 19 | 20 | Inv2 |

| Date | | | | |
|---|---|---|---|---|
| 4/22/1997 | 9 | 12 | 15 | Inv2 |
| 4/22/1997 | 8 | 9 | 17 | Inv2 |
| 4/22/1997 | 12 | 18 | 21 | Inv2 |
| 4/22/1997 | 10 | 15 | 17 | Inv2 |
| 4/22/1997 | 4 | 5 | 7 | Inv2 |
| 4/22/1997 | 4 | 4 | 7 | Inv2 |
| 4/22/1997 | 1 | 1 | 2 | Inv2 |
| 4/19/2001 | 92 | 111 | 119 | Inv2 |
| 4/19/2001 | 78 | 85 | 74 | Inv2 |
| 4/19/2001 | 142 | 126 | 120 | Inv2 |
| 4/19/2001 | 120 | 129 | 121 | Inv2 |
| 4/19/2001 | 64 | 68 | 79 | Inv2 |
| 4/19/2001 | 92 | 101 | 78 | Inv2 |
| 4/19/2001 | 74 | 62 | 94 | Inv2 |
| 4/19/2001 | 89 | 69 | 67 | Inv2 |
| 4/19/2001 | 85 | 87 | 97 | Inv2 |
| 4/19/2001 | 71 | 58 | 55 | Inv2 |
| 5/3/2001 | 161 | 143 | 123 | Inv2 |
| 5/3/2001 | 132 | 141 | 124 | Inv2 |
| 5/3/2001 | 88 | 69 | 70 | Inv2 |
| 5/3/2001 | 77 | 65 | 55 | Inv2 |
| 5/3/2001 | 72 | 71 | 80 | Inv2 |
| 5/3/2001 | 62 | 73 | 58 | Inv2 |
| 5/3/2001 | 73 | 80 | 78 | Inv2 |
| 5/3/2001 | 89 | 76 | 85 | Inv2 |
| 5/3/2001 | 58 | 74 | 85 | Inv2 |
| 5/3/2001 | 83 | 90 | 71 | Inv2 |
| 5/21/2001 | 167 | 175 | 191 | Inv2 |

| Date | | | | |
|---|---|---|---|---|
| 5/21/2001 | 109 | 94 | 121 | Inv2 |
| 5/21/2001 | 141 | 138 | 144 | Inv2 |
| 5/21/2001 | 118 | 117 | 145 | Inv2 |
| 5/21/2001 | 188 | 179 | 176 | Inv2 |
| 5/21/2001 | 164 | 160 | 170 | Inv2 |
| 5/21/2001 | 166 | 155 | 161 | Inv2 |
| 5/21/2001 | 166 | 164 | 183 | Inv2 |
| 5/21/2001 | 165 | 189 | 166 | Inv2 |
| 5/21/2001 | 157 | 163 | 159 | Inv2 |
| 7/9/2001 | 180 | 182 | 207 | Inv2 |
| 7/9/2001 | 178 |  | 179 | Inv2 |
| 7/9/2001 | 125 | 109 | 133 | Inv2 |
| 7/9/2001 | 81 | 81 | 91 | Inv2 |
| 7/9/2001 | 113 | 121 | 105 | Inv2 |
| 7/9/2001 | 117 | 100 | 104 | Inv2 |
| 7/9/2001 | 113 | 89 | 112 | Inv2 |
| 7/9/2001 | 72 | 78 | 72 | Inv2 |
| 7/31/2001 | 108 | 114 | 114 | Inv2 |
| 7/31/2001 | 85 | 81 |  | Inv2 |
| 7/31/2001 | 63 | 68 | 83 | Inv2 |
| 7/31/2001 | 94 | 84 | 73 | Inv2 |
| 7/31/2001 | 66 | 108 | 67 | Inv2 |
| 7/31/2001 | 61 | 93 | 73 | Inv2 |
| 7/31/2001 | 56 | 55 | 52 | Inv2 |
| 7/31/2001 | 47 | 38 | 37 | Inv2 |
| 7/31/2001 | 46 | 38 | 54 | Inv2 |
| 7/31/2001 | 40 | 30 | 30 | Inv2 |
| 8/31/2001 | 191 | 193 | 199 | Inv2 |

| Date | | | | |
|---|---|---|---|---|
| 8/31/2001 | 272 | 208 | 241 | Inv2 |
| 8/31/2001 | 271 | 281 | 290 | Inv2 |
| 8/31/2001 | 234 | 223 | 275 | Inv2 |
| 8/31/2001 | 221 | 238 | 225 | Inv2 |
| 8/31/2001 | 260 | 239 | 230 | Inv2 |
| 8/31/2001 | 208 | 225 | 211 | Inv2 |
| 8/31/2001 | 245 | 248 | 227 | Inv2 |
| 10/8/2001 | 105 | 104 | 98 | Inv2 |
| 10/8/2001 | 97 | 78 | 87 | Inv2 |
| 10/8/2001 | 55 | 66 | 61 | Inv2 |
| 10/8/2001 | 56 | 48 | 41 | Inv2 |
| 10/8/2001 | 47 | 44 |  | Inv2 |
| 10/8/2001 | 31 | 39 | 42 | Inv2 |
| 10/8/2001 | 34 | 28 | 24 | Inv2 |
| 10/8/2001 | 16 | 26 | 14 | Inv2 |
| 10/8/2001 | 26 | 36 | 21 | Inv2 |
| 10/8/2001 | 164 | 175 | 158 | Inv2 |
| 10/8/2001 | 139 | 170 | 156 | Inv2 |
| 10/30/2001 | 89 | 83 | 69 | Inv2 |
| 10/30/2001 | 32 | 28 | 44 | Inv2 |
| 10/30/2001 | 25 | 31 | 30 | Inv2 |
| 10/30/2001 | 38 | 29 | 31 | Inv2 |
| 10/30/2001 | 31 | 27 | 33 | Inv2 |
| 10/30/2001 | 183 | 183 | 196 | Inv2 |
| 10/30/2001 | 155 | 166 | 168 | Inv2 |
| 10/30/2001 | 116 | 124 |  | Inv2 |
| 10/30/2001 | 147 | 134 | 114 | Inv2 |
| 10/30/2001 | 108 | 115 | 127 | Inv2 |

| Date | | | | |
|---|---|---|---|---|
| 10/31/2001 | 343 | 310 | 334 | Inv2 |
| 10/31/2001 | 236 | 232 | 242 | Inv2 |
| 10/31/2001 | 347 | 269 | 287 | Inv2 |
| 10/31/2001 | 362 | 293 | 335 | Inv2 |
| 10/31/2001 | 314 | | | Inv2 |
| 10/31/2001 | 310 | 291 | 284 | Inv2 |
| 10/31/2001 | 350 | 329 | 323 | Inv2 |
| 10/31/2001 | 324 | 294 | 362 | Inv2 |
| 10/31/2001 | 324 | 332 | 306 | Inv2 |
| 10/31/2001 | 343 | 329 | 303 | Inv2 |
| 11/27/2002 | 8 | 7 | 6 | Inv3 |
| 11/27/2002 | 4 | 3 | 3 | Inv3 |
| 11/27/2002 | 42 | 47 | 36 | Inv3 |
| 11/27/2002 | 32 | 33 | 33 | Inv3 |
| 11/27/2002 | 34 | 44 | 45 | Inv3 |
| 11/27/2002 | 32 | 30 | 27 | Inv3 |
| 11/27/2002 | 26 | 23 | 28 | Inv3 |
| 11/27/2002 | 32 | 37 | 32 | Inv3 |
| 11/27/2002 | 39 | 34 | 35 | Inv3 |
| 11/27/2002 | 37 | 38 | 29 | Inv3 |
| 7/14/2000 | 97 | 93 | 115 | Inv4 |
| 7/14/2000 | 102 | 123 | 111 | Inv4 |
| 7/14/2000 | 64 | 77 | 64 | Inv4 |
| 7/14/2000 | 47 | 50 | 60 | Inv4 |
| 7/14/2000 | 36 | 39 | 26 | Inv4 |
| 7/14/2000 | 296 | 292 | 235 | Inv4 |
| 7/14/2000 | 230 | 283 | 296 | Inv4 |
| 7/25/2000 | 97 | 93 | 115 | Inv4 |

| Date | | | | |
|---|---|---|---|---|
| 7/25/2000 | 102 | 123 | 111 | Inv4 |
| 7/25/2000 | 64 | 77 | 64 | Inv4 |
| 7/25/2000 | 47 | 50 | 60 | Inv4 |
| 7/25/2000 | 36 | 39 | 26 | Inv4 |
| 7/25/2000 | 296 | 292 | 235 | Inv4 |
| 7/25/2000 | 230 | 283 | 296 | Inv4 |
| 8/4/2000 | 177 | 146 | 190 | Inv4 |
| 8/4/2000 | 169 | 186 | 168 | Inv4 |
| 8/4/2000 | 119 | 115 | 114 | Inv4 |
| 8/4/2000 | 111 | 94 | 93 | Inv4 |
| 8/4/2000 | 76 | 73 | 63 | Inv4 |
| 8/4/2000 | 178 | 183 | 187 | Inv4 |
| 8/4/2000 | 63 | 49 | 67 | Inv4 |
| 5/7/1996 | 192 | 207 | 201 | Inv5 |
| 5/7/1996 | 185 | 178 | 202 | Inv5 |
| 5/7/1996 | 46 | 64 | 62 | Inv5 |
| 5/7/1996 | 81 | 87 | 79 | Inv5 |
| 5/7/1996 | 87 | 105 | 92 | Inv5 |
| 5/7/1996 | 154 | 168 | 186 | Inv5 |
| 5/7/1996 | 190 | 159 | 172 | Inv5 |
| 5/7/1996 | 51 | 68 | 41 | Inv5 |
| 5/7/1996 | 34 | 43 | 32 | Inv5 |
| 5/7/1996 | 59 | 59 | 49 | Inv5 |
| 5/13/1996 | 87 | 79 | 76 | Inv5 |
| 5/13/1996 | 56 | 45 | 50 | Inv5 |
| 5/13/1996 | 8 | 8 | 7 | Inv5 |
| 5/13/1996 | 3 | 3 | 4 | Inv5 |
| 5/13/1996 | 24 | 24 | 29 | Inv5 |

| Date | | | | |
|---|---|---|---|---|
| 5/13/1996 | 14 | 10 | 11 | Inv5 |
| 5/13/1996 | 54 | 41 | 36 | Inv5 |
| 5/13/1996 | 4 | 5 | 12 | Inv5 |
| 5/13/1996 | 28 | 21 | 29 | Inv5 |
| 5/13/1996 | 17 | 13 | 20 | Inv5 |
| 5/13/1996 | 238 | 240 | 247 | Inv5 |
| 5/13/1996 | 236 | 232 | 235 | Inv5 |
| 5/13/1996 | 163 | 170 | 177 | Inv5 |
| 5/13/1996 | 211 | 205 | 204 | Inv5 |
| 5/13/1996 | 115 | 109 | 174 | Inv5 |
| 5/13/1996 | 81 | 83 | 74 | Inv5 |
| 5/13/1996 | 39 | 39 | 34 | Inv5 |
| 5/13/1996 | 177 | 178 | 163 | Inv5 |
| 5/13/1996 | 190 | 229 | 241 | Inv5 |
| 6/1/1996 | 43 | 55 | 34 | Inv5 |
| 6/1/1996 | 106 | 87 | 103 | Inv5 |
| 6/1/1996 | 66 | 60 | 64 | Inv5 |
| 6/1/1996 | 37 | 36 | 41 | Inv5 |
| 6/1/1996 | 43 | 50 | 45 | Inv5 |
| 6/1/1996 | 56 | 45 | 59 | Inv5 |
| 6/1/1996 | 114 | 119 | 101 | Inv5 |
| 6/1/1996 | 63 | 67 | 60 | Inv5 |
| 6/1/1996 | 26 | 29 | 27 | Inv5 |
| 7/16/1996 | 40 | 58 | 44 | Inv5 |
| 7/16/1996 | 40 | 44 | 45 | Inv5 |
| 7/16/1996 | 197 | 209 | 197 | Inv5 |
| 7/16/1996 | 30 | 29 | 40 | Inv5 |
| 7/16/1996 | 34 | 30 | 29 | Inv5 |

| Date | | | | |
|---|---|---|---|---|
| 7/16/1996 | 46 | 58 | 49 | Inv5 |
| 7/16/1996 | 46 | 57 | 46 | Inv5 |
| 7/16/1996 | 114 | 107 | 122 | Inv5 |
| 7/16/1996 | 32 | 32 | 29 | Inv5 |
| 7/16/1996 | 19 | 14 | 16 | Inv5 |
| 7/30/1996 | 34 | 48 | 51 | Inv5 |
| 7/30/1996 | 36 | 38 | 29 | Inv5 |
| 7/30/1996 | 34 | 41 | 30 | Inv5 |
| 7/30/1996 | 114 | 129 | 123 | Inv5 |
| 7/30/1996 | 75 | 85 | 96 | Inv5 |
| 7/30/1996 | 45 | 44 | 55 | Inv5 |
| 7/30/1996 | 31 | 39 | 26 | Inv5 |
| 7/30/1996 | 26 | 37 | 38 | Inv5 |
| 7/30/1996 | 90 | 89 | 96 | Inv5 |
| 7/30/1996 | 54 | 55 | 61 | Inv5 |
| 6/5/2000 | 133 | 127 | 106 | Inv6 |
| 6/5/2000 | 170 | 167 | 163 | Inv6 |
| 6/5/2000 | 123 | 119 | 102 | Inv6 |
| 6/5/2000 | 89 | 101 | 83 | Inv6 |
| 6/5/2000 | 59 | 51 | 65 | Inv6 |
| 6/5/2000 | 51 | 55 | 68 | Inv6 |
| 6/5/2000 | 47 | 59 | 62 | Inv6 |
| 6/5/2000 | 62 | 68 | 57 | Inv6 |
| 6/5/2000 | 396 | 379 | 384 | Inv6 |
| 6/5/2000 | 376 | 360 | 353 | Inv6 |
| 10/2/2000 | 37 | 48 | 35 | Inv7 |
| 10/2/2000 | 38 | 36 | 31 | Inv7 |
| 10/2/2000 | 52 | 57 | 47 | Inv7 |

| Date | | | | |
|---|---|---|---|---|
| 10/2/2000 | 57 | 65 | 72 | Inv7 |
| 10/2/2000 | 25 | 22 | 22 | Inv7 |
| 10/2/2000 | 50 | 51 | 47 | Inv7 |
| 10/2/2000 | 38 | 50 | 43 | Inv7 |
| 10/2/2000 | 39 | 60 | 48 | Inv7 |
| 10/2/2000 | 41 | 34 | 38 | Inv7 |
| 10/2/2000 | 19 | 19 | 20 | Inv7 |
| 12/14/2000 | 197 | 200 | 202 | Inv7 |
| 12/14/2000 | 73 | 69 | 82 | Inv7 |
| 12/14/2000 | 62 | 63 | 62 | Inv7 |
| 12/14/2000 | 61 | 60 | 55 | Inv7 |
| 12/14/2000 | 38 | 47 | 41 | Inv7 |
| 12/14/2000 | 49 | 42 | 37 | Inv7 |
| 12/14/2000 | 32 | 37 | 36 | Inv7 |
| 12/14/2000 | 220 | 222 | 211 | Inv7 |
| 12/14/2000 | 182 | 175 | 170 | Inv7 |
| 12/14/2000 | 168 | 185 | 179 | Inv7 |
| 2/19/2001 | 49 | 44 | 26 | Inv7 |
| 2/19/2001 | 32 | 38 | 40 | Inv7 |
| 2/19/2001 | 22 | 11 | 21 | Inv7 |
| 2/19/2001 | 24 | 23 | 24 | Inv7 |
| 2/19/2001 | 14 | 25 | 25 | Inv7 |
| 2/19/2001 | 27 | 24 |  | Inv7 |
| 2/19/2001 | 15 | 15 |  | Inv7 |
| 2/19/2001 | 30 | 28 | 21 | Inv7 |
| 2/19/2001 | 36 | 37 |  | Inv7 |
| 2/19/2001 | 20 | 15 | 18 | Inv7 |
| 3/12/2001 | 83 | 71 | 94 | Inv7 |

| Date | | | | |
|---|---|---|---|---|
| 3/12/2001 | 85 | 92 | 82 | Inv7 |
| 3/12/2001 | 94 | 75 | 75 | Inv7 |
| 3/12/2001 | 75 | 54 | 73 | Inv7 |
| 3/12/2001 | 53 | 36 | 37 | Inv7 |
| 3/12/2001 | 70 | 73 | 49 | Inv7 |
| 4/2/2001 | 29 | 42 | 31 | Inv7 |
| 4/2/2001 | 43 | 32 | 29 | Inv7 |
| 4/2/2001 | 21 | 31 | 20 | Inv7 |
| 4/2/2001 | 27 | 37 | 29 | Inv7 |
| 4/2/2001 | 35 | 25 | 28 | Inv7 |
| 4/2/2001 | 43 | 34 | 22 | Inv7 |
| 4/2/2001 | 17 | 29 | 41 | Inv7 |
| 4/2/2001 | 19 | 23 | 10 | Inv7 |
| 4/2/2001 | 19 | 24 | 13 | Inv7 |
| 4/2/2001 | 31 | 19 | 26 | Inv7 |
| 5/3/2001 | 80 | 78 | 95 | Inv7 |
| 5/3/2001 | 87 | 86 | 83 | Inv7 |
| 5/3/2001 | 59 | 62 | 53 | Inv7 |
| 5/3/2001 | 61 | 52 | 52 | Inv7 |
| 5/3/2001 | 51 | 47 | 44 | Inv7 |
| 5/3/2001 | 64 | 54 | 51 | Inv7 |
| 5/3/2001 | 41 | 33 | 45 | Inv7 |
| 5/3/2001 | 59 | 59 | 61 | Inv7 |
| 5/3/2001 | 42 | 48 | 33 | Inv7 |
| 5/3/2001 | 69 | 76 | 71 | Inv7 |
| 5/25/2001 | 86 | 78 | 72 | Inv7 |
| 5/25/2001 | 50 | 76 | 63 | Inv7 |
| 5/25/2001 | 71 | 81 | 68 | Inv7 |

| Date | | | | |
|---|---|---|---|---|
| 5/25/2001 | 68 | 54 | 66 | Inv7 |
| 5/25/2001 | 62 | 55 | 55 | Inv7 |
| 5/25/2001 | 81 | 66 | 67 | Inv7 |
| 5/25/2001 | 29 | 38 | 46 | Inv7 |
| 5/25/2001 | 54 | 53 | 51 | Inv7 |
| 5/25/2001 | 38 | 46 | 38 | Inv7 |
| 5/25/2001 | 40 | 36 | 36 | Inv7 |
| 6/25/2001 | 100 | 90 | | Inv7 |
| 6/25/2001 | 78 | 72 | | Inv7 |
| 6/25/2001 | 97 | 107 | 93 | Inv7 |
| 6/25/2001 | 58 | 61 | 57 | Inv7 |
| 6/25/2001 | 68 | 51 | 63 | Inv7 |
| 6/25/2001 | 61 | 39 | 49 | Inv7 |
| 6/25/2001 | 57 | 42 | 39 | Inv7 |
| 6/25/2001 | 43 | 47 | 34 | Inv7 |
| 6/25/2001 | 47 | 55 | 49 | Inv7 |
| 6/25/2001 | 61 | 61 | 52 | Inv7 |
| 7/13/2001 | 100 | 138 | 140 | Inv7 |
| 7/13/2001 | 135 | 128 | 129 | Inv7 |
| 7/13/2001 | 157 | 180 | 160 | Inv7 |
| 7/13/2001 | 173 | 155 | 180 | Inv7 |
| 7/13/2001 | 132 | 150 | 124 | Inv7 |
| 7/13/2001 | 129 | 128 | 119 | Inv7 |
| 7/13/2001 | 130 | 137 | 135 | Inv7 |
| 7/13/2001 | 119 | 100 | 110 | Inv7 |
| 7/13/2001 | 115 | 132 | 127 | Inv7 |
| 9/27/1999 | 80 | 74 | 70 | Inv8 |
| 9/27/1999 | 65 | 53 | 64 | Inv8 |

| Date | | | | |
|---|---|---|---|---|
| 9/27/1999 | 64 | 56 | 65 | Inv8 |
| 9/27/1999 | 64 | 67 | 53 | Inv8 |
| 9/27/1999 | 74 | 86 | 69 | Inv8 |
| 9/27/1999 | 65 | 65 | 65 | Inv8 |
| 12/27/1999 | 14 | 20 | 21 | Inv8 |
| 12/27/1999 | 31 | 32 | 25 | Inv8 |
| 12/27/1999 | 25 | 34 | 25 | Inv8 |
| 12/27/1999 | 30 | 37 | 46 | Inv8 |
| 12/27/1999 | 25 | 32 | 35 | Inv8 |
| 12/27/1999 | 35 | 33 | 37 | Inv8 |
| 12/27/1999 | 42 | 39 | 41 | Inv8 |
| 12/27/1999 | 50 | 58 | 55 | Inv8 |
| 12/27/1999 | 44 | 37 | 41 | Inv8 |
| 12/27/1999 | 73 | 50 | 67 | Inv8 |
| 2/23/2000 | 58 | 50 | 55 | Inv8 |
| 2/23/2000 | 59 | 55 | 61 | Inv8 |
| 2/23/2000 | 48 | 58 | 46 | Inv8 |
| 2/23/2000 | 42 | 39 | 37 | Inv8 |
| 2/23/2000 | 49 | 54 | 44 | Inv8 |
| 2/23/2000 | 45 | 45 | 44 | Inv8 |
| 2/23/2000 | 60 | 58 | 59 | Inv8 |
| 2/23/2000 | 52 | 53 | 54 | Inv8 |
| 2/23/2000 | 41 | 38 | 44 | Inv8 |
| 3/15/2000 | 79 | 68 | 72 | Inv8 |
| 3/15/2000 | 73 | 70 | 79 | Inv8 |
| 3/15/2000 | 210 | 217 | | Inv8 |
| 3/15/2000 | 13 | 16 | 19 | Inv8 |
| 3/15/2000 | 21 | 23 | 22 | Inv8 |

| Date | | | | |
|---|---|---|---|---|
| 3/15/2000 | 18 | 15 | 21 | Inv8 |
| 3/15/2000 | 12 | 23 | 13 | Inv8 |
| 3/15/2000 | 8 | 10 | 12 | Inv8 |
| 3/15/2000 | 18 | 22 | 11 | Inv8 |
| 3/15/2000 | 11 | 14 | 13 | Inv8 |
| 3/15/2000 | 14 | 11 | 16 | Inv8 |
| 3/15/2000 | 6 | 14 | 9 | Inv8 |
| 3/15/2000 | 14 | 16 | 17 | Inv8 |
| 3/16/2000 | 8 | 9 | 7 | Inv8 |
| 3/16/2000 | 70 | 49 | 51 | Inv8 |
| 3/16/2000 | 5 | 6 | 8 | Inv8 |
| 3/16/2000 | 52 | 25 | 18 | Inv8 |
| 3/16/2000 | 3 | 1 | 2 | Inv8 |
| 3/16/2000 | 41 | 35 | 31 | Inv8 |
| 3/16/2000 | 11 | 5 | 11 | Inv8 |
| 3/16/2000 | 52 | 51 | 46 | Inv8 |
| 3/16/2000 | 5 | 8 | 7 | Inv8 |
| 3/16/2000 | 63 | 53 | 47 | Inv8 |
| 3/16/2000 | 29 | 19 | 17 | Inv8 |
| 3/16/2000 | 16 | 15 | 16 | Inv8 |
| 3/16/2000 | 24 | 23 | 18 | Inv8 |
| 3/16/2000 | 25 | 20 | 17 | Inv8 |
| 4/3/2000 | 28 | 30 | 32 | Inv8 |
| 4/3/2000 | 37 | 40 | 32 | Inv8 |
| 4/3/2000 | 42 | 37 | 40 | Inv8 |
| 4/3/2000 | 32 | 23 | 34 | Inv8 |
| 4/3/2000 | 35 | 34 | 34 | Inv8 |
| 4/3/2000 | 40 | 44 | 35 | Inv8 |

| Date | | | | |
|---|---|---|---|---|
| 4/3/2000 | 27 | 43 | 39 | Inv8 |
| 4/3/2000 | 31 | 29 | 26 | Inv8 |
| 4/3/2000 | 35 | 31 | 36 | Inv8 |
| 4/3/2000 | 16 | 37 | 33 | Inv8 |
| 4/10/2000 | 169 | 176 | 150 | Inv8 |
| 4/10/2000 | 64 | 53 | 50 | Inv8 |
| 4/10/2000 | 35 | 37 | 47 | Inv8 |
| 4/10/2000 | 232 | 243 | 261 | Inv8 |
| 4/10/2000 | 186 | 167 | 178 | Inv8 |
| 4/10/2000 | 68 | 66 | 61 | Inv8 |
| 4/10/2000 | 92 | 83 | 78 | Inv8 |
| 4/10/2000 | 74 | 60 | 68 | Inv8 |
| 4/10/2000 | 45 | 56 | 45 | Inv8 |
| 4/10/2000 | 19 | 19 | 17 | Inv8 |
| 5/23/2000 | 57 | 76 | 70 | Inv8 |
| 5/23/2000 | 73 | 66 | 56 | Inv8 |
| 5/23/2000 | 88 | 78 | 89 | Inv8 |
| 5/23/2000 | 117 | 119 | 105 | Inv8 |
| 5/23/2000 | 106 | 115 | 105 | Inv8 |
| 5/23/2000 | 66 | 81 | 69 | Inv8 |
| 5/23/2000 | 65 | 74 | 73 | Inv8 |
| 5/23/2000 | 66 | 51 | 48 | Inv8 |
| 5/23/2000 | 48 | 68 | 53 | Inv8 |
| 5/23/2000 | 85 | 86 | 82 | Inv8 |
| 5/23/2000 | 112 | 123 | | Inv8 |
| 5/23/2000 | 132 | 118 | 111 | Inv8 |
| 5/23/2000 | 52 | 66 | 59 | Inv8 |
| 5/23/2000 | 64 | 63 | 72 | Inv8 |

| Date | | | | |
|---|---|---|---|---|
| 5/23/2000 | 60 | 62 | 54 | Inv8 |
| 5/23/2000 | 55 | 52 | 54 | Inv8 |
| 5/23/2000 | 80 | 78 | 72 | Inv8 |
| 5/23/2000 | 106 | 106 | 112 | Inv8 |
| 5/23/2000 | 108 | 113 | 129 | Inv8 |
| 6/30/1992 | 155 | 150 | 162 | Inv9 |
| 6/30/1992 | 98 | 80 | 67 | Inv9 |
| 6/30/1992 | 107 | 110 | | Inv9 |
| 6/30/1992 | 91 | 64 | 62 | Inv9 |
| 6/30/1992 | 42 | 54 | 46 | Inv9 |
| 6/30/1992 | 27 | 21 | | Inv9 |
| 6/30/1992 | 9 | 10 | 12 | Inv9 |
| 6/30/1992 | 72 | 67 | | Inv9 |
| 6/30/1992 | 15 | 23 | | Inv9 |
| 6/30/1992 | 60 | 76 | | Inv9 |
| 6/30/1992 | 53 | 58 | 42 | Inv9 |
| 6/30/1992 | 150 | 136 | | Inv9 |
| 6/30/1992 | 84 | 79 | 68 | Inv9 |
| 10/21/1992 | 262 | 277 | 245 | Inv9 |
| 10/21/1992 | 128 | 128 | 133 | Inv9 |
| 10/21/1992 | 93 | 93 | 89 | Inv9 |
| 10/21/1992 | 30 | 40 | 37 | Inv9 |
| 10/21/1992 | 28 | 33 | 38 | Inv9 |
| 10/21/1992 | 209 | 193 | 198 | Inv9 |
| 10/21/1992 | 135 | 135 | 130 | Inv9 |
| 10/21/1992 | 100 | 97 | 108 | Inv9 |
| 10/21/1992 | 51 | 57 | 53 | Inv9 |
| 10/21/1992 | 48 | 50 | 41 | Inv9 |

| Date | | | | |
|---|---|---|---|---|
| 10/21/1992 | 308 | 319 | 315 | Inv9 |
| 5/14/1992 | 189 | 182 | 180 | Inv9 |
| 5/14/1992 | 76 | 87 | 74 | Inv9 |
| 5/14/1992 | 29 | 26 | 25 | Inv9 |
| 5/14/1992 | 82 | 72 | 93 | Inv9 |
| 5/14/1992 | 46 | 36 | 57 | Inv9 |
| 5/14/1992 | 46 | 29 | 34 | Inv9 |
| 5/14/1992 | 24 | 23 | 23 | Inv9 |
| 5/14/1992 | 96 | 105 | 119 | Inv9 |
| 5/14/1992 | 68 | 60 | 60 | Inv9 |
| 5/14/1992 | 170 | 186 | 175 | Inv9 |
| 4/23/1992 | 266 | 247 | 262 | Inv9 |
| 4/23/1992 | 170 | 151 | 156 | Inv9 |
| 4/23/1992 | 66 | 66 | 56 | Inv9 |
| 4/23/1992 | 22 | 13 | 27 | Inv9 |
| 4/23/1992 | 1 | 1 | 4 | Inv9 |
| 4/23/1992 | 10 | 13 | 12 | Inv9 |
| 4/23/1992 | 320 | 311 | 312 | Inv9 |
| 4/23/1992 | 194 | 192 | 203 | Inv9 |
| 4/23/1992 | 238 | 228 | 215 | Inv9 |
| 4/23/1992 | 94 | 81 | 79 | Inv9 |
| 4/23/1992 | 17 | 22 | 36 | Inv9 |
| 4/23/1992 | 14 | 8 | 13 | Inv9 |
| 4/23/1992 | 14 | 18 | 11 | Inv9 |
| 4/23/1992 | 5 | 3 | 6 | Inv9 |
| 4/23/1992 | 28 | 29 | 26 | Inv9 |
| 4/23/1992 | 33 | 20 | 37 | Inv9 |
| 4/23/1992 | 14 | 39 | 32 | Inv9 |

**Outside Lab 1 Colonies**

| date | col1 | col2 | col3 |
|---|---|---|---|
| 2/4/2010 | 54 | 55 | 59 |
|  | 47 | 60 | 47 |
| 2/5/2010 | 55 | 60 | 53 |
|  | 58 | 54 | 59 |
|  | 17 | 17 | 15 |
| 2/12/2012 | 65 | 64 | 55 |
| 2/15/2012 | 64 | 57 | 73 |
|  | 84 | 109 | 89 |
| 2/17/2012 | 64 | 64 | 62 |
|  | 68 | 57 | 68 |
| 2/19/2012 | 66 | 65 | 78 |
|  | 71 | 72 | 80 |
|  | 61 | 61 | 77 |
|  | 66 | 70 | 66 |
| 2/22/2012 | 96 | 102 | 104 |
| 2/24/2012 | 93 | 94 | 102 |
|  | 70 | 72 | 81 |
| 2/26/2012 | 70 | 70 | 78 |
| 3/8/2012 | 72 | 80 | 70 |
|  | 74 | 72 | 83 |
|  | 109 | 97 | 89 |
|  | 30 | 53 | 23 |
| 3/15/2012 | 77 | 82 | 76 |
|  | 72 | 80 | 70 |
|  | 74 | 67 | 84 |

| Date | | | |
|---|---|---|---|
| 4/6/2012 | 75 | 78 | 88 |
| 4/7/2012 | 85 | 80 | 85 |
| | 100 | 105 | 98 |
| | 64 | 76 | 60 |
| 4/8/2012 | 81 | 89 | 79 |
| 4/15/2012 | 56 | 67 | 57 |
| | 42 | 51 | 40 |
| 4/16/2012 | 77 | 60 | 67 |
| 4/19/2012 | 83 | 83 | 84 |
| | 68 | 63 | 55 |
| 4/20/2012 | 82 | 81 | 91 |
| | 49 | 45 | 57 |
| 4/21/2012 | 84 | 75 | 83 |
| | 63 | 71 | 76 |
| 4/22/2012 | 77 | 73 | 75 |
| 4/26/2012 | 89 | 72 | 85 |
| | 74 | 82 | 83 |
| 4/30/2012 | 83 | 77 | 80 |
| 5/4/2012 | 74 | 74 | 69 |
| | 66 | 65 | 75 |
| | 48 | 64 | 55 |
| 5/10/2012 | 69 | 63 | 57 |
| 5/12/2012 | 80 | 82 | 80 |
| | 47 | 58 | 55 |
| 5/21/2012 | 72 | 63 | 70 |

**RTS Coulters**

| Date | cou1 | cou2 | cou3 |
|---|---|---|---|
| 10/20/1997 | 531 | 508 | 541 |
| | 650 | 626 | 595 |
| | 460 | 455 | 468 |
| | 550 | 530 | 538 |
| | 466 | 468 | 452 |
| | 567 | 555 | 521 |
| | 558 | 581 | 636 |
| | 567 | 563 | 537 |
| | 594 | 550 | 543 |
| | 611 | 599 | 507 |
| 10/31/1997 | 548 | 490 | 532 |
| | 295 | 270 | 257 |
| | 693 | 622 | 586 |
| | 429 | 456 | 408 |
| | 414 | 407 | 406 |
| | 581 | 551 | 550 |
| | 535 | 507 | 543 |
| | 491 | 493 | 460 |
| | 358 | 384 | 370 |
| | 376 | 355 | 340 |
| 11/10/1997 | 539 | 543 | 579 |
| | 628 | 619 | 587 |
| | 678 | 703 | 705 |
| | 582 | 549 | 543 |
| | 626 | 702 | 604 |
| | 732 | 713 | 743 |
|  | 785 | 781 | 634 |
|  | 441 | 450 | 469 |
|  | 544 | 550 | 562 |
|  | 556 | 557 | 550 |
|  | 857 | 855 | 870 |
|  | 832 | 859 | 860 |
|  | 827 | 814 | 755 |
|  | 854 | 852 | 882 |
|  | 721 | 733 | 760 |
|  | 859 | 845 | 827 |
|  | 884 | 872 | 796 |
|  | 910 | 880 | 893 |
|  | 836 | 865 | 809 |
|  | 735 | 755 | 754 |
| 11/24/1997 | 568 | 555 | 533 |
|  | 570 | 512 | 500 |
|  | 543 | 562 | 545 |
|  | 672 | 650 | 660 |
|  | 635 | 649 | 655 |
|  | 557 | 549 | 572 |
|  | 636 | 609 | 634 |
|  | 585 | 542 | 524 |
|  | 698 | 675 | 680 |
|  | 498 | 512 | 475 |
| 12/4/1997 | 498 | 504 | 532 |
|  | 769 | 790 | 711 |
|  | 799 | 785 | 765 |
|  | 630 | 645 | 659 |
|  | 669 | 659 | 660 |
|  | 765 | 745 | 730 |
|  | 650 | 672 | 721 |
|  | 814 | 805 | 767 |
|  | 732 | 719 | 674 |
|  | 814 | 769 | 742 |
| 12/15/1997 | 631 | 639 | 676 |
|  | 641 | 603 | 639 |
|  | 561 | 592 | 594 |
|  | 659 | 617 | 607 |
|  | 644 | 637 | 634 |
|  | 698 | 711 | 685 |
|  | 718 | 695 | 661 |
|  | 617 | 594 | 633 |
|  | 717 | 681 | 669 |
|  | 643 | 581 | 585 |
| 1/19/1998 | 535 | 535 | 530 |
|  | 620 | 625 | 628 |
|  | 476 | 519 | 526 |
|  | 602 | 624 | 589 |
|  | 482 | 489 | 510 |
|  | 562 | 555 | 547 |
|  | 495 | 485 | 482 |
|  | 561 | 567 | 570 |
|  | 490 | 515 | 505 |
|  | 489 | 482 | 484 |
| 2/2/1998 | 290 | 310 | 324 |
|  | 514 | 536 | 510 |
|            | 539 | 535 | 487 |
|            | 484 | 464 | 469 |
|            | 438 | 436 | 422 |
|            | 499 | 486 | 503 |
|            | 517 | 537 | 493 |
|            | 439 | 464 | 467 |
|            | 442 | 429 | 440 |
| 2/13/1998  | 578 | 548 | 552 |
|            | 579 | 551 | 585 |
|            | 641 | 624 | 654 |
|            | 591 | 586 | 591 |
|            | 536 | 550 | 533 |
|            | 585 | 588 | 550 |
|            | 483 | 469 | 487 |
|            | 609 | 604 | 617 |
|            | 590 | 582 | 550 |
|            | 558 | 581 | 544 |
| 2/6/1998   | 679 | 688 | 644 |
|            | 480 | 458 | 469 |
|            | 554 | 539 | 513 |
|            | 586 | 588 | 611 |
|            | 594 | 564 | 549 |
|            | 650 | 625 | 638 |
|            | 451 | 456 | 440 |
|            | 494 | 486 | 527 |
|            | 440 | 432 | 443 |
|            | 573 | 580 | 595 |
| 2/16/1998  | 616 | 658 | 615 |
|  | 632 | 638 | 606 |
|  | 581 | 590 | 620 |
|  | 548 | 536 | 517 |
|  | 509 | 525 | 496 |
|  | 519 | 528 | 514 |
|  | 455 | 498 | 444 |
|  | 506 | 520 | 531 |
|  | 537 | 545 | 558 |
|  | 577 | 572 | 542 |
| 2/20/1998 | 647 | 633 | 621 |
|  | 621 | 619 | 651 |
|  | 700 | 684 | 645 |
|  | 610 | 605 | 611 |
|  | 602 | 617 | 633 |
|  | 703 | 733 | 739 |
|  | 640 | 628 | 664 |
|  | 753 | 735 | 721 |
|  | 749 | 738 | 698 |
|  | 675 | 669 | 645 |
| 3/2/1998 | 274 | 279 | 261 |
|  | 292 | 317 | 320 |
|  | 293 | 282 | 270 |
|  | 311 | 322 | 291 |
|  | 309 | 295 | 328 |
|  | 332 | 334 | 337 |
|  | 318 | 317 | 323 |
|  | 315 | 299 | 305 |
|  | 303 | 283 | 283 |
|            | 263  | 254  | 249  |
| 3/6/1998   | 638  | 592  | 608  |
|            | 661  | 713  | 639  |
|            | 689  | 716  | 687  |
|            | 622  | 592  | 590  |
|            | 679  | 691  | 673  |
|            | 606  | 581  | 573  |
|            | 556  | 567  | 576  |
|            | 744  | 748  | 720  |
|            | 563  | 573  | 579  |
|            | 633  | 595  | 609  |
| 3/9/1998   | 630  | 631  | 650  |
|            | 777  | 791  | 796  |
|            | 792  | 783  | 791  |
|            | 680  | 656  | 654  |
|            | 752  | 741  | 734  |
|            | 783  | 765  | 743  |
|            | 803  | 818  | 827  |
|            | 750  | 784  | 764  |
|            | 854  | 828  | 823  |
| 3/13/1998  | 1398 | 1378 | 1344 |
|            | 1486 | 1469 | 1463 |
|            | 1671 | 1696 | 1651 |
|            | 1613 | 1622 | 1588 |
|            | 1823 | 1832 | 1845 |
|            | 1695 | 1651 | 1708 |
|            | 1568 | 1593 | 1551 |
|            | 1692 | 1663 | 1709 |
|           | 1562 | 1525 | 1540 |
| 3/23/1998 | 886  | 876  | 890  |
|           | 997  | 983  | 972  |
|           | 1051 | 1040 | 1047 |
|           | 1051 | 1023 | 1025 |
|           | 949  | 969  | 947  |
|           | 1065 | 1066 | 1031 |
|           | 1012 | 1059 | 1064 |
|           | 1044 | 1016 | 976  |
|           | 1053 | 1057 | 1014 |
|           | 926  | 957  | 903  |
| 3/27/1998 | 464  | 461  | 449  |
|           | 488  | 512  | 479  |
|           | 617  | 609  | 619  |
|           | 664  | 668  | 651  |
|           | 647  | 654  | 632  |
|           | 587  | 572  | 569  |
|           | 687  | 689  | 617  |
|           | 626  | 595  | 624  |
|           | 640  | 622  | 606  |
|           | 645  | 652  | 628  |
| 4/13/1998 | 893  | 903  | 863  |
|           | 416  | 402  | 478  |
|           | 767  | 771  | 802  |
|           | 845  | 802  | 793  |
|           | 668  | 575  | 578  |
|           | 874  | 876  | 858  |
|           | 874  | 854  | 807  |
|  | 825 | 767 | 777 |
|  | 426 | 425 | 413 |
| 4/17/1998 | 468 | 463 | 478 |
|  | 458 | 485 | 481 |
|  | 484 | 457 | 450 |
|  | 439 | 422 | 446 |
|  | 419 | 436 | 411 |
|  | 482 | 454 | 494 |
|  | 517 | 515 | 525 |
|  | 560 | 563 | 593 |
|  | 434 | 472 | 468 |
|  | 388 | 382 | 368 |
| 4/24/1998 | 776 | 760 | 711 |
|  | 701 | 667 | 684 |
|  | 695 | 711 | 690 |
|  | 725 | 750 | 759 |
|  | 878 | 892 | 868 |
|  | 719 | 688 | 661 |
|  | 704 | 702 | 755 |
|  | 772 | 780 | 744 |
|  | 664 | 636 | 612 |
|  | 835 | 806 | 810 |
| 4/27/1998 | 468 | 463 | 448 |
|  | 414 | 413 | 427 |
|  | 480 | 469 | 476 |
|  | 448 | 475 | 458 |
|  | 432 | 424 | 447 |
|  | 429 | 409 | 408 |
|  | 437 | 425 | 438 |
|  | 375 | 366 | 394 |
|  | 423 | 456 | 418 |
|  | 405 | 400 | 412 |
| 5/1/1998 | 589 | 529 | 512 |
|  | 564 | 560 | 585 |
|  | 627 | 605 | 593 |
|  | 518 | 461 | 460 |
|  | 695 | 667 | 695 |
|  | 730 | 721 | 749 |
|  | 621 | 599 | 613 |
|  | 672 | 687 | 666 |
|  | 576 | 523 | 533 |
|  | 458 | 421 | 475 |
| 5/8/1998 | 570 | 584 | 588 |
|  | 734 | 654 | 715 |
|  | 593 | 527 | 549 |
|  | 531 | 478 | 468 |
|  | 668 | 601 | 639 |
|  | 668 | 717 | 727 |
|  | 542 | 552 | 526 |
|  | 737 | 703 | 696 |
|  | 471 | 411 | 401 |
|  | 531 | 535 | 503 |
| 5/11/1998 | 511 | 506 | 469 |
|  | 460 | 417 | 418 |
|  | 432 | 452 | 463 |
|  | 501 | 487 | 482 |
|            | 428 | 420 | 444 |
|            | 422 | 407 | 427 |
|            | 398 | 402 | 419 |
|            | 414 | 408 | 441 |
|            | 403 | 405 | 413 |
|            | 455 | 429 | 427 |
| 5/15/1998  | 298 | 252 | 244 |
|            | 300 | 285 | 279 |
|            | 370 | 338 | 351 |
|            | 274 | 236 | 250 |
|            | 286 | 232 | 221 |
|            | 465 | 421 | 449 |
|            | 474 | 384 | 416 |
|            | 424 | 369 | 378 |
|            | 235 | 211 | 205 |
|            | 348 | 333 | 317 |
| 5/18/1998  | 480 | 490 | 499 |
|            | 392 | 415 | 442 |
|            | 311 | 325 | 309 |
|            | 375 | 390 | 392 |
|            | 350 | 329 | 341 |
|            | 425 | 419 | 433 |
|            | 280 | 270 | 262 |
|            | 355 | 342 | 352 |
|            | 285 | 270 | 272 |
|            | 475 | 490 | 411 |
| 5/22/1998  | 321 | 317 | 316 |
|            | 354 | 320 | 355 |
|   | 356 | 352 | 346 |
|   | 370 | 350 | 355 |
|   | 324 | 322 | 316 |
|   | 326 | 330 | 318 |
|   | 328 | 318 | 312 |
|   | 309 | 327 | 323 |
|   | 327 | 320 | 319 |
|   | 314 | 311 | 330 |
| 5/25/1998 | 838 | 818 | 849 |
|   | 842 | 818 | 812 |
|   | 929 | 940 | 967 |
|   | 857 | 835 | 799 |
|   | 847 | 823 | 802 |
|   | 765 | 789 | 767 |
|   | 861 | 865 | 801 |
|   | 827 | 869 | 849 |
|   | 903 | 910 | 941 |
|   | 854 | 818 | 854 |
| 5/29/1998 | 755 | 710 | 725 |
|   | 733 | 744 | 695 |
|   | 687 | 690 | 645 |
|   | 714 | 726 | 719 |
|   | 828 | 847 | 837 |
|   | 667 | 613 | 611 |
|   | 656 | 693 | 675 |
|   | 678 | 627 | 664 |
|   | 740 | 728 | 738 |
|   | 846 | 847 | 885 |

| Date | | | |
|---|---|---|---|
| 6/19/1998 | 753 | 759 | 734 |
| | 769 | 780 | 752 |
| | 832 | 856 | 842 |
| | 861 | 868 | 867 |
| | 766 | 708 | 704 |
| | 872 | 862 | 869 |
| | 869 | 875 | 892 |
| | 855 | 862 | 872 |
| | 924 | 908 | 904 |
| | 769 | 777 | 748 |
| 6/22/1998 | 741 | 765 | 767 |
| | 778 | 775 | 814 |
| | 805 | 823 | 811 |
| | 809 | 783 | 825 |
| | 774 | 746 | 778 |
| | 775 | 787 | 800 |
| | 815 | 816 | 828 |
| | 816 | 823 | 787 |
| | 821 | 825 | 807 |
| | 837 | 822 | 774 |
| 6/26/1998 | 781 | 730 | 765 |
| | 695 | 687 | 726 |
| | 677 | 672 | 633 |
| | 698 | 695 | 702 |
| | 833 | 865 | 865 |
| | 591 | 612 | 625 |
| | 752 | 770 | 787 |
| | 593 | 580 | 597 |
|           | 659  | 651  | 658  |
|           | 659  | 618  | 636  |
| 7/3/1998  | 821  | 850  | 830  |
|           | 865  | 832  | 834  |
|           | 829  | 790  | 791  |
|           | 805  | 782  | 792  |
|           | 903  | 893  | 896  |
|           | 725  | 719  | 749  |
|           | 903  | 933  | 882  |
|           | 793  | 746  | 730  |
|           | 807  | 822  | 816  |
|           | 831  | 819  | 811  |
| 7/6/1998  | 746  | 740  | 752  |
|           | 753  | 782  | 798  |
|           | 791  | 796  | 780  |
|           | 749  | 751  | 761  |
|           | 785  | 729  | 713  |
|           | 767  | 777  | 785  |
|           | 682  | 685  | 687  |
|           | 691  | 654  | 667  |
|           | 677  | 648  | 659  |
|           | 674  | 654  | 634  |
| 7/17/1998 | 636  | 607  | 620  |
|           | 669  | 675  | 659  |
|           | 712  | 680  | 722  |
|           | 960  | 1040 | 1021 |
|           | 1170 | 1243 | 1077 |
|           | 1263 | 1191 | 1202 |
|  | 861 | 842 | 806 |
|  | 1219 | 1249 | 1269 |
|  | 880 | 880 | 866 |
|  | 1279 | 1266 | 1220 |
| 7/20/1998 | 475 | 469 | 443 |
|  | 432 | 426 | 456 |
|  | 565 | 578 | 557 |
|  | 474 | 469 | 441 |
|  | 464 | 482 | 486 |
|  | 130 | 118 | 118 |
|  | 462 | 457 | 463 |
|  | 475 | 476 | 476 |
|  | 481 | 472 | 491 |
|  | 476 | 483 | 495 |
| 7/24/1998 | 734 | 740 | 724 |
|  | 863 | 851 | 861 |
|  | 1000 | 1064 | 1032 |
|  | 825 | 857 | 867 |
|  | 850 | 829 | 834 |
|  | 764 | 771 | 779 |
|  | 965 | 959 | 974 |
|  | 487 | 499 | 495 |
|  | 815 | 834 | 831 |
|  | 448 | 454 | 421 |
| 7/27/1998 | 815 | 815 | 834 |
|  | 793 | 787 | 796 |
|  | 720 | 738 | 738 |
|  | 777 | 822 | 774 |
|            | 856  | 821  | 799  |
|            | 831  | 853  | 824  |
|            | 653  | 679  | 670  |
|            | 683  | 678  | 687  |
|            | 779  | 759  | 768  |
|            | 921  | 932  | 927  |
| 7/28/1998  | 3329 | 3257 | 3268 |
|            | 3121 | 3243 | 3214 |
|            | 2696 | 2537 | 2605 |
|            | 2401 | 2459 | 2437 |
|            | 2646 | 2537 | 2605 |
|            | 2401 | 2454 | 2437 |
| 7/29/1998  | 2264 | 2118 | 2205 |
|            | 533  | 551  | 535  |
|            | 3376 | 3344 | 3256 |
| 7/31/1998  | 580  | 574  | 545  |
|            | 654  | 603  | 624  |
|            | 599  | 614  | 603  |
|            | 585  | 597  | 541  |
|            | 655  | 606  | 630  |
|            | 662  | 602  | 637  |
|            | 618  | 633  | 581  |
|            | 574  | 558  | 560  |
|            | 597  | 583  | 575  |
|            | 615  | 593  | 619  |
| 8/3/1998   | 696  | 698  | 700  |
|            | 668  | 650  | 645  |
|            | 632  | 630  | 625  |
|            | 575  | 561  | 549  |
|            | 718  | 708  | 709  |
|            | 715  | 708  | 706  |
|            | 740  | 737  | 726  |
|            | 728  | 716  | 733  |
|            | 685  | 693  | 688  |
|            | 763  | 749  | 755  |
| 8/11/1998  | 3382 | 3234 | 3202 |
|            | 2717 | 2600 | 2620 |
|            | 3810 | 3745 | 3719 |
|            | 4507 | 4409 | 4343 |
| 8/17/1998  | 443  | 441  | 410  |
|            | 443  | 432  | 408  |
|            | 371  | 374  | 370  |
|            | 428  | 452  | 409  |
|            | 386  | 393  | 400  |
|            | 428  | 411  | 404  |
|            | 418  | 403  | 393  |
|            | 393  | 415  | 379  |
|            | 378  | 370  | 375  |
|            | 345  | 388  | 393  |
| 8/21/1998  | 514  | 536  | 490  |
|            | 626  | 606  | 612  |
|            | 781  | 634  | 652  |
|            | 505  | 517  | 482  |
|            | 511  | 492  | 482  |
|            | 620  | 602  | 614  |
|            | 588  | 578  | 569  |
|            | 556  | 536  | 521  |
|            | 456  | 478  | 477  |
|            | 461  | 452  | 442  |
| 8/24/1998  | 655  | 615  | 645  |
|            | 639  | 673  | 656  |
|            | 776  | 762  | 786  |
|            | 556  | 551  | 518  |
|            | 585  | 540  | 540  |
|            | 568  | 529  | 527  |
|            | 734  | 756  | 712  |
|            | 573  | 544  | 546  |
|            | 569  | 546  | 529  |
|            | 552  | 546  | 509  |
| 8/27/1998  | 456  | 432  | 441  |
|            | 520  | 556  | 531  |
|            | 475  | 485  | 461  |
|            | 572  | 536  | 559  |
|            | 595  | 606  | 585  |
|            | 499  | 512  | 526  |
|            | 602  | 617  | 622  |
|            | 566  | 555  | 545  |
|            | 523  | 535  | 517  |
|            | 495  | 505  | 485  |
| 9/2/1998   | 4070 | 4120 | 4175 |
|            | 2861 | 2779 | 2740 |
|            | 2325 | 2340 | 2318 |
|            | 3305 | 3297 | 3291 |
| 9/4/1998   | 5110 | 5125 | 5213 |
|  | 5007 | 5107 | 5123 |
|  | 4813 | 4918 | 4925 |
|  | 4439 | 4556 | 4579 |
|  | 4322 | 4429 | 4372 |
| 9/18/1998 | 7786 | 7540 | 7677 |
|  | 6682 | 6639 | 6407 |
|  | 2841 | 2755 | 2649 |
|  | 2842 | 2855 | 2811 |
|  | 3561 | 3647 | 3466 |
| 9/24/1998 | 5504 | 5457 | 5442 |
|  | 4553 | 4591 | 4454 |
|  | 3616 | 3360 | 3327 |
|  | 2823 | 2894 | 2891 |
|  | 2141 | 2114 | 2198 |
|  | 2394 | 2289 | 2201 |
| 9/25/1998 | 6394 | 6298 | 6309 |
|  | 6256 | 6082 | 6272 |
|  | 4104 | 4076 | 3986 |
|  | 2993 | 3056 | 2949 |
|  | 2691 | 2485 | 2510 |
|  | 1151 | 1142 | 1156 |
| 9/29/1998 | 420 | 435 | 455 |
|  | 465 | 440 | 421 |
|  | 425 | 411 | 401 |
|  | 398 | 388 | 375 |
|  | 435 | 422 | 417 |
|  | 455 | 426 | 436 |
|  | 375 | 381 | 394 |
|            | 355  | 372  | 381  |
|            | 402  | 398  | 385  |
|            | 454  | 418  | 432  |
| 10/12/1998 | 472  | 426  | 413  |
|            | 475  | 429  | 414  |
|            | 416  | 400  | 411  |
|            | 648  | 632  | 622  |
|            | 606  | 592  | 631  |
|            | 542  | 535  | 558  |
|            | 429  | 455  | 435  |
|            | 494  | 533  | 509  |
|            | 593  | 624  | 609  |
|            | 517  | 517  | 510  |
| 10/5/1998  | 720  | 693  | 680  |
|            | 685  | 675  | 679  |
|            | 475  | 462  | 478  |
|            | 601  | 620  | 611  |
|            | 593  | 569  | 568  |
|            | 568  | 573  | 593  |
|            | 531  | 525  | 506  |
|            | 555  | 503  | 514  |
|            | 514  | 532  | 557  |
|            | 529  | 510  | 524  |
| 10/6/1998  | 6510 | 6439 | 6552 |
|            | 8013 | 8060 | 8091 |
|            | 3379 | 3183 | 3272 |
|            | 4480 | 4372 | 4497 |
|            | 3423 | 3524 | 3374 |
|            | 5222 | 5024 | 5003 |
| 10/7/1998  | 3830 | 3833 | 3740 |
|            | 5729 | 5881 | 5876 |
|            | 2568 | 2444 | 2384 |
|            | 5224 | 5234 | 5053 |
|            | 3412 | 3363 | 3374 |
|            | 3116 | 3086 | 3028 |
| 10/14/1998 | 3830 | 3833 | 3740 |
|            | 5729 | 5881 | 5876 |
|            | 2568 | 2444 |      |
|            | 5224 | 5234 | 5053 |
|            | 3412 | 3363 | 3374 |
|            | 3116 | 3086 | 3028 |
| 10/14/1998 | 587  | 576  | 569  |
|            | 619  | 666  | 648  |
|            | 595  | 576  | 563  |
|            | 619  | 617  | 605  |
|            | 637  | 634  | 648  |
|            | 558  | 548  | 543  |
|            | 591  | 626  | 597  |
|            | 478  | 482  | 464  |
|            | 563  | 591  | 558  |
|            | 517  | 548  | 569  |
| 10/16/1998 | 485  | 502  | 498  |
|            | 497  | 462  | 444  |
|            | 411  | 399  | 436  |
|            | 428  | 409  | 425  |
|            | 389  | 389  | 374  |

| Date | | | |
|---|---|---|---|
| | 495 | 472 | 458 |
| 10/23/1998 | 673 | 633 | 695 |
| | 691 | 730 | 740 |
| | 723 | 744 | 713 |
| | 844 | 832 | 865 |
| | 720 | 702 | 664 |
| | 640 | 702 | 671 |
| | 935 | 915 | 945 |
| | 841 | 782 | 771 |
| | 683 | 645 | 654 |
| | 508 | 522 | 505 |
| 10/26/1998 | 736 | 739 | 682 |
| | 706 | 741 | 744 |
| | 798 | 771 | 767 |
| | 775 | 785 | 798 |
| | 791 | 773 | 759 |
| | 899 | 892 | 902 |
| | 654 | 648 | 633 |
| | 907 | 948 | 928 |
| | 982 | 992 | 989 |
| | 991 | 961 | 961 |
| 11/3/1998 | 418 | 422 | 425 |
| | 501 | 447 | 503 |
| | 487 | 473 | 463 |
| | 461 | 454 | 466 |
| | 478 | 448 | 450 |
| | 460 | 488 | 475 |
| | 434 | 442 | 428 |
| | 376 | 344 | 352 |
| | 414 | 422 | 420 |
| | 385 | 373 | 402 |
| 11/7/1998 | 415 | 404 | 383 |
| | 383 | 352 | 373 |
| | 346 | 337 | 356 |
| | 363 | 361 | 372 |
| | 362 | 365 | 395 |
| | 354 | 395 | 377 |
| | 359 | 366 | 344 |
| | 385 | 368 | 353 |
| | 368 | 350 | 362 |
| | 444 | 455 | 473 |
| 11/13/1998 | 484 | 471 | 465 |
| | 520 | 490 | 507 |
| | 539 | 541 | 535 |
| | 583 | 570 | 541 |
| | 541 | 548 | 567 |
| | 488 | 481 | 445 |
| | 533 | 584 | 591 |
| | 432 | 432 | 426 |
| | 644 | 669 | 626 |
| | 649 | 624 | 638 |
| 11/16/1998 | 666 | 618 | 638 |
| | 634 | 638 | 611 |
| | 589 | 565 | 548 |
| | 653 | 613 | 647 |
| | 569 | 562 | 580 |
|  | 624 | 605 | 604 |
|  | 535 | 524 | 558 |
|  | 566 | 543 | 531 |
|  | 691 | 688 | 677 |
|  | 637 | 628 | 618 |
| 11/23/1998 | 501 | 524 | 474 |
|  | 475 | 472 | 441 |
|  | 437 | 458 | 424 |
|  | 616 | 630 | 622 |
|  | 367 | 334 | 338 |
|  | 447 | 441 | 432 |
|  | 530 | 511 | 520 |
|  | 552 | 549 | 564 |
|  | 512 | 490 | 537 |
|  | 484 | 466 | 452 |
| 12/4/1998 | 252 | 262 | 270 |
|  | 245 | 221 | 259 |
|  | 276 | 236 | 249 |
|  | 302 | 279 | 271 |
|  | 284 | 294 | 291 |
|  | 404 | 416 | 426 |
|  | 299 | 295 | 336 |
|  | 244 | 244 | 231 |
|  | 231 | 225 | 228 |
|  | 274 | 288 | 300 |
| 12/7/1998 | 452 | 422 | 439 |
|  | 492 | 475 | 444 |
|  | 513 | 496 | 488 |
|  | 443 | 432 | 430 |
|  | 549 | 568 | 570 |
|  | 548 | 566 | 571 |
|  | 515 | 506 | 472 |
|  | 490 | 548 | 543 |
|  | 412 | 395 | 393 |
|  | 451 | 438 | 426 |
| 12/8/1998 | 371 | 349 | 361 |
|  | 355 | 359 | 372 |
|  | 340 | 325 | 322 |
|  | 309 | 322 | 330 |
|  | 237 | 250 | 245 |
|  | 319 | 333 | 323 |
|  | 366 | 342 | 353 |
|  | 497 | 480 | 495 |
|  | 445 | 460 | 432 |
|  | 421 | 409 | 415 |
| 12/9/1998 | 543 | 516 | 541 |
|  | 500 | 502 | 513 |
|  | 460 | 482 | 434 |
|  | 531 | 542 | 504 |
|  | 531 | 493 | 488 |
|  | 542 | 530 | 518 |
|  | 512 | 473 | 457 |
|  | 514 | 494 | 487 |
|  | 454 | 439 | 444 |
|  | 469 | 490 | 450 |
| 12/20/1999 | 545 | 567 | 551 |
|            | 552 | 543 | 541 |
|            | 468 | 455 | 449 |
|            | 649 | 683 | 666 |
|            | 562 | 523 | 550 |
|            | 605 | 590 | 603 |
|            | 588 | 574 | 614 |
|            | 750 | 767 | 739 |
|            | 789 | 772 | 769 |
|            | 721 | 719 | 710 |
|            | 630 | 666 | 639 |
|            | 612 | 631 | 645 |
|            | 565 | 574 | 591 |
|            | 638 | 610 | 580 |
| 12/21/1998 | 770 | 793 | 789 |
|            | 721 | 756 | 721 |
|            | 751 | 777 | 757 |
|            | 721 | 694 | 690 |
|            | 743 | 754 | 768 |
|            | 759 | 766 | 724 |
|            | 753 | 740 | 713 |
|            | 750 | 754 | 775 |
|            | 707 | 687 | 674 |
|            | 620 | 625 | 620 |
| 1/1/1999   | 518 | 550 | 569 |
|            | 482 | 500 | 501 |
|            | 458 | 485 | 510 |
|            | 541 | 483 | 488 |
|            | 531 | 525 | 523 |
|  | 609 | 657 | 652 |
|  | 506 | 524 | 485 |
|  | 432 | 440 | 428 |
|  | 497 | 453 | 507 |
|  | 453 | 475 | 508 |
| 1/8/1999 | 425 | 454 | 451 |
|  | 446 | 417 | 425 |
|  | 463 | 463 | 456 |
|  | 412 | 429 | 430 |
|  | 404 | 429 | 411 |
|  | 420 | 417 | 401 |
|  | 430 | 421 | 413 |
|  | 386 | 388 | 393 |
|  | 443 | 464 | 437 |
|  | 453 | 469 | 479 |
| 1/15/1999 | 488 | 482 | 478 |
|  | 510 | 472 | 463 |
|  | 517 | 505 | 506 |
|  | 468 | 458 | 455 |
|  | 528 | 527 | 524 |
|  | 622 | 648 | 609 |
|  | 442 | 422 | 432 |
|  | 467 | 438 | 438 |
|  | 481 | 452 | 522 |
|  | 487 | 516 | 482 |
| 1/18/1999 | 505 | 478 | 482 |
|  | 438 | 424 | 432 |
|  | 435 | 397 | 381 |
|  | 379 | 394 | 422 |
|  | 418 | 417 | 397 |
|  | 365 | 345 | 313 |
|  | 476 | 495 | 462 |
|  | 478 | 439 | 449 |
|  | 511 | 488 | 490 |
|  | 467 | 427 | 399 |
| 1/22/1999 | 345 | 335 | 376 |
|  | 467 | 461 | 415 |
|  | 380 | 384 | 377 |
|  | 525 | 537 | 519 |
|  | 474 | 452 | 491 |
|  | 434 | 413 | 448 |
|  | 465 | 445 | 466 |
|  | 465 | 491 | 453 |
|  | 449 | 459 | 467 |
|  | 411 | 426 | 445 |
| 1/29/1999 | 664 | 644 | 696 |
|  | 716 | 726 | 721 |
|  | 651 | 675 | 664 |
|  | 643 | 681 | 636 |
|  | 700 | 690 | 717 |
|  | 618 | 618 | 657 |
|  | 684 | 691 | 653 |
|  | 536 | 505 | 557 |
|  | 642 | 626 | 633 |
|  | 541 | 558 | 569 |
| 2/5/1999 | 672 | 677 | 708 |
|  | 758 | 734 | 726 |
|  | 759 | 777 | 757 |
|  | 752 | 745 | 733 |
|  | 675 | 690 | 691 |
|  | 700 | 677 | 662 |
|  | 704 | 685 | 674 |
|  | 663 | 698 | 681 |
|  | 661 | 662 | 650 |
|  | 693 | 671 | 681 |
| 2/15/1999 | 711 | 745 | 735 |
|  | 852 | 837 | 860 |
|  | 811 | 799 | 789 |
|  | 735 | 745 | 760 |
|  | 809 | 811 | 835 |
|  | 799 | 805 | 811 |
|  | 760 | 741 | 752 |
|  | 715 | 721 | 726 |
|  | 760 | 775 | 745 |
|  | 801 | 812 | 823 |
| 2/19/1999 | 645 | 677 | 655 |
|  | 611 | 620 | 637 |
|  | 677 | 689 | 667 |
|  | 707 | 699 | 701 |
|  | 635 | 649 | 617 |
|  | 690 | 681 | 678 |
|  | 660 | 670 | 652 |
|  | 601 | 622 | 634 |
|  | 640 | 656 | 654 |
|            | 685 | 695 | 690 |
| 2/22/1999  | 760 | 747 | 730 |
|            | 745 | 744 | 736 |
|            | 690 | 701 | 685 |
|            | 712 | 722 | 735 |
|            | 747 | 765 | 742 |
|            | 675 | 665 | 681 |
|            | 640 | 655 | 631 |
|            | 672 | 662 | 652 |
|            | 719 | 725 | 707 |
|            | 790 | 811 | 795 |
| 2/26/1999  | 833 | 750 | 777 |
|            | 781 | 737 | 741 |
|            | 717 | 758 | 759 |
|            | 785 | 760 | 747 |
|            | 750 | 706 | 715 |
|            | 693 | 670 | 660 |
|            | 678 | 645 | 676 |
|            | 713 | 704 | 716 |
|            | 622 | 610 | 608 |
|            | 703 | 703 | 731 |
| 3/1/1999   | 640 | 666 | 685 |
|            | 701 | 722 | 735 |
|            | 735 | 742 | 722 |
|            | 760 | 750 | 742 |
|            | 775 | 790 | 769 |
|            | 711 | 729 | 731 |
|            | 699 | 719 | 698 |
|  | 765 | 777 | 759 |
|  | 745 | 755 | 761 |
|  | 762 | 772 | 760 |
|  | 685 | 672 | 669 |
|  | 701 | 735 | 723 |
|  | 795 | 780 | 776 |
|  | 740 | 722 | 729 |
|  | 811 | 788 | 792 |
| 3/5/1999 | 480 | 495 | 478 |
|  | 450 | 477 | 458 |
|  | 575 | 589 | 567 |
|  | 435 | 449 | 429 |
|  | 445 | 455 | 462 |
|  | 540 | 559 | 547 |
|  | 560 | 550 | 542 |
|  | 610 | 625 | 609 |
|  | 675 | 685 | 687 |
| 3/12/1999 | 620 | 635 | 645 |
|  | 590 | 575 | 585 |
|  | 560 | 550 | 541 |
|  | 595 | 570 | 583 |
|  | 630 | 609 | 607 |
|  | 590 | 615 | 580 |
|  | 635 | 650 | 669 |
|  | 610 | 645 | 637 |
|  | 575 | 562 | 548 |
|  | 520 | 525 | 535 |
| 3/22/1999 | 610 | 635 | 590 |
| | 495 | 535 | 545 |
| | 570 | 552 | 535 |
| | 625 | 619 | 607 |
| | 499 | 521 | 539 |
| | 577 | 565 | 572 |
| | 537 | 555 | 542 |
| | 609 | 612 | 592 |
| | 572 | 559 | 567 |
| | 562 | 585 | 585 |
| 4/5/1999 | 712 | 725 | 730 |
| | 622 | 635 | 627 |
| | 678 | 775 | 665 |
| | 573 | 542 | 535 |
| | 545 | 562 | 569 |
| | 523 | 523 | 533 |
| | 507 | 511 | 531 |
| | 545 | 521 | 510 |
| 4/13/1999 | 186 | 187 | 189 |
| | 191 | 188 | 174 |
| | 284 | 289 | 297 |
| | 263 | 256 | 259 |
| | 352 | 324 | 360 |
| | 350 | 362 | 332 |
| | 320 | 333 | 342 |
| | 237 | 232 | 239 |
| 5/7/1999 | 425 | 435 | 420 |
| | 510 | 490 | 475 |
| | 405 | 375 | 395 |
|  | 452 | 437 | 414 |
|  | 358 | 372 | 360 |
|  | 411 | 419 | 432 |
|  | 437 | 465 | 422 |
|  | 375 | 365 | 392 |
|  | 419 | 402 | 398 |
|  | 399 | 412 | 375 |
| 5/21/1999 | 475 | 465 | 439 |
|  | 501 | 512 | 522 |
|  | 529 | 475 | 495 |
|  | 395 | 425 | 430 |
|  | 520 | 539 | 540 |
|  | 427 | 439 | 444 |
|  | 402 | 422 | 435 |
|  | 511 | 529 | 535 |
|  | 478 | 459 | 465 |
|  | 395 | 378 | 385 |
| 5/27/1999 | 427 | 439 | 450 |
|  | 511 | 527 | 530 |
|  | 411 | 398 | 422 |
|  | 378 | 409 | 415 |
|  | 425 | 440 | 421 |
|  | 471 | 485 | 492 |
|  | 332 | 350 | 371 |
|  | 409 | 419 | 429 |
|  | 452 | 429 | 435 |
|  | 322 | 350 | 342 |
| 6/11/1999 | 569 | 600 | 585 |
|  | 611 | 625 | 635 |
|  | 517 | 532 | 522 |
|  | 485 | 472 | 495 |
|  | 549 | 565 | 572 |
|  | 585 | 595 | 569 |
|  | 476 | 489 | 492 |
|  | 505 | 509 | 522 |
|  | 566 | 542 | 555 |
|  | 562 | 575 | 559 |
| 6/21/1999 | 512 | 492 | 507 |
|  | 488 | 475 | 497 |
|  | 522 | 533 | 511 |
|  | 462 | 473 | 485 |
|  | 555 | 549 | 535 |
|  | 509 | 522 | 512 |
|  | 432 | 466 | 449 |
|  | 546 | 565 | 539 |
|  | 479 | 485 | 465 |
|  | 529 | 545 | 561 |
| 6/28/1999 | 610 | 635 | 624 |
|  | 585 | 592 | 603 |
|  | 532 | 549 | 521 |
|  | 609 | 622 | 611 |
|  | 529 | 545 | 535 |
|  | 635 | 611 | 607 |
|  | 585 | 572 | 565 |
|  | 592 | 575 | 562 |
|  | 630 | 655 | 631 |
|  | 666 | 673 | 679 |
| 7/1/1999 | 609 | 588 | 611 |
|  | 652 | 641 | 627 |
|  | 666 | 632 | 645 |
|  | 612 | 608 | 631 |
|  | 745 | 765 | 722 |
|  | 644 | 651 | 632 |
|  | 702 | 722 | 732 |
|  | 650 | 675 | 662 |
| 7/12/1999 | 650 | 666 | 646 |
|  | 637 | 649 | 622 |
|  | 640 | 653 | 660 |
|  | 610 | 662 | 635 |
|  | 678 | 659 | 665 |
|  | 718 | 692 | 707 |
|  | 635 | 609 | 611 |
|  | 735 | 719 | 726 |
|  | 657 | 672 | 659 |
|  | 670 | 685 | 678 |
| 7/16/1999 | 505 | 519 | 529 |
|  | 485 | 495 | 480 |
|  | 476 | 510 | 498 |
|  | 415 | 435 | 422 |
|  | 509 | 517 | 511 |
|  | 435 | 419 | 410 |
|  | 456 | 444 | 429 |
|  | 411 | 409 | 425 |
|  | 449 | 421 | 439 |
|  | 385 | 412 | 375 |
| 7/26/1999 | 728 | 711 | 735 |
|  | 750 | 741 | 721 |
|  | 810 | 833 | 809 |
|  | 755 | 762 | 748 |
|  | 697 | 712 | 688 |
|  | 735 | 755 | 765 |
|  | 772 | 762 | 781 |
|  | 819 | 822 | 835 |
|  | 727 | 709 | 742 |
|  | 856 | 869 | 888 |
| 7/29/1999 | 540 | 565 | 562 |
|  | 511 | 525 | 533 |
|  | 542 | 555 | 532 |
|  | 509 | 522 | 531 |
|  | 488 | 492 | 499 |
|  | 569 | 582 | 569 |
|  | 611 | 609 | 619 |
|  | 532 | 511 | 529 |
|  | 550 | 562 | 542 |
|  | 495 | 485 | 502 |
| 7/30/1999 | 610 | 592 | 609 |
|  | 558 | 572 | 581 |
|  | 532 | 539 | 555 |
|  | 559 | 562 | 541 |
|  | 571 | 562 | 575 |
|  | 523 | 539 | 519 |
|  | 499 | 507 | 521 |
|            | 537 | 527 | 511 |
|            | 567 | 560 | 539 |
|            | 507 | 485 | 499 |
| 8/19/1999  | 236 | 226 | 228 |
|            | 257 | 259 | 267 |
|            | 264 | 265 | 260 |
|            | 249 | 240 | 248 |
| 8/26/1999  | 750 | 731 | 722 |
|            | 692 | 685 | 699 |
|            | 645 | 635 | 619 |
|            | 657 | 666 | 661 |
|            | 688 | 671 | 657 |
|            | 712 | 729 | 730 |
|            | 736 | 772 | 762 |
|            | 690 | 717 | 709 |
|            | 733 | 745 | 759 |
|            | 769 | 751 | 749 |
| 8/30/1999  | 652 | 673 | 669 |
|            | 701 | 685 | 711 |
|            | 623 | 639 | 644 |
|            | 645 | 666 | 653 |
|            | 629 | 642 | 647 |
|            | 602 | 613 | 629 |
|            | 677 | 652 | 669 |
|            | 712 | 701 | 719 |
|            | 698 | 687 | 679 |
|            | 735 | 719 | 709 |
| 9/10/1999  | 643 | 666 | 652 |
|  | 601 | 636 | 645 |
|  | 622 | 649 | 619 |
|  | 645 | 662 | 639 |
|  | 685 | 672 | 679 |
|  | 561 | 586 | 592 |
|  | 595 | 611 | 622 |
|  | 695 | 707 | 687 |
|  | 669 | 639 | 652 |
|  | 631 | 663 | 622 |
| 9/13/1999 | 712 | 711 | 702 |
|  | 741 | 753 | 720 |
|  | 1503 | 1524 | 1511 |
|  | 1547 | 1518 | 1539 |
|  | 1509 | 1556 | 1519 |
|  | 1401 | 1426 | 1435 |
|  | 110 | 113 | 119 |
|  | 122 | 107 | 112 |
|  | 129 | 145 | 135 |
|  | 95 | 105 | 111 |
| 10/1/1999 | 772 | 761 | 756 |
|  | 666 | 655 | 677 |
|  | 701 | 711 | 722 |
|  | 656 | 631 | 634 |
|  | 732 | 745 | 739 |
|  | 741 | 756 | 762 |
|  | 635 | 659 | 662 |
|  | 672 | 657 | 659 |
|  | 713 | 732 | 742 |
|  | 699 | 710 | 729 |
| 10/4/1999 | 499 | 488 | 502 |
|  | 436 | 456 | 462 |
|  | 522 | 532 | 542 |
|  | 536 | 542 | 539 |
|  | 561 | 572 | 585 |
|  | 437 | 452 | 462 |
|  | 501 | 533 | 529 |
|  | 490 | 471 | 478 |
|  | 522 | 535 | 542 |
|  | 531 | 555 | 563 |
| 9/24/1999 | 589 | 598 | 571 |
|  | 611 | 627 | 631 |
|  | 541 | 559 | 561 |
|  | 629 | 642 | 629 |
|  | 667 | 656 | 672 |
|  | 542 | 561 | 559 |
|  | 620 | 635 | 642 |
|  | 529 | 549 | 557 |
|  | 607 | 598 | 622 |
|  | 511 | 509 | 507 |
| 9/29/1999 | 411 | 431 | 435 |
|  | 471 | 461 | 459 |
|  | 389 | 362 | 372 |
|  | 332 | 321 | 341 |
|  | 441 | 456 | 465 |
|  | 432 | 444 | 456 |
|  | 409 | 422 | 436 |
|  | 381 | 392 | 401 |
|  | 356 | 365 | 369 |
|  | 403 | 372 | 385 |
| 9/27/1999 | 609 | 595 | 575 |
|  | 578 | 566 | 571 |
|  | 613 | 622 | 635 |
|  | 575 | 560 | 582 |
|  | 535 | 519 | 511 |
|  | 569 | 555 | 539 |
|  | 591 | 599 | 602 |
|  | 621 | 635 | 607 |
|  | 545 | 551 | 539 |
|  | 511 | 503 | 529 |
| 10/15/1999 | 527 | 509 | 512 |
|  | 545 | 562 | 571 |
|  | 551 | 563 | 549 |
|  | 490 | 485 | 472 |
|  | 567 | 579 | 582 |
|  | 579 | 561 | 555 |
|  | 611 | 592 | 589 |
|  | 542 | 562 | 559 |
|  | 499 | 522 | 535 |
|  | 572 | 559 | 563 |
| 10/28/1999 | 546 | 523 | 543 |
|  | 606 | 619 | 627 |
|  | 521 | 533 | 529 |
|  | 611 | 601 | 622 |
|  | 575 | 585 | 595 |
|            | 535 | 540 | 549 |
|            | 621 | 603 | 613 |
|            | 550 | 560 | 569 |
| 11/8/1999  | 411 | 427 | 437 |
|            | 520 | 545 | 539 |
|            | 535 | 521 | 552 |
|            | 482 | 493 | 475 |
|            | 436 | 427 | 419 |
|            | 501 | 492 | 485 |
|            | 562 | 550 | 572 |
|            | 449 | 462 | 475 |
|            | 555 | 579 | 567 |
|            | 409 | 399 | 417 |
| 11/12/1999 | 589 | 608 | 570 |
|            | 617 | 624 | 607 |
|            | 575 | 595 | 606 |
|            | 512 | 514 | 536 |
|            | 588 | 566 | 568 |
|            | 545 | 529 | 520 |
|            | 494 | 463 | 445 |
|            | 504 | 480 | 479 |
|            | 491 | 479 | 491 |
|            | 416 | 429 | 420 |
| 11/15/1999 | 511 | 490 | 485 |
|            | 507 | 482 | 499 |
|            | 573 | 548 | 566 |
|            | 563 | 574 | 558 |
|            | 493 | 483 | 472 |
|            | 456 | 440 | 465 |
|            | 459 | 429 | 431 |
|            | 411 | 419 | 395 |
|            | 385 | 399 | 379 |
|            | 339 | 350 | 341 |
| 12/6/1999  | 801 | 784 | 832 |
|            | 829 | 836 | 827 |
|            | 759 | 747 | 732 |
|            | 745 | 713 | 696 |
|            | 824 | 782 | 807 |
|            | 808 | 792 | 771 |
|            | 728 | 732 | 702 |
|            | 648 | 648 | 651 |
|            | 725 | 723 | 713 |
|            | 656 | 647 | 634 |
| 12/10/1999 | 635 | 649 | 660 |
|            | 711 | 722 | 745 |
|            | 672 | 666 | 659 |
|            | 782 | 799 | 786 |
|            | 881 | 862 | 872 |
|            | 835 | 852 | 821 |
|            | 811 | 829 | 835 |
|            | 756 | 742 | 769 |
|            | 701 | 695 | 714 |
|            | 649 | 661 | 639 |
|            | 852 | 829 | 840 |
|            | 633 | 615 | 609 |
|            | 719 | 709 | 715 |
|  | 695 | 727 | 711 |
| **12/20/1999** | 545 | 567 | 551 |
|  | 552 | 543 | 541 |
|  | 468 | 455 | 449 |
|  | 649 | 683 | 666 |
|  | 561 | 623 | 550 |
|  | 605 | 590 | 603 |
|  | 588 | 574 | 614 |
|  | 750 | 767 | 739 |
|  | 789 | 772 | 769 |
|  | 721 | 719 | 710 |
|  | 630 | 666 | 639 |
|  | 612 | 631 | 645 |
|  | 565 | 574 | 591 |
|  | 638 | 610 | 580 |
| **1/7/2000** | 525 | 540 | 523 |
|  | 560 | 574 | 550 |
|  | 251 | 249 | 248 |
|  | 446 | 424 | 430 |
|  | 396 | 373 | 396 |
|  | 452 | 432 | 429 |
|  | 423 | 443 | 437 |
|  | 383 | 367 | 385 |
|  | 375 | 379 | 408 |
|  | 401 | 398 | 399 |
|  | 345 | 365 | 340 |
|  | 395 | 367 | 385 |
|  | 436 | 449 | 458 |

| Date | | | |
|---|---|---|---|
| 1/13/2000 | 575 | 549 | 562 |
| | 492 | 488 | 475 |
| | 511 | 501 | 493 |
| | 562 | 549 | 537 |
| | 589 | 599 | 582 |
| | 575 | 565 | 559 |
| | 670 | 665 | 669 |
| | 603 | 619 | 592 |
| | 585 | 599 | 572 |
| | 435 | 425 | 411 |
| | 611 | 623 | 609 |
| | 607 | 603 | 589 |
| | 629 | 630 | 641 |
| | 688 | 698 | 679 |
| 1/14/2000 | 598 | 607 | 588 |
| | 575 | 566 | 559 |
| | 457 | 435 | 461 |
| | 522 | 535 | 542 |
| | 545 | 519 | 527 |
| | 488 | 475 | 462 |
| | 562 | 575 | 569 |
| | 513 | 522 | 530 |
| | 501 | 511 | 507 |
| | 475 | 465 | 470 |
| | 495 | 489 | 512 |
| | 445 | 439 | 452 |
| | 432 | 418 | 420 |
| | 455 | 429 | 439 |

| Date | | | |
|---|---|---|---|
| 1/20/2000 | 575 | 566 | 569 |
| | 582 | 572 | 568 |
| | 595 | 582 | 601 |
| | 555 | 565 | 549 |
| | 579 | 599 | 582 |
| | 611 | 623 | 607 |
| | 635 | 619 | 622 |
| | 601 | 609 | 611 |
| | 551 | 561 | 571 |
| | 569 | 557 | 547 |
| | 529 | 535 | 542 |
| | 565 | 570 | 568 |
| | 582 | 592 | 579 |
| | 575 | 562 | 555 |
| 2/7/2000 | 839 | 822 | 847 |
| | 727 | 710 | 734 |
| | 729 | 724 | 751 |
| | 785 | 793 | 768 |
| | 737 | 733 | 703 |
| | 824 | 832 | 843 |
| | 847 | 866 | 833 |
| | 746 | 732 | 719 |
| | 759 | 729 | 732 |
| | 703 | 695 | 697 |
| 2/14/2000 | 3976 | 3957 | 3874 |
| | 3353 | 3415 | 3299 |
| | 3681 | 3505 | 3456 |
| | 4438 | 4297 | 4389 |
|  | 5023 | 5112 | 4990 |
|  | 1899 | 1881 | 1865 |
|  | 1732 | 1694 | 1680 |
|  | 4388 | 4329 | 4362 |
|  | 3995 | 4190 | 4129 |
|  | 8816 | 8800 | 8604 |
|  | 9579 | 9516 | 9499 |
|  | 16382 | 16276 | 16852 |
|  | 13252 | 13462 | 13987 |
| 2/15/2000 | 43327 | 43020 |  |
|  | 31417 | 33740 | 33823 |
|  | 31542 | 31213 | 31563 |
|  | 43403 | 42487 | 42676 |
|  | 45592 | 47132 | 46036 |
|  | 3615 | 3711 | 3610 |
|  | 3834 | 3818 | 3800 |
|  | 1278 | 1472 | 1408 |
|  | 1316 | 1349 | 1233 |
|  | 1862 | 2125 | 1973 |
|  | 2155 | 2148 | 1934 |
|  | 296 | 320 | 330 |
|  | 309 | 365 | 294 |
| 2/17/2000 | 4745 | 4924 | 4732 |
|  | 3712 | 3811 | 3790 |
|  | 4012 | 3922 | 3939 |
|  | 4811 | 4927 | 4993 |
|  | 4110 | 4219 | 4279 |
|  | 2493 | 2337 | 2338 |
|  | 2417 | 2491 | 2364 |
|  | 5322 | 5286 | 5218 |
|  | 5198 | 5239 | 5203 |
|  | 4120 | 4132 | 4107 |
|  | 4217 | 4107 | 4198 |
|  | 13006 | 13153 | 13122 |
|  | 12895 | 12798 | 12957 |
| 2/18/2000 | 3976 | 3957 | 3874 |
|  | 3353 | 3415 | 3299 |
|  | 3681 | 3505 | 3456 |
|  | 4438 | 4297 | 4389 |
|  | 5023 | 5112 | 4990 |
|  | 1849 | 1881 | 1865 |
|  | 1732 | 1694 | 1680 |
|  | 4388 | 4329 | 4362 |
|  | 3995 | 4190 | 4129 |
|  | 8816 | 8800 | 8604 |
|  | 9579 | 9516 | 9499 |
|  | 16382 | 16276 | 16852 |
|  | 13252 | 13462 | 13897 |
| 2/28/2000 | 511 | 526 | 532 |
|  | 546 | 530 | 519 |
|  | 520 | 540 | 593 |
|  | 475 | 495 | 482 |
|  | 531 | 521 | 547 |
|  | 480 | 499 | 501 |
|  | 507 | 519 | 489 |
|  | 465 | 472 | 482 |
|            | 555 | 532 | 521 |
|            | 411 | 408 | 429 |
|            | 460 | 469 | 472 |
|            | 422 | 410 | 404 |
| 3/16/2000  | 512 | 530 | 526 |
|            | 532 | 541 | 552 |
|            | 540 | 569 | 561 |
|            | 548 | 532 | 539 |
|            | 511 | 531 | 522 |
|            | 542 | 555 | 562 |
|            | 575 | 589 | 579 |
|            | 450 | 479 | 462 |
|            | 537 | 559 | 547 |
|            | 571 | 590 | 582 |
| 3/23/2000  | 478 | 468 | 456 |
|            | 543 | 543 | 530 |
|            | 490 | 475 | 471 |
|            | 466 | 449 | 432 |
|            | 365 | 365 | 360 |
|            | 196 | 192 | 169 |
| 3/24/2000  | 532 | 549 | 559 |
|            | 574 | 560 | 552 |
|            | 579 | 582 | 563 |
|            | 575 | 559 | 571 |
|            | 532 | 555 | 543 |
|            | 565 | 575 | 562 |
|            | 511 | 529 | 539 |
|            | 498 | 475 | 481 |
|  | 519 | 545 | 529 |
|  | 550 | 566 | 542 |
| 3/30/2000 | 512 | 539 | 547 |
|  | 480 | 490 | 469 |
|  | 575 | 590 | 595 |
|  | 435 | 452 | 442 |
|  | 2490 | 2565 | 2511 |
|  | 5100 | 5223 | 5112 |
|  | 4617 | 4479 | 4579 |
|  | 16111 | 15892 | 16389 |
| 3/31/2000 | 501 | 512 | 523 |
|  | 475 | 455 | 449 |
|  | 411 | 393 | 401 |
|  | 375 | 382 | 392 |
|  | 520 | 535 | 511 |
|  | 535 | 555 | 549 |
|  | 575 | 565 | 582 |
|  | 589 | 575 | 580 |
|  | 566 | 552 | 539 |
|  | 592 | 599 | 572 |
| 4/3/2000 | 475 | 492 | 487 |
|  | 512 | 509 | 521 |
|  | 535 | 549 | 555 |
|  | 556 | 551 | 575 |
|  | 2485 | 2366 | 2392 |
|  | 5011 | 4990 | 4819 |
|  | 4103 | 4089 | 4232 |
|  | 14930 | 15320 | 15112 |

| Date | | | |
|---|---|---|---|
| 4/10/2000 | 540 | 569 | 562 |
| | 555 | 549 | 562 |
| | 515 | 529 | 535 |
| | 475 | 489 | 467 |
| | 495 | 512 | 488 |
| | 501 | 509 | 507 |
| 6/9/2000 | 157 | 164 | 193 |
| | 155 | 173 | 156 |
| | 154 | 136 | 152 |
| | 131 | 123 | 119 |
| | 161 | 152 | 166 |
| | 139 | 176 | 126 |
| | 132 | 142 | 155 |
| | 168 | 176 | 155 |
| 6/19/2000 | 592 | 569 | 575 |
| | 601 | 623 | 617 |
| | 509 | 523 | 535 |
| | 542 | 567 | 572 |
| | 511 | 532 | 542 |
| | 635 | 652 | 666 |
| | 562 | 589 | 555 |
| | 611 | 589 | 598 |
| | 581 | 592 | 571 |
| | 610 | 639 | 621 |
| 6/20/2000 | 750 | 769 | 742 |
| | 645 | 685 | 672 |
| | 711 | 739 | 742 |
| | 675 | 666 | 672 |
|            | 766 | 742 | 752 |
|            | 639 | 652 | 647 |
|            | 709 | 721 | 735 |
|            | 741 | 729 | 721 |
|            | 762 | 777 | 759 |
|            | 756 | 742 | 766 |
| 6/23/2000  | 568 | 588 | 571 |
|            | 579 | 593 | 569 |
|            | 592 | 575 | 561 |
|            | 531 | 511 | 520 |
|            | 576 | 556 | 587 |
|            | 566 | 589 | 572 |
|            | 609 | 632 | 621 |
|            | 545 | 519 | 562 |
|            | 575 | 559 | 561 |
|            | 615 | 589 | 602 |
| 6/26/2000  | 611 | 629 | 633 |
|            | 590 | 631 | 619 |
|            | 582 | 571 | 585 |
|            | 662 | 679 | 649 |
|            | 562 | 582 | 592 |
|            | 637 | 639 | 655 |
|            | 581 | 609 | 622 |
|            | 629 | 639 | 641 |
|            | 693 | 688 | 672 |
|            | 763 | 777 | 749 |
| 7/17/2000  | 549 | 562 | 571 |
|            | 501 | 522 | 519 |
|  | 528 | 511 | 532 |
|  | 560 | 569 | 542 |
|  | 507 | 521 | 511 |
|  | 482 | 472 | 495 |
|  | 582 | 595 | 579 |
|  | 537 | 552 | 542 |
|  | 477 | 468 | 492 |
|  | 544 | 569 | 555 |
| 7/17/2000 | 612 | 648 | 631 |
|  | 652 | 639 | 627 |
|  | 612 | 633 | 649 |
|  | 656 | 677 | 642 |
|  | 607 | 592 | 580 |
|  | 657 | 671 | 662 |
|  | 641 | 639 | 621 |
|  | 526 | 536 | 546 |
|  | 630 | 659 | 662 |
|  | 599 | 619 | 640 |
|  | 613 | 637 | 651 |
|  | 673 | 685 | 666 |
|  | 581 | 574 | 583 |
|  | 679 | 659 | 651 |
| 7/24/2000 | 407 | 403 | 401 |
|  | 420 | 410 | 447 |
|  | 520 | 535 | 541 |
|  | 559 | 572 | 566 |
|  | 511 | 519 | 531 |
|  | 551 | 561 | 535 |
|  | 521 | 537 | 518 |
|  | 532 | 541 | 553 |
|  | 481 | 472 | 461 |
|  | 495 | 499 | 519 |
|  | 512 | 507 | 532 |
|  | 455 | 439 | 429 |
|  | 595 | 609 | 589 |
|  | 575 | 581 | 562 |
| 7/31/2000 | 589 | 571 | 565 |
|  | 522 | 539 | 551 |
|  | 573 | 585 | 580 |
|  | 519 | 533 | 521 |
|  | 566 | 519 | 539 |
|  | 559 | 541 | 521 |
|  | 621 | 609 | 611 |
|  | 529 | 546 | 566 |
|  | 495 | 505 | 517 |
|  | 611 | 600 | 621 |
| 8/4/2000 | 511 | 529 | 530 |
|  | 575 | 585 | 562 |
|  | 532 | 519 | 541 |
|  | 475 | 495 | 512 |
|  | 533 | 569 | 542 |
|  | 509 | 521 | 517 |
|  | 592 | 581 | 572 |
|  | 535 | 549 | 562 |
|  | 502 | 519 | 535 |
|  | 435 | 475 | 462 |
|            | 544 | 565 | 575 |
|            | 517 | 529 | 545 |
|            | 498 | 475 | 525 |
|            | 479 | 509 | 513 |
| 8/7/2000   | 659 | 672 | 681 |
|            | 613 | 635 | 622 |
|            | 679 | 689 | 661 |
|            | 652 | 632 | 619 |
|            | 617 | 632 | 642 |
|            | 677 | 690 | 670 |
|            | 609 | 631 | 621 |
|            | 649 | 666 | 672 |
|            | 633 | 642 | 655 |
|            | 661 | 649 | 657 |
|            | 671 | 689 | 666 |
|            | 639 | 659 | 642 |
|            | 622 | 619 | 601 |
|            | 688 | 671 | 659 |
| 8/11/2000  | 537 | 555 | 562 |
|            | 511 | 509 | 531 |
|            | 569 | 577 | 552 |
|            | 511 | 529 | 537 |
|            | 581 | 572 | 562 |
|            | 601 | 592 | 589 |
|            | 532 | 549 | 562 |
|            | 532 | 547 | 561 |
|            | 517 | 509 | 528 |
|            | 581 | 599 | 572 |
|  | 549 | 567 | 550 |
|  | 489 | 499 | 513 |
|  | 533 | 561 | 547 |
|  | 561 | 574 | 539 |
| 8/4/2000 | 507 | 519 | 531 |
|  | 532 | 557 | 525 |
|  | 536 | 542 | 562 |
|  | 569 | 576 | 582 |
|  | 622 | 607 | 619 |
|  | 551 | 572 | 561 |
|  | 582 | 571 | 591 |
|  | 571 | 581 | 592 |
|  | 566 | 519 | 539 |
|  | 569 | 532 | 547 |
|  | 539 | 555 | 567 |
|  | 572 | 599 | 581 |
|  | 511 | 514 | 492 |
|  | 541 | 531 | 561 |
| 9/18/2000 | 449 | 432 | 416 |
|  | 472 | 482 | 461 |
|  | 412 | 435 | 460 |
|  | 371 | 356 | 383 |
|  | 428 | 456 | 439 |
|  | 433 | 452 | 449 |
|  | 476 | 461 | 452 |
|  | 516 | 529 | 541 |
|  | 423 | 450 | 462 |
|  | 389 | 390 | 415 |
|  | 471 | 461 | 455 |
|  | 511 | 526 | 539 |
|  | 490 | 472 | 463 |
|  | 590 | 571 | 581 |
| 10/6/2000 | 601 | 565 | 593 |
|  | 594 | 585 | 572 |
|  | 662 | 616 | 622 |
|  | 591 | 572 | 599 |
|  | 531 | 552 | 549 |
|  | 562 | 572 | 581 |
|  | 600 | 619 | 583 |
|  | 532 | 545 | 563 |
|  | 637 | 659 | 629 |
|  | 581 | 561 | 549 |
|  | 566 | 593 | 573 |
|  | 609 | 623 | 633 |
|  | 513 | 500 | 529 |
|  | 600 | 647 | 636 |
| 10/9/2000 | 577 | 592 | 563 |
|  | 611 | 607 | 653 |
|  | 581 | 593 | 617 |
|  | 633 | 645 | 619 |
|  | 511 | 537 | 549 |
|  | 544 | 562 | 573 |
|  | 666 | 672 | 693 |
|  | 601 | 572 | 633 |
|  | 511 | 529 | 541 |
|  | 532 | 555 | 562 |
|  | 513 | 549 | 562 |
|  | 562 | 539 | 547 |
|  | 560 | 542 | 522 |
|  | 680 | 669 | 671 |
| 10/16/2000 | 478 | 429 | 444 |
|  | 501 | 519 | 535 |
|  | 418 | 439 | 452 |
|  | 431 | 422 | 445 |
|  | 511 | 539 | 517 |
|  | 488 | 472 | 462 |
|  | 459 | 463 | 440 |
|  | 535 | 511 | 545 |
|  | 455 | 462 | 471 |
|  | 435 | 455 | 469 |
|  | 501 | 518 | 507 |
|  | 447 | 467 | 440 |
|  | 466 | 472 | 444 |
|  | 503 | 489 | 492 |
| 10/16/2000 | 541 | 564 | 543 |
|  | 493 | 509 | 489 |
|  | 523 | 522 | 511 |
|  | 555 | 543 | 563 |
|  | 509 | 521 | 512 |
|  | 623 | 611 | 643 |
|  | 598 | 607 | 588 |
|  | 537 | 531 | 520 |
|  | 553 | 540 | 532 |
|  | 517 | 508 | 521 |
|            | 574  | 563  | 559  |
|            | 526  | 513  | 520  |
|            | 563  | 579  | 588  |
|            | 600  | 593  | 588  |
| 12/18/2000 | 1232 | 1209 | 1265 |
|            | 2715 | 2620 | 2702 |
|            | 4688 | 4336 | 4734 |
|            | 3389 | 3356 | 3387 |
|            | 3674 | 3592 | 3633 |
|            | 4846 | 4822 | 4888 |
|            | 4414 | 4549 | 4327 |
|            | 4116 | 4010 | 4114 |
|            | 5151 | 5125 | 5068 |
|            | 4263 | 4058 | 4013 |
| 1/31/2001  | 680  | 667  | 672  |
|            | 639  | 621  | 645  |
|            | 609  | 617  | 629  |
|            | 666  | 675  | 689  |
|            | 721  | 742  | 719  |
|            | 690  | 710  | 693  |
|            | 666  | 650  | 645  |
|            | 701  | 709  | 713  |
|            | 675  | 667  | 658  |
|            | 635  | 611  | 623  |
|            | 685  | 670  | 693  |
|            | 711  | 729  | 730  |
|            | 721  | 735  | 749  |
|            | 735  | 759  | 750  |

| Date | | | |
|---|---|---|---|
| 3/8/2001 | 566 | 573 | 582 |
| | 619 | 645 | 634 |
| | 667 | 662 | 639 |
| | 630 | 619 | 636 |
| | 664 | 649 | 655 |
| | 533 | 522 | 519 |
| | 644 | 632 | 621 |
| 3/12/2001 | 187 | 165 | 172 |
| | 111 | 109 | 119 |
| | 250 | 261 | 243 |
| | 263 | 249 | 270 |
| | 240 | 238 | 227 |
| | 261 | 275 | 270 |
| | 240 | 238 | 227 |
| | 160 | 171 | 181 |
| | 111 | 120 | 109 |
| | 241 | 238 | 229 |
| | 232 | 241 | 247 |
| | 243 | 261 | 251 |
| | 229 | 231 | 221 |
| | 241 | 227 | 251 |
| 3/19/2001 | 512 | 534 | 545 |
| | 555 | 523 | 539 |
| | 516 | 525 | 533 |
| | 577 | 566 | 553 |
| | 511 | 509 | 495 |
| | 583 | 568 | 562 |
| 3/23/2001 | 1130 | 1145 | 1165 |

| Date | | | |
|---|---|---|---|
| | 440 | 421 | 435 |
| | 1100 | 1082 | 1075 |
| | 998 | 1005 | 1021 |
| | 1217 | 1178 | 1169 |
| | 960 | 982 | 971 |
| | 982 | 1015 | 999 |
| 3/26/2001 | 612 | 632 | 643 |
| | 633 | 621 | 654 |
| | 634 | 654 | 666 |
| | 635 | 619 | 644 |
| | 579 | 598 | 609 |
| | 599 | 641 | 642 |
| | 598 | 601 | 582 |
| 3/30/2001 | 385 | 395 | 361 |
| | 1211 | 1235 | 1224 |
| | 1192 | 1185 | 1197 |
| | 1209 | 1215 | 1222 |
| | 1175 | 1187 | 1195 |
| | 1190 | 1181 | 1192 |
| | 1165 | 1180 | 1172 |
| | 365 | 380 | 377 |
| | 1127 | 1135 | 1145 |
| | 1207 | 1229 | 1217 |
| | 1177 | 1189 | 1162 |
| | 1150 | 1169 | 1141 |
| | 1135 | 1169 | 1147 |
| | 1081 | 1065 | 1047 |
| 4/6/2001 | 580 | 565 | 571 |
|  | 572 | 578 | 561 |
|  | 527 | 530 | 544 |
|  | 566 | 571 | 561 |
|  | 572 | 571 | 565 |
|  | 554 | 551 | 569 |
|  | 533 | 521 | 511 |
| 4/16/2001 | 841 | 834 | 858 |
|  | 566 | 563 | 524 |
|  | 799 | 800 | 861 |
|  | 599 | 588 | 592 |
|  | 593 | 584 | 578 |
|  | 668 | 721 | 724 |
|  | 582 | 603 | 625 |
|  | 633 | 636 | 678 |
|  | 587 | 575 | 602 |
|  | 537 | 618 | 573 |
|  | 612 | 655 | 579 |
|  | 551 | 571 | 525 |
|  | 494 | 445 | 491 |
| 4/16/2001 | 5953 | 5976 | 5793 |
|  | 7161 | 7069 | 7125 |
|  | 6363 | 6635 | 6434 |
|  | 6533 | 6834 | 6663 |
|  | 5070 | 5248 | 5268 |
|  | 6514 | 6804 | 6550 |
|  | 5702 | 5893 | 5957 |
| 5/14/2001 | 7247 | 7144 | 7193 |
|  | 8432 | 8315 | 8160 |
|  | 6304 | 6484 | 6415 |
|  | 6741 | 6821 | 6661 |
|  | 5462 | 5530 | 5384 |
|  | 5955 | 5865 | 5708 |
|  | 5792 | 5813 | 5739 |
| 5/14/2001 | 764 | 676 | 684 |
|  | 697 | 719 | 669 |
|  | 773 | 784 | 726 |
|  | 938 | 903 | 856 |
|  | 802 | 789 | 815 |
|  | 804 | 784 | 794 |
|  | 759 | 791 | 807 |
|  | 634 | 601 | 618 |
|  | 665 | 655 | 671 |
|  | 624 | 646 | 672 |
|  | 638 | 594 | 589 |
|  | 688 | 729 | 666 |
|  | 536 | 523 | 515 |
|  | 399 | 424 | 420 |
| 6/19/2001 | 715 | 713 | 693 |
|  | 725 | 697 | 677 |
|  | 647 | 571 | 503 |
|  | 533 | 669 | 677 |
|  | 535 | 499 | 510 |
|  | 514 | 494 | 515 |
|  | 510 | 532 | 530 |
| 6/22/2001 | 2250 | 2358 | 2232 |
|  | 2217 | 2136 | 2083 |
|            | 2806 | 2883 | 2796 |
|            | 3084 | 3175 | 3054 |
|            | 1948 | 1948 | 1914 |
|            | 1901 | 1858 | 1764 |
|            | 1856 | 1786 | 1726 |
| 7/16/2001  | 2503 | 2631 | 2542 |
|            | 2411 | 2478 | 2399 |
|            | 2616 | 2806 | 2650 |
|            | 2692 | 2730 | 2659 |
|            | 2224 | 2289 | 2213 |
|            | 2206 | 2335 | 2294 |
|            | 2497 | 2317 | 2410 |
|            | 2087 | 2138 | 2051 |

| Other Investigators Coulters | | | | | |
|---|---|---|---|---|---|
| Date | cou1 | cou2 | cou3 | Investigator | |
| 12/7/2001 | 5379 | 5074 | 5007 | Inv1 | |
| 12/7/2001 | 4820 | 4843 | 4898 | Inv1 | |
| 12/7/2001 | 4830 | 4939 | 4866 | Inv1 | |
| 12/7/2001 | 5488 | 5562 | 5577 | Inv1 | |
| 12/7/2001 | 5255 | 5229 | 5452 | Inv1 | |
| 12/7/2001 | 5881 | 5881 | 5850 | Inv1 | |
| 12/7/2001 | 6288 | 6288 | 6277 | Inv1 | |
| 12/7/2001 | 5703 | 5603 | 5520 | Inv1 | |
| 12/7/2001 | 5151 | 5198 | 5125 | Inv1 | |
| 12/7/2001 | 5162 | 5229 | 5172 | Inv1 | |
| 12/13/2001 | 7102 | 7258 | 7002 | Inv1 | |
| 12/13/2001 | 6646 | 6650 | 6596 | Inv1 | |
| 12/13/2001 | 6934 | 6715 | 6809 | Inv1 | |
| 12/13/2001 | 6879 | 6990 | 6923 | Inv1 | |
| 12/13/2001 | 5581 | 5429 | 5541 | Inv1 | |
| 12/13/2001 | 5312 | 5394 | 5422 | Inv1 | |
| 12/13/2001 | 4668 | 4736 | 4650 | Inv1 | |
| 12/13/2001 | 4893 | 4822 | 4833 | Inv1 | |
| 12/13/2001 | 4669 | 4566 | 4677 | Inv1 | |
| 12/13/2001 | 3833 | 4043 | 3965 | Inv1 | |
| 12/25/2001 | 3734 | 3806 | 3721 | Inv1 | |
| 12/25/2001 | 3536 | 3609 | 3624 | Inv1 | |
| 12/25/2001 | 3491 | 3534 | 3616 | Inv1 | |
| 12/25/2001 | 3155 | 2958 | 2959 | Inv1 | |
| 12/25/2001 | 3757 | 3875 | 3852 | Inv1 | |
| 12/25/2001 | 3619 | 3632 | 3576 | Inv1 | |
| 12/25/2001 | 3871 | 3947 | 3987 | Inv1 | |
| 12/25/2001 | 3813 | 3806 | 3829 | Inv1 | |
| 12/25/2001 | 3645 | 3609 | 3718 | Inv1 | |
| 12/25/2001 | 3678 | 3550 | 3594 | Inv1 | |
| 1/31/2002 | 177 | 174 | 196 | Inv1 | |
| 1/31/2002 | 154 | 150 | 128 | Inv1 | |
| 1/31/2002 | 171 | 138 | 159 | Inv1 | |
| 1/31/2002 | 156 | 156 | 154 | Inv1 | |
| 1/31/2002 | 122 | 111 | 133 | Inv1 | |
| 1/31/2002 | 153 | 140 | 142 | Inv1 | |
| 1/31/2002 | 129 | 127 | 153 | Inv1 | |
| 1/31/2002 | 147 | 151 | 147 | Inv1 | |
| 1/31/2002 | 141 | 108 | 114 | Inv1 | |

| Date | | | | |
|---|---|---|---|---|
| 1/31/2002 | 128 | 95 | 124 | Inv1 |
| 1/31/2002 | 147 | 132 | 135 | Inv1 |
| 1/31/2002 | 222 | 210 | 244 | Inv1 |
| 1/31/2002 | 220 | 221 | 226 | Inv1 |
| 1/31/2002 | 200 | 221 | 176 | Inv1 |
| 1/31/2002 | 162 | 143 | 155 | Inv1 |
| 1/31/2002 | 176 | 184 | 183 | Inv1 |
| 1/31/2002 | 203 | 178 | 198 | Inv1 |
| 1/31/2002 | 179 | 177 | 174 | Inv1 |
| 1/31/2002 | 151 | 178 | 195 | Inv1 |
| 1/31/2002 | 155 | 169 | 174 | Inv1 |
| 1/31/2002 | 131 | 154 | 158 | Inv1 |
| 1/31/2002 | 139 | 128 | 160 | Inv1 |
| 2/1/2002 | 448 | 436 | 478 | Inv1 |
| 2/1/2002 | 342 | 384 | 355 | Inv1 |
| 2/1/2002 | 303 | 309 | 289 | Inv1 |
| 2/1/2002 | 322 | 283 | 318 | Inv1 |
| 2/1/2002 | 275 | 291 | 287 | Inv1 |
| 2/1/2002 | 279 | 279 | 267 | Inv1 |
| 2/1/2002 | 254 | 246 | 221 | Inv1 |
| 2/1/2002 | 207 | 235 | 222 | Inv1 |
| 2/1/2002 | 199 | 222 | 201 | Inv1 |
| 2/1/2002 | 194 | 199 | 202 | Inv1 |
| 2/1/2002 | 184 | 205 | 215 | Inv1 |
| 5/21/2002 | 4939 | 4827 | 4904 | Inv1 |
| 4/1/2005 | 737 | 701 |  | Inv1 |
| 4/1/2005 | 986 | 1017 | 965 | Inv1 |
| 4/1/2005 | 989 | 1025 |  | Inv1 |
| 4/1/2005 | 561 | 576 | 540 | Inv1 |
| 4/1/2005 | 601 | 558 | 538 | Inv1 |
| 4/1/2005 | 577 | 550 | 524 | Inv1 |
| 4/1/2005 | 557 | 565 | 550 | Inv1 |
| 4/1/2005 | 554 | 547 | 520 | Inv1 |
| 4/1/2005 | 619 | 604 | 612 | Inv1 |
| 4/1/2005 | 2466 | 2397 | 2480 | Inv1 |
| 4/1/2005 | 221 | 234 | 277 | Inv1 |
| 4/1/2005 | 221 | 184 | 217 | Inv1 |
| 4/1/2005 | 244 | 258 | 262 | Inv1 |
| 4/14/2005 | 275 | 256 | 284 | Inv1 |
| 4/14/2005 | 305 | 295 | 294 | Inv1 |
| 4/14/2005 | 374 | 392 | 386 | Inv1 |

| | | | | | |
|---|---|---|---|---|---|
| 4/14/2005 | 463 | 465 | 454 | Inv1 | |
| 5/17/2005 | 1450 | 1439 | 1504 | Inv1 | |
| 5/17/2005 | 1364 | 1336 | 1312 | Inv1 | |
| 5/17/2005 | 1691 | 1673 | 1723 | Inv1 | |
| 5/17/2005 | 1605 | 1632 | 1549 | Inv1 | |
| 5/17/2005 | 2149 | 2169 | 2077 | Inv1 | |
| 5/17/2005 | 1818 | 1784 | 1746 | Inv1 | |
| 5/17/2005 | 1173 | 1122 | 1197 | Inv1 | |
| 5/17/2005 | 1436 | 1369 | 1430 | Inv1 | |
| 5/17/2005 | 1063 | 1083 | 1028 | Inv1 | |
| 5/17/2005 | 1100 | 1084 | 1146 | Inv1 | |
| 5/17/2005 | 696 | 706 | 728 | Inv1 | |
| 5/17/2005 | 680 | 661 | 707 | Inv1 | |
| 5/21/2005 | 1705 | 1748 | 1713 | Inv1 | |
| 5/21/2005 | 1760 | 1795 | 1790 | Inv1 | |
| 5/21/2005 | 1241 | 1229 | 1116 | Inv1 | |
| 5/21/2005 | 1959 | 1918 | 2044 | Inv1 | |
| 5/21/2005 | 1064 | 1104 | 1072 | Inv1 | |
| 5/21/2005 | 1729 | 1752 | 1638 | Inv1 | |
| 5/21/2005 | 1776 | 1732 | 1792 | Inv1 | |
| 5/21/2005 | 1429 | 1317 | 1415 | Inv1 | |
| 5/21/2005 | 1736 | 1711 | 1613 | Inv1 | |
| 5/21/2005 | 1467 | 1444 | 1484 | Inv1 | |
| 5/21/2005 | 1465 | 1528 | 1578 | Inv1 | |
| 9/10/1999 | 8282 | 8315 | 8221 | Inv8 | |
| 9/10/1999 | 7355 | 7404 | | Inv8 | |
| 9/10/1999 | 7962 | 7852 | | Inv8 | |
| 9/10/1999 | 5530 | 5520 | | Inv8 | |
| 9/10/1999 | 5541 | 5494 | | Inv8 | |
| 9/10/1999 | 8409 | 8165 | | Inv8 | |
| 9/10/1999 | 8760 | 8755 | | Inv8 | |
| 9/10/1999 | 7748 | 7688 | 7726 | Inv8 | |
| 9/10/1999 | 5734 | 5871 | 5984 | Inv8 | |
| 9/10/1999 | 6484 | 6474 | 6240 | Inv8 | |
| 9/10/1999 | 5987 | 5395 | | Inv8 | |
| 9/10/1999 | 6426 | 6203 | | Inv8 | |
| 9/10/1999 | 3539 | 2225 | | Inv8 | |
| 9/10/1999 | 3327 | 4678 | 6607 | Inv8 | |
| 9/10/1999 | 5110 | 3497 | | Inv8 | |
| 9/10/1999 | 8254 | 8354 | 4875 | Inv8 | |
| 9/10/1999 | 4838 | 4756 | 3112 | Inv8 | |

| Date | | | | |
|---|---|---|---|---|
| 9/10/1999 | 7797 | 6087 | 5855 | Inv8 |
| 9/10/1999 | 4053 | 4060 | 7204 | Inv8 |
| 9/10/1999 | 3752 | 4007 | 4013 | Inv8 |
| 9/10/1999 | 5577 | 5504 | 5640 | Inv8 |
| 9/10/1999 | 8598 | 4575 | 4653 | Inv8 |
| 9/10/1999 | 4435 | 6666 | 6113 | Inv8 |
| 9/10/1999 | 9784 | 6314 | 4755 | Inv8 |
| 9/10/1999 | 4144 | 6415 | 6735 | Inv8 |
| 12/6/1999 | 801 | 784 | 832 | Inv8 |
| 12/6/1999 | 829 | 836 | 827 | Inv8 |
| 12/6/1999 | 759 | 747 | 732 | Inv8 |
| 12/6/1999 | 745 | 713 | 696 | Inv8 |
| 12/6/1999 | 824 | 782 | 807 | Inv8 |
| 12/6/1999 | 808 | 792 | 771 | Inv8 |
| 12/6/1999 | 728 | 752 | 702 | Inv8 |
| 12/6/1999 | 648 | 648 | 651 | Inv8 |
| 12/6/1999 | 725 | 723 | 713 | Inv8 |
| 12/6/1999 | 656 | 647 | 634 | Inv8 |
| 2/13/1995 | 148 | 179 | 143 | Inv2 |
| 2/13/1995 | 149 | 163 | 136 | Inv2 |
| 2/13/1995 | 162 | 173 | 149 | Inv2 |
| 2/13/1995 | 159 | 124 | 146 | Inv2 |
| 2/13/1995 | 146 | 155 | 165 | Inv2 |
| 2/13/1995 | 153 | 142 | 164 | Inv2 |
| 2/13/1995 | 133 | 154 | 159 | Inv2 |
| 4/4/1995 | 334 | 327 | | Inv2 |
| 4/4/1995 | 294 | 300 | | Inv2 |
| 4/4/1995 | 342 | 347 | | Inv2 |
| 4/4/1995 | 347 | 368 | | Inv2 |
| 4/4/1995 | 368 | 401 | 358 | Inv2 |
| 4/4/1995 | 314 | 297 | | Inv2 |
| 4/4/1995 | 368 | 361 | | Inv2 |
| 4/4/1995 | 352 | 362 | | Inv2 |
| 6/27/1995 | 505 | 474 | | Inv2 |
| 6/27/1995 | 524 | 477 | | Inv2 |
| 6/27/1995 | 495 | 474 | | Inv2 |
| 6/27/1995 | 467 | 458 | | Inv2 |
| 6/27/1995 | 446 | 497 | | Inv2 |
| 6/27/1995 | 451 | 448 | | Inv2 |
| 6/27/1995 | 515 | 485 | | Inv2 |
| 6/27/1995 | 469 | 458 | | Inv2 |

| 6/27/1995 | 490 | 466 |     | Inv2 |
|-----------|-----|-----|-----|------|
| 6/27/1995 | 441 | 435 |     | Inv2 |
| 7/24/1995 | 495 | 480 |     | Inv2 |
| 7/24/1995 | 481 | 508 |     | Inv2 |
| 7/24/1995 | 523 | 568 |     | Inv2 |
| 7/24/1995 | 545 | 511 |     | Inv2 |
| 7/24/1995 | 493 | 546 |     | Inv2 |
| 7/24/1995 | 498 | 541 |     | Inv2 |
| 7/24/1995 | 532 | 521 |     | Inv2 |
| 7/24/1995 | 476 | 448 |     | Inv2 |
| 7/24/1995 | 484 | 440 |     | Inv2 |
| 7/24/1995 | 451 | 402 |     | Inv2 |
| 7/24/1995 | 50  | 56  | 84  | Inv2 |
| 7/24/1995 | 57  | 37  | 55  | Inv2 |
| 7/24/1995 | 32  | 29  | 34  | Inv2 |
| 7/24/1995 | 109 | 126 | 135 | Inv2 |
| 7/24/1995 | 63  | 58  | 62  | Inv2 |
| 7/24/1995 | 57  | 52  | 49  | Inv2 |
| 7/24/1995 | 55  | 45  | 54  | Inv2 |
| 7/24/1995 | 153 | 163 | 153 | Inv2 |
| 7/24/1995 | 263 | 253 | 267 | Inv2 |
| 7/24/1995 | 38  | 23  | 35  | Inv2 |
| 9/25/1995 | 471 | 503 |     | Inv2 |
| 9/25/1995 | 593 | 612 |     | Inv2 |
| 9/25/1995 | 563 | 576 |     | Inv2 |
| 9/25/1995 | 543 | 558 |     | Inv2 |
| 9/25/1995 | 564 | 544 |     | Inv2 |
| 9/25/1995 | 580 | 562 |     | Inv2 |
| 9/25/1995 | 542 | 591 |     | Inv2 |
| 9/25/1995 | 521 | 511 |     | Inv2 |
| 9/25/1995 | 536 | 578 |     | Inv2 |
| 9/25/1995 | 556 | 571 |     | Inv2 |
| 10/1/1995 | 493 | 490 |     | Inv2 |
| 10/1/1995 | 507 | 517 |     | Inv2 |
| 10/1/1995 | 517 | 520 |     | Inv2 |
| 10/1/1995 | 591 | 588 |     | Inv2 |
| 10/1/1995 | 537 | 529 |     | Inv2 |
| 10/1/1995 | 505 | 538 |     | Inv2 |
| 10/1/1995 | 543 | 541 |     | Inv2 |
| 10/1/1995 | 492 | 500 |     | Inv2 |
| 10/1/1995 | 426 | 458 |     | Inv2 |

| Date | | | | | |
|---|---|---|---|---|---|
| 10/1/1995 | 481 | 453 | | Inv2 | |
| 10/1/1995 | 1312 | 1323 | 1247 | Inv2 | |
| 10/1/1995 | 1216 | 1210 | 1237 | Inv2 | |
| 10/1/1995 | 1218 | 1179 | 1234 | Inv2 | |
| 10/1/1995 | 1145 | 1208 | 1166 | Inv2 | |
| 10/1/1995 | 1142 | 1128 | 1088 | Inv2 | |
| 10/1/1995 | 1153 | 1122 | 1136 | Inv2 | |
| 10/1/1995 | 1245 | 1111 | 1217 | Inv2 | |
| 10/1/1995 | 1233 | 1233 | 1204 | Inv2 | |
| 10/1/1995 | 1153 | 1174 | 1170 | Inv2 | |
| 10/1/1995 | 998 | 977 | 994 | Inv2 | |
| 10/1/1995 | 696 | 698 | 662 | Inv2 | |
| 10/1/1995 | 931 | 971 | 889 | Inv2 | |
| 10/1/1995 | 969 | 927 | 897 | Inv2 | |
| 10/1/1995 | 859 | 929 | 842 | Inv2 | |
| 10/1/1995 | 741 | 738 | | Inv2 | |
| 10/1/1995 | 956 | 950 | | Inv2 | |
| 10/1/1995 | 987 | 1021 | | Inv2 | |
| 10/1/1995 | 964 | 1005 | | Inv2 | |
| 10/1/1995 | 1016 | 1036 | | Inv2 | |
| 10/1/1995 | 792 | 735 | | Inv2 | |
| 1/12/1996 | 936 | 959 | 958 | Inv2 | |
| 1/12/1996 | 1012 | 988 | 959 | Inv2 | |
| 1/12/1996 | 986 | 943 | 1019 | Inv2 | |
| 1/12/1996 | 940 | 888 | 916 | Inv2 | |
| 1/12/1996 | 874 | 896 | 869 | Inv2 | |
| 1/12/1996 | 901 | 894 | 899 | Inv2 | |
| 1/12/1996 | 947 | 977 | 984 | Inv2 | |
| 1/12/1996 | 876 | 868 | 893 | Inv2 | |
| 1/12/1996 | 932 | 984 | 916 | Inv2 | |
| 1/12/1996 | 900 | 862 | 895 | Inv2 | |
| 1/12/1996 | 694 | 764 | 718 | Inv2 | |
| 1/12/1996 | 767 | 872 | 767 | Inv2 | |
| 1/12/1996 | 685 | 690 | 626 | Inv2 | |
| 1/12/1996 | 682 | 667 | 596 | Inv2 | |
| 1/12/1996 | 529 | 560 | 544 | Inv2 | |
| 1/12/1996 | 585 | 675 | 598 | Inv2 | |
| 1/12/1996 | 676 | 674 | 709 | Inv2 | |
| 1/12/1996 | 604 | 608 | 591 | Inv2 | |
| 1/12/1996 | 641 | 638 | 603 | Inv2 | |
| 1/12/1996 | 516 | 514 | 520 | Inv2 | |

| Date | | | | |
|---|---|---|---|---|
| 2/1/1996 | 684 | 600 | 641 | Inv2 |
| 2/1/1996 | 570 | 665 | 613 | Inv2 |
| 2/1/1996 | 602 | 631 | 565 | Inv2 |
| 2/1/1996 | 555 | 515 | 602 | Inv2 |
| 2/1/1996 | 487 | 571 | 521 | Inv2 |
| 2/1/1996 | 581 | 503 | 550 | Inv2 |
| 2/1/1996 | 604 | 536 | 611 | Inv2 |
| 2/1/1996 | 537 | 497 | 491 | Inv2 |
| 2/1/1996 | 532 | 578 | 491 | Inv2 |
| 2/1/1996 | 499 | 498 | 466 | Inv2 |
| 2/1/1996 | 516 | 469 | 504 | Inv2 |
| 2/1/1996 | 502 | 459 | 504 | Inv2 |
| 2/1/1996 | 425 | 381 | 404 | Inv2 |
| 2/1/1996 | 351 | 393 | 367 | Inv2 |
| 2/1/1996 | 364 | 349 | 376 | Inv2 |
| 2/1/1996 | 402 | 402 | 400 | Inv2 |
| 2/1/1996 | 459 | 450 | 392 | Inv2 |
| 2/1/1996 | 384 | 463 | 392 | Inv2 |
| 2/1/1996 | 424 | 371 | 407 | Inv2 |
| 2/1/1996 | 370 | 388 | 343 | Inv2 |
| 4/15/1996 | 882 | 854 | 849 | Inv2 |
| 4/15/1996 | 874 | 925 | 876 | Inv2 |
| 4/15/1996 | 687 | 648 | 661 | Inv2 |
| 4/15/1996 | 648 | 703 | 690 | Inv2 |
| 4/15/1996 | 663 | 647 | 672 | Inv2 |
| 4/15/1996 | 829 | 822 | 792 | Inv2 |
| 4/15/1996 | 942 | 872 | 892 | Inv2 |
| 4/15/1996 | 717 | 672 | 640 | Inv2 |
| 4/15/1996 | 651 | 674 | 656 | Inv2 |
| 4/15/1996 | 709 | 693 | 708 | Inv2 |
| 1/23/1997 | 1310 | 1315 | 1236 | Inv2 |
| 1/23/1997 | 1413 | 1358 | 1396 | Inv2 |
| 1/23/1997 | 1215 | 1230 | 1285 | Inv2 |
| 1/23/1997 | 1430 | 1307 | 1137 | Inv2 |
| 1/23/1997 | 1177 | 1167 | 1061 | Inv2 |
| 1/23/1997 | 1583 | 1591 | 1405 | Inv2 |
| 1/23/1997 | 1549 | 1458 | 1418 | Inv2 |
| 1/23/1997 | 1205 | 1150 | 1115 | Inv2 |
| 1/23/1997 | 1344 | 1239 | 1148 | Inv2 |
| 1/23/1997 | 1311 | 1494 | 1420 | Inv2 |
| 4/15/1997 | 1114 | 1061 | 1111 | Inv2 |
| 4/15/1997 | 1217 | 1124 | 1132 | Inv2 |
| 4/15/1997 | 1028 | 1044 | 1099 | Inv2 |
| 4/15/1997 | 956 | 1028 | 971 | Inv2 |
| 4/15/1997 | 940 | 1003 | 963 | Inv2 |
| 4/15/1997 | 1092 | 1098 | 1051 | Inv2 |
| 4/15/1997 | 1032 | 1053 | 995 | Inv2 |
| 4/15/1997 | 954 | 874 | 952 | Inv2 |
| 4/15/1997 | 967 | 869 | 944 | Inv2 |
| 4/15/1997 | 1079 | 1104 | 1045 | Inv2 |
| 4/18/1997 | 1206 | 1253 | 1312 | Inv2 |
| 4/18/1997 | 1201 | 1257 | 1241 | Inv2 |
| 4/18/1997 | 1043 | 1094 | 1048 | Inv2 |
| 4/18/1997 | 1188 | 1136 | 1138 | Inv2 |
| 4/18/1997 | 1084 | 996 | 957 | Inv2 |
| 4/18/1997 | 1270 | 1255 | 1224 | Inv2 |
| 4/18/1997 | 1266 | 1169 | 1214 | Inv2 |
| 4/18/1997 | 1158 | 1164 | 1144 | Inv2 |
| 4/18/1997 | 1227 | 1155 | 1181 | Inv2 |
| 4/18/1997 | 1211 | 1173 | 1176 | Inv2 |
| 4/18/1997 | 645 | 670 | 708 | Inv2 |
| 4/18/1997 | 726 | 732 | 729 | Inv2 |
| 4/18/1997 | 802 | 747 | 835 | Inv2 |
| 4/18/1997 | 1201 | 1056 | 1214 | Inv2 |
| 4/18/1997 | 1168 | 1198 | 1171 | Inv2 |
| 4/18/1997 | 1177 | 1172 | 1167 | Inv2 |
| 4/21/1997 | 415 | 418 | 390 | Inv2 |
| 4/21/1997 | 220 | 270 | 242 | Inv2 |
| 4/21/1997 | 554 | 597 | 561 | Inv2 |
| 4/21/1997 | 521 | 520 | 526 | Inv2 |
| 4/21/1997 | 502 | 504 | 505 | Inv2 |
| 4/21/1997 | 514 | 569 | 550 | Inv2 |
| 4/21/1997 | 527 | 565 | 508 | Inv2 |
| 4/21/1997 | 527 | 518 | 505 | Inv2 |
| 4/21/1997 | 935 | 976 | 962 | Inv2 |
| 4/21/1997 | 596 | 611 | 569 | Inv2 |
| 4/22/1997 | 1010 | 1119 | 1040 | Inv2 |
| 4/22/1997 | 1085 | 1086 | 1080 | Inv2 |
| 4/22/1997 | 988 | 888 | 936 | Inv2 |
| 4/22/1997 | 924 | 924 | 908 | Inv2 |
| 4/22/1997 | 1031 | 1029 | 1008 | Inv2 |
| 4/22/1997 | 1014 | 1040 | 1015 | Inv2 |
| 4/22/1997 | 1021 | 967 | 968 | Inv2 |
| 4/22/1997 | 898 | 926 | 910 | Inv2 |
| 4/22/1997 | 834 | 887 | 882 | Inv2 |
| 4/22/1997 | 631 | 570 | 582 | Inv2 |
| 4/22/1997 | 567 | 589 | 579 | Inv2 |
| 4/22/1997 | 619 | 512 | 578 | Inv2 |
| 4/22/1997 | 613 | 675 | 590 | Inv2 |
| 4/22/1997 | 652 | 602 | 598 | Inv2 |
| 4/22/1997 | 479 | 469 | 468 | Inv2 |
| 4/22/1997 | 472 | 460 | 437 | Inv2 |
| 4/22/1997 | 396 | 422 | 427 | Inv2 |
| 4/22/1997 | 434 | 397 | 454 | Inv2 |
| 10/8/2000 | 3180 | 3022 | 3065 | Inv2 |
| 10/8/2000 | 3150 | 3088 | 2988 | Inv2 |
| 10/8/2000 | 2754 | 2832 | 2782 | Inv2 |
| 10/8/2000 | 2741 | 2722 | 2617 | Inv2 |
| 10/8/2000 | 2423 | 2324 | 2351 | Inv2 |
| 10/8/2000 | 2795 | 2742 | 2702 | Inv2 |
| 10/8/2000 | 2897 | 2809 | 2705 | Inv2 |
| 10/8/2000 | 2687 | 2701 | 2653 | Inv2 |
| 10/8/2000 | 2766 | 2746 | 2680 | Inv2 |
| 10/8/2000 | 2528 | 2428 | 2495 | Inv2 |
| 4/23/2001 | 7967 | 7945 | 8022 | Inv2 |
| 4/23/2001 | 7404 | 7497 | 7513 | Inv2 |
| 4/23/2001 | 8321 | 8426 | 8221 | Inv2 |
| 4/23/2001 | 7644 | 7486 | 7622 | Inv2 |
| 4/23/2001 | 7535 | 7595 | 7432 | Inv2 |
| 4/23/2001 | 7009 | 7128 | 7026 | Inv2 |
| 4/23/2001 | 7540 | 7458 | 7458 | Inv2 |
| 4/23/2001 | 7117 | 7107 | 7177 | Inv2 |
| 4/23/2001 | 7204 | 7236 | 7144 | Inv2 |
| 4/23/2001 | 6896 | 6826 | 6875 | Inv2 |
| 5/7/2001 | 8985 | 8822 | 8917 | Inv2 |
| 5/7/2001 | 9664 | 9567 | 9312 | Inv2 |
| 5/7/2001 | 9476 | 9505 | 9323 | Inv2 |
| 5/7/2001 | 7770 | 7907 | 7819 | Inv2 |
| 5/7/2001 | 7901 | 7830 | 7715 | Inv2 |
| 5/7/2001 | 7270 | 7241 | 7095 | Inv2 |
| 5/7/2001 | 7231 | 7150 | 7128 | Inv2 |
| 5/7/2001 | 7590 | 7508 | 7655 | Inv2 |
| 5/7/2001 | 6853 | 6826 | 6447 | Inv2 |

| 5/7/2001  | 7188  | 6886  | 6972  | Inv2 |
|-----------|-------|-------|-------|------|
| 5/10/2001 | 9369  | 9103  | 9227  | Inv2 |
| 5/10/2001 | 9727  | 9585  | 9579  | Inv2 |
| 5/10/2001 | 8693  | 8509  | 9579  | Inv2 |
| 5/10/2001 | 7737  | 7611  | 7600  | Inv2 |
| 5/10/2001 | 7020  | 6816  | 6977  | Inv2 |
| 5/10/2001 | 8526  | 8421  | 8221  | Inv2 |
| 5/10/2001 | 7227  | 7173  | 7246  | Inv2 |
| 5/10/2001 | 6389  | 6438  | 6286  | Inv2 |
| 5/10/2001 | 6780  | 6713  | 6284  | Inv2 |
| 5/10/2001 | 6926  | 7072  | 6991  | Inv2 |
| 5/14/2001 | 2486  | 2494  | 2319  | Inv2 |
| 5/14/2001 | 3707  | 3712  | 3810  | Inv2 |
| 5/14/2001 | 3070  | 3236  | 3239  | Inv2 |
| 5/14/2001 | 3140  | 3161  | 3048  | Inv2 |
| 5/14/2001 | 3254  | 3223  | 3238  | Inv2 |
| 5/14/2001 | 2847  | 2849  | 2832  | Inv2 |
| 5/14/2001 | 2182  | 2146  | 2177  | Inv2 |
| 5/14/2001 | 2243  | 2091  | 2165  | Inv2 |
| 5/14/2001 | 2004  | 1930  | 1969  | Inv2 |
| 5/16/2001 | 6330  | 6426  | 6698  | Inv2 |
| 5/16/2001 | 5781  | 5650  | 5363  | Inv2 |
| 5/16/2001 | 6516  | 6725  | 6880  | Inv2 |
| 5/16/2001 | 7150  | 7274  | 7095  | Inv2 |
| 5/16/2001 | 6426  | 6325  | 6203  | Inv2 |
| 5/16/2001 | 6768  | 6687  | 6843  | Inv2 |
| 5/16/2001 | 6977  | 6993  | 6768  | Inv2 |
| 5/16/2001 | 6693  | 6816  | 6789  | Inv2 |
| 5/16/2001 | 7150  | 7464  | 7241  | Inv2 |
| 5/18/2001 | 8537  | 8249  | 8526  | Inv2 |
| 5/18/2001 | 8326  | 8454  | 8210  | Inv2 |
| 5/18/2001 | 8210  | 8365  | 8121  | Inv2 |
| 5/18/2001 | 8448  | 8393  | 8443  | Inv2 |
| 5/18/2001 | 7721  | 7978  | 7764  | Inv2 |
| 5/18/2001 | 8554  | 8889  | 8856  | Inv2 |
| 5/18/2001 | 7578  | 7742  | 7578  | Inv2 |
| 5/18/2001 | 8559  | 8497  | 8371  | Inv2 |
| 5/18/2001 | 8138  | 8326  | 8309  | Inv2 |
| 5/18/2001 | 7247  | 7036  | 7074  | Inv2 |
| 5/20/2001 | 10709 | 10861 | 10791 | Inv2 |
| 5/20/2001 | 9465  | 9403  | 9510  | Inv2 |

| | | | | | |
|---|---|---|---|---|---|
| 5/20/2001 | 9522 | 9493 | 9664 | Inv2 | |
| 5/20/2001 | 9836 | 9973 | 10089 | Inv2 | |
| 5/20/2001 | 10072 | 9853 | 9939 | Inv2 | |
| 5/20/2001 | 10100 | 10117 | 10267 | Inv2 | |
| 5/20/2001 | 10072 | 10383 | 10204 | Inv2 | |
| 5/20/2001 | 8985 | 8957 | 8766 | Inv2 | |
| 5/20/2001 | 9193 | 9222 | 9193 | Inv2 | |
| 5/20/2001 | 8811 | 8759 | 8626 | Inv2 | |
| 5/21/2001 | 4045 | 4068 | 4157 | Inv2 | |
| 5/21/2001 | 3408 | 3532 | 3463 | Inv2 | |
| 5/21/2001 | 4809 | 4918 | 4724 | Inv2 | |
| 5/21/2001 | 4955 | 4289 | 4451 | Inv2 | |
| 5/21/2001 | 4404 | 4404 | 4672 | Inv2 | |
| 5/21/2001 | 5520 | 5468 | 5692 | Inv2 | |
| 5/21/2001 | 5676 | 5750 | 5682 | Inv2 | |
| 5/21/2001 | 3131 | 3271 | 3201 | Inv2 | |
| 5/21/2001 | 3326 | 3438 | 3417 | Inv2 | |
| 5/21/2001 | 3347 | 3522 | 3587 | Inv2 | |
| 7/2/2001 | 9779 | 9910 | 9659 | Inv2 | |
| 7/2/2001 | 9996 | 10014 | 10072 | Inv2 | |
| 7/2/2001 | 9369 | 9004 | 8934 | Inv2 | |
| 7/2/2001 | 9687 | 9505 | 9647 | Inv2 | |
| 7/2/2001 | 9080 | 9335 | 9210 | Inv2 | |
| 7/2/2001 | 9643 | 9596 | 9431 | Inv2 | |
| 7/2/2001 | 8609 | 8598 | 8421 | Inv2 | |
| 7/2/2001 | 8744 | 8699 | 8716 | Inv2 | |
| 7/20/2001 | 4827 | 4878 | 4582 | Inv2 | |
| 7/20/2001 | 5348 | 5384 | 5312 | Inv2 | |
| 7/20/2001 | 4752 | 4898 | 4672 | Inv2 | |
| 7/20/2001 | 5120 | 4944 | 4934 | Inv2 | |
| 7/20/2001 | 5457 | 5625 | 5562 | Inv2 | |
| 7/20/2001 | 5265 | 5084 | | Inv2 | |
| 7/20/2001 | 4929 | 4878 | 4753 | Inv2 | |
| 7/20/2001 | 5115 | 4965 | 5094 | Inv2 | |
| 7/20/2001 | 4615 | 4711 | 4583 | Inv2 | |
| 7/20/2001 | 4662 | 4596 | 4647 | Inv2 | |
| 7/23/2001 | 8660 | 8632 | | Inv2 | |
| 7/23/2001 | 8221 | 8643 | | Inv2 | |
| 7/23/2001 | 8321 | 8571 | | Inv2 | |
| 7/23/2001 | 7231 | 7155 | | Inv2 | |
| 7/23/2001 | 7769 | 7770 | | Inv2 | |

| Date | | | | | |
|---|---|---|---|---|---|
| 7/23/2001 | 3920 | 3990 | 3970 | Inv2 | |
| 7/23/2001 | 3793 | 3876 | 3873 | Inv2 | |
| 7/23/2001 | 4388 | 4288 | 4308 | Inv2 | |
| 7/23/2001 | 4728 | 5001 | 4888 | Inv2 | |
| 7/26/2001 | 1533 | 1634 | 1624 | Inv2 | |
| 7/26/2001 | 1642 | 1653 | 1705 | Inv2 | |
| 7/26/2001 | 1246 | 1154 | 1201 | Inv2 | |
| 7/26/2001 | 1699 | 1627 | 1667 | Inv2 | |
| 7/26/2001 | 991 | 995 | 905 | Inv2 | |
| 7/26/2001 | 1257 | 1323 | 1224 | Inv2 | |
| 7/26/2001 | 1397 | 1350 | 1405 | Inv2 | |
| 7/26/2001 | 1485 | 1383 | 1367 | Inv2 | |
| 7/26/2001 | 1381 | 1331 | 1364 | Inv2 | |
| 7/30/2001 | 6039 | 5960 | 6050 | Inv2 | |
| 7/30/2001 | 7639 | 7858 | 7775 | Inv2 | |
| 8/2/2001 | 2503 | 2631 | 2542 | Inv2 | |
| 8/2/2001 | 2411 | 2478 | 2399 | Inv2 | |
| 8/2/2001 | 2616 | 2806 | 2650 | Inv2 | |
| 8/2/2001 | 2692 | 2730 | 2659 | Inv2 | |
| 8/2/2001 | 2224 | 2289 | 2213 | Inv2 | |
| 8/2/2001 | 2206 | 2335 | 2294 | Inv2 | |
| 8/2/2001 | 2497 | 2317 | 2410 | Inv2 | |
| 8/2/2001 | 2087 | 2138 | 2051 | Inv2 | |
| 9/27/2001 | 3642 | 3420 | 3521 | Inv2 | |
| 9/27/2001 | 3584 | 3611 | 3603 | Inv2 | |
| 9/27/2001 | 3942 | 3986 | 3899 | Inv2 | |
| 9/27/2001 | 3691 | 3715 | 3671 | Inv2 | |
| 9/27/2001 | 3879 | 3764 | 4033 | Inv2 | |
| 9/27/2001 | 3381 | 3457 | 3429 | Inv2 | |
| 9/27/2001 | 3177 | 3184 | 3164 | Inv2 | |
| 9/27/2001 | 2683 | 2560 | 2568 | Inv2 | |
| 9/27/2001 | 2595 | 2538 | 2661 | Inv2 | |
| 9/27/2001 | 2687 | 2770 | 2835 | Inv2 | |
| 9/28/2001 | 1800 | 1788 | 1776 | Inv2 | |
| 9/28/2001 | 2014 | 2205 | 2183 | Inv2 | |
| 9/28/2001 | 2211 | 2205 | 2183 | Inv2 | |
| 9/28/2001 | 2453 | 2449 | 2468 | Inv2 | |
| 9/28/2001 | 1632 | 1610 | 1632 | Inv2 | |
| 9/28/2001 | 1928 | 1938 | 1806 | Inv2 | |
| 9/28/2001 | 1433 | 1581 | 1539 | Inv2 | |
| 10/2/2001 | 711 | 743 | 727 | Inv2 | |
| 10/2/2001 | 1000 | 848 | 965 | Inv2 | |
| 10/2/2001 | 3100 | 2959 | 2903 | Inv2 | |
| 10/4/2001 | 10129 | 10465 | 10459 | Inv2 | |
| 10/4/2001 | 10726 | 10984 | 10996 | Inv2 | |
| 10/4/2001 | 7649 | 7852 | 7524 | Inv2 | |
| 10/4/2001 | 5488 | 5462 | 5520 | Inv2 | |
| 10/4/2001 | 5229 | 5198 | 5193 | Inv2 | |
| 10/4/2001 | 4846 | 4617 | 4692 | Inv2 | |
| 10/4/2001 | 4801 | 4713 | 4873 | Inv2 | |
| 10/4/2001 | 4555 | 4307 | 4461 | Inv2 | |
| 10/4/2001 | 4617 | 4532 | 4573 | Inv2 | |
| 10/4/2001 | 4435 | 4323 | | Inv2 | |
| 10/4/2001 | 1651 | 1732 | 1679 | Inv2 | |
| 10/4/2001 | 3028 | 3240 | 3186 | Inv2 | |
| 10/4/2001 | 2119 | 2305 | 2171 | Inv2 | |
| 10/11/2001 | 6671 | 6821 | 6821 | Inv2 | |
| 10/11/2001 | 7770 | 7945 | 7764 | Inv2 | |
| 10/11/2001 | 8309 | 8249 | 8271 | Inv2 | |
| 10/11/2001 | 7802 | 7704 | 7721 | Inv2 | |
| 10/11/2001 | 8249 | 8127 | 8188 | Inv2 | |
| 10/11/2001 | 8409 | 8504 | 8426 | Inv2 | |
| 10/15/2001 | 2545 | 2567 | 2587 | Inv2 | |
| 10/15/2001 | 2452 | 2471 | 2458 | Inv2 | |
| 10/15/2001 | 2429 | 2309 | 2446 | Inv2 | |
| 10/15/2001 | 2352 | 2353 | 2288 | Inv2 | |
| 10/15/2001 | 2350 | 2377 | 2367 | Inv2 | |
| 10/15/2001 | 2184 | 2096 | 2178 | Inv2 | |
| 10/15/2001 | 2133 | 2242 | 2163 | Inv2 | |
| 10/15/2001 | 2491 | 2436 | 2396 | Inv2 | |
| 10/15/2001 | 2286 | 2298 | 2245 | Inv2 | |
| 10/15/2001 | 2289 | 2281 | 2433 | Inv2 | |
| 10/15/2001 | 3728 | 3228 | 3519 | Inv2 | |
| 10/15/2001 | 3580 | 3380 | 3388 | Inv2 | |
| 10/15/2001 | 3057 | 2847 | 3022 | Inv2 | |
| 10/15/2001 | 2959 | 2896 | 2876 | Inv2 | |
| 10/15/2001 | 2960 | 2909 | 3043 | Inv2 | |
| 10/15/2001 | 2585 | 2441 | 2631 | Inv2 | |
| 10/15/2001 | 2474 | 2527 | 2520 | Inv2 | |
| 10/15/2001 | 2286 | 2264 | 2259 | Inv2 | |
| 10/15/2001 | 2451 | 2486 | 2497 | Inv2 | |
| 10/15/2001 | 2596 | 2467 | 2466 | Inv2 | |
| 10/16/2001 | 2011 | 2052 | 2012 | Inv2 | |
| 10/16/2001 | 1476 | 1567 | 1502 | Inv2 | |
| 10/16/2001 | 1536 | 1309 | 1393 | Inv2 | |
| 10/24/2001 | 1543 | 1499 | | Inv2 | |
| 10/24/2001 | 1551 | 1470 | 1545 | Inv2 | |
| 10/24/2001 | 1058 | 1008 | 989 | Inv2 | |
| 10/24/2001 | 856 | 892 | 867 | Inv2 | |
| 10/24/2001 | 872 | 852 | 887 | Inv2 | |
| 10/26/2001 | 1259 | 1259 | 1318 | Inv2 | |
| 10/26/2001 | 1212 | 1179 | 1241 | Inv2 | |
| 10/26/2001 | 981 | 970 | 939 | Inv2 | |
| 10/26/2001 | 1022 | 1088 | 1066 | Inv2 | |
| 10/26/2001 | 1093 | 1038 | 1056 | Inv2 | |
| 10/26/2001 | 1413 | 1471 | 1471 | Inv2 | |
| 10/26/2001 | 1111 | 1041 | 1077 | Inv2 | |
| 10/26/2001 | 1057 | 1078 | 1010 | Inv2 | |
| 10/26/2001 | 1041 | 957 | 1016 | Inv2 | |
| 10/26/2001 | 891 | 884 | 937 | Inv2 | |
| 10/29/2001 | 1155 | 1144 | 1191 | Inv2 | |
| 10/29/2001 | 1404 | 1533 | 1438 | Inv2 | |
| 10/29/2001 | 993 | 971 | 942 | Inv2 | |
| 10/29/2001 | 917 | 895 | 899 | Inv2 | |
| 10/29/2001 | 746 | 777 | 764 | Inv2 | |
| 10/29/2001 | 1180 | 1121 | 1285 | Inv2 | |
| 10/29/2001 | 1026 | 932 | 997 | Inv2 | |
| 10/29/2001 | 939 | 925 | 905 | Inv2 | |
| 10/29/2001 | 760 | 732 | 720 | Inv2 | |
| 10/29/2001 | 698 | 702 | 706 | Inv2 | |
| 10/30/2001 | 3728 | 3228 | 3519 | Inv2 | |
| 10/30/2001 | 3580 | 3380 | 3388 | Inv2 | |
| 10/30/2001 | 3057 | 2847 | 3022 | Inv2 | |
| 10/30/2001 | 2959 | 2896 | 2876 | Inv2 | |
| 10/30/2001 | 2960 | 2909 | 3043 | Inv2 | |
| 10/30/2001 | 2585 | 2441 | 2631 | Inv2 | |
| 10/30/2001 | 2474 | 2527 | 2520 | Inv2 | |
| 10/30/2001 | 2286 | 2264 | 2259 | Inv2 | |
| 10/30/2001 | 2451 | 2486 | 2497 | Inv2 | |
| 10/30/2001 | 2596 | 2467 | 2466 | Inv2 | |
| 11/2/2001 | 2383 | 2564 | 2471 | Inv2 | |
| 11/2/2001 | 2899 | 2843 | 2909 | Inv2 | |
| 11/2/2001 | 1808 | 1840 | 1922 | Inv2 | |
| 11/2/2001 | 1761 | 1852 | 1789 | Inv2 | |
| 11/2/2001 | 1777 | 1721 | 1768 | Inv2 | |
| 11/2/2001 | 3159 | 3060 | 3081 | Inv2 | |
| 11/2/2001 | 2791 | 2736 | 2752 | Inv2 | |
| 11/2/2001 | 2193 | 2202 | 2202 | Inv2 | |
| 11/2/2001 | 1754 | 1799 | 1716 | Inv2 | |
| 11/2/2001 | 1489 | 1471 | 1509 | Inv2 | |
| 11/2/2001 | 5650 | 5562 | 5619 | Inv2 | |
| 11/2/2001 | 4183 | 3992 | 3862 | Inv2 | |
| 11/2/2001 | 3123 | 3212 | 3205 | Inv2 | |
| 11/2/2001 | 3531 | 3428 | 3545 | Inv2 | |
| 11/2/2001 | 2962 | 3047 | 3039 | Inv2 | |
| 11/5/2001 | 3438 | 3571 | 3566 | Inv2 | |
| 11/5/2001 | 3715 | 3784 | 3761 | Inv2 | |
| 11/5/2001 | 2164 | 1991 | 3761 | Inv2 | |
| 11/5/2001 | 1749 | 1714 | | Inv2 | |
| 11/5/2001 | 3733 | 3756 | 3801 | Inv2 | |
| 11/5/2001 | 3391 | 3243 | 3336 | Inv2 | |
| 11/5/2001 | 3322 | 3294 | 3405 | Inv2 | |
| 11/5/2001 | 3770 | 3693 | 3863 | Inv2 | |
| 11/5/2001 | 3675 | 3410 | 3525 | Inv2 | |
| 11/5/2001 | 3440 | 3353 | 3605 | Inv2 | |
| 11/5/2001 | 3503 | 3468 | 3397 | Inv2 | |
| 11/5/2001 | 3538 | 3371 | 3465 | Inv2 | |
| 12/20/2001 | 424 | 411 | 437 | Inv2 | |
| 12/20/2001 | 494 | 445 | 499 | Inv2 | |
| 12/20/2001 | 414 | 410 | 397 | Inv2 | |
| 12/20/2001 | 376 | 368 | 353 | Inv2 | |
| 12/20/2001 | 297 | 296 | 332 | Inv2 | |
| 12/20/2001 | 290 | 302 | 293 | Inv2 | |
| 12/20/2001 | 260 | 258 | 271 | Inv2 | |
| 12/20/2001 | 270 | 250 | 227 | Inv2 | |
| 12/20/2001 | 226 | 233 | 249 | Inv2 | |
| 12/20/2001 | 258 | 231 | 212 | Inv2 | |
| 1/10/2000 | 4784 | 4924 | 4924 | Inv7 | |
| 1/10/2000 | 4660 | 4691 | 4534 | Inv7 | |
| 1/10/2000 | 4222 | 4365 | 4164 | Inv7 | |
| 1/10/2000 | 4812 | 4949 | 4845 | Inv7 | |
| 1/10/2000 | 5105 | 5043 | 4862 | Inv7 | |
| 1/10/2000 | 3755 | 3782 | 3766 | Inv7 | |
| 1/10/2000 | 4908 | 5022 | 4747 | Inv7 | |

| Date | Val1 | Val2 | Val3 | Inv |
|---|---|---|---|---|
| 1/10/2000 | 4334 | 4888 | 4929 | Inv7 |
| 10/2/2000 | 304 | 266 | 288 | Inv7 |
| 10/2/2000 | 395 | 341 | 307 | Inv7 |
| 10/2/2000 | 414 | 346 | 336 | Inv7 |
| 10/2/2000 | 379 | 408 | 331 | Inv7 |
| 10/2/2000 | 360 | 367 | 316 | Inv7 |
| 10/2/2000 | 354 | 290 | 269 | Inv7 |
| 10/2/2000 | 246 | 233 | 198 | Inv7 |
| 10/2/2000 | 319 | 322 | 333 | Inv7 |
| 10/2/2000 | 298 | 294 | 314 | Inv7 |
| 10/2/2000 | 373 | 328 | 335 | Inv7 |
| 12/15/2000 | 158 | 160 | 175 | Inv7 |
| 12/15/2000 | 170 | 161 | 152 | Inv7 |
| 12/15/2000 | 172 | 134 | 114 | Inv7 |
| 12/15/2000 | 126 | 94 | 125 | Inv7 |
| 12/15/2000 | 150 | 124 | 81 | Inv7 |
| 12/15/2000 | 159 | 157 | 168 | Inv7 |
| 12/15/2000 | 144 | 109 | 117 | Inv7 |
| 12/15/2000 | 150 | 125 | 146 | Inv7 |
| 12/15/2000 | 204 | 194 | 219 | Inv7 |
| 12/15/2000 | 226 | 215 | 215 | Inv7 |
| 12/23/2000 | 5892 | 5839 | 5793 | Inv7 |
| 12/23/2000 | 6119 | 6114 | 6118 | Inv7 |
| 12/23/2000 | 5792 | 5755 | 5850 | Inv7 |
| 12/23/2000 | 6634 | 6682 | 6469 | Inv7 |
| 12/23/2000 | 6891 | 7004 | 6795 | Inv7 |
| 12/23/2000 | 6623 | 6693 | 6666 | Inv7 |
| 12/23/2000 | 6378 | 6447 | 6511 | Inv7 |
| 12/23/2000 | 6783 | 6870 | 6735 | Inv7 |
| 12/23/2000 | 6768 | 6511 | 6618 | Inv7 |
| 12/23/2000 | 5671 | 5713 | 5698 | Inv7 |
| 12/23/2000 | 5022 | 5058 | 5146 | Inv7 |
| 12/23/2000 | 4965 | 4955 | 5063 | Inv7 |
| 12/23/2000 | 6490 | 5844 | 6018 | Inv7 |
| 1/5/2001 | 3958 | 3820 | 3830 | Inv7 |
| 1/5/2001 | 4853 | 4705 | 4686 | Inv7 |
| 1/5/2001 | 5244 | 5281 | 5162 | Inv7 |
| 1/5/2001 | 5671 | 5807 | 5802 | Inv7 |
| 1/5/2001 | 5270 | 5462 | 5447 | Inv7 |
| 1/5/2001 | 5807 | 5771 | 5929 | Inv7 |
| 1/11/2001 | 10262 | 9899 | 10314 | Inv7 |

| Date | | | | |
|---|---|---|---|---|
| 1/11/2001 | 9266 | 9939 | 9670 | Inv7 |
| 1/11/2001 | 8794 | 9261 | 9369 | Inv7 |
| 1/11/2001 | 9329 | 9561 |  | Inv7 |
| 1/11/2001 | 9928 | 10037 | 9876 | Inv7 |
| 1/11/2001 | 9188 | 9392 | 9408 | Inv7 |
| 1/11/2001 | 10896 | 10896 | 11031 | Inv7 |
| 1/11/2001 | 9539 | 10072 | 9842 | Inv7 |
| 1/11/2001 | 8177 | 7726 | 7699 | Inv7 |
| 1/11/2001 | 8263 | 8165 | 8326 | Inv7 |
| 1/11/2001 | 7709 | 7660 | 7688 | Inv7 |
| 1/11/2001 | 6282 | 6171 | 6261 | Inv7 |
| 1/11/2001 | 7671 | 7568 | 7682 | Inv7 |
| 1/11/2001 | 7655 | 7529 | 7429 | Inv7 |
| 1/11/2001 | 7726 | 7693 | 7573 | Inv7 |
| 1/11/2001 | 6256 | 6235 | 6336 | Inv7 |
| 1/13/2001 | 3995 | 4004 | 4023 | Inv7 |
| 1/13/2001 | 3050 | 3128 | 3087 | Inv7 |
| 1/13/2001 | 3183 | 3310 | 3296 | Inv7 |
| 1/13/2001 | 3886 | 3794 | 3817 | Inv7 |
| 1/13/2001 | 3363 | 3383 | 3317 | Inv7 |
| 1/13/2001 | 4069 | 3998 | 3934 | Inv7 |
| 1/14/2001 | 6891 | 7301 | 8315 | Inv7 |
| 1/14/2001 | 5850 | 5863 | 5659 | Inv7 |
| 1/14/2001 | 5128 | 5348 | 5165 | Inv7 |
| 1/14/2001 | 6302 | 6425 | 6501 | Inv7 |
| 1/14/2001 | 6376 | 6559 | 6596 | Inv7 |
| 1/14/2001 | 5499 | 5541 | 5270 | Inv7 |
| 1/14/2001 | 8393 | 8677 | 8265 | Inv7 |
| 1/14/2001 | 7858 | 7901 | 8210 | Inv7 |
| 1/14/2001 | 4802 | 4588 | 4550 | Inv7 |
| 1/14/2001 | 4918 | 5099 | 4981 | Inv7 |
| 1/14/2001 | 4784 | 4816 | 4841 | Inv7 |
| 1/14/2001 | 4341 | 4425 | 4321 | Inv7 |
| 1/15/2001 | 3308 | 3203 | 3169 | Inv7 |
| 1/15/2001 | 5224 | 5084 | 5156 | Inv7 |
| 1/15/2001 | 4585 | 4913 | 4794 | Inv7 |
| 1/15/2001 | 5187 | 4913 | 4893 | Inv7 |
| 1/15/2001 | 4709 | 4805 | 4758 | Inv7 |
| 1/15/2001 | 5089 | 5379 | 5198 | Inv7 |
| 1/15/2001 | 2888 | 2932 | 2920 | Inv7 |
| 1/15/2001 | 3470 | 3324 | 3510 | Inv7 |

| | | | | | |
|---|---|---|---|---|---|
| 1/15/2001 | 3702 | 3672 | 3608 | Inv7 | |
| 1/15/2001 | 3293 | 3378 | 3306 | Inv7 | |
| 1/15/2001 | 4094 | 3993 | 4008 | Inv7 | |
| 1/15/2001 | 3905 | 3803 | 3840 | Inv7 | |
| 2/16/2001 | 992 | 963 | 945 | Inv7 | |
| 2/16/2001 | 1008 | 950 | 987 | Inv7 | |
| 2/16/2001 | 630 | 670 | 626 | Inv7 | |
| 2/16/2001 | 969 | 704 | 451 | Inv7 | |
| 2/16/2001 | 600 | 592 | 610 | Inv7 | |
| 2/16/2001 | 516 | 469 | 576 | Inv7 | |
| 2/16/2001 | 527 | 515 | 513 | Inv7 | |
| 2/16/2001 | 638 | 556 | 655 | Inv7 | |
| 2/16/2001 | 2116 | 2088 | 2134 | Inv7 | |
| 2/16/2001 | 2006 | 1928 | 1846 | Inv7 | |
| 2/16/2001 | 1511 | 1462 | 1454 | Inv7 | |
| 2/16/2001 | 1085 | 1512 | 1512 | Inv7 | |
| 2/16/2001 | 1720 | 1673 | 1688 | Inv7 | |
| 2/16/2001 | 1850 | 1845 | 1758 | Inv7 | |
| 2/16/2001 | 1433 | 1514 | 1548 | Inv7 | |
| 2/16/2001 | 1457 | 1515 | 1357 | Inv7 | |
| 2/16/2001 | 1472 | 1489 | 1511 | Inv7 | |
| 2/19/2001 | 3473 | 3378 | 3365 | Inv7 | |
| 2/19/2001 | 3769 | 3862 | 3883 | Inv7 | |
| 2/19/2001 | 2594 | 2682 | 2596 | Inv7 | |
| 2/19/2001 | 2616 | 2535 | 2493 | Inv7 | |
| 2/19/2001 | 3196 | 3153 | 3156 | Inv7 | |
| 2/19/2001 | 2979 | 3024 | 3078 | Inv7 | |
| 2/19/2001 | 2578 | 2630 | 2637 | Inv7 | |
| 2/19/2001 | 2809 | 2778 | 2828 | Inv7 | |
| 2/19/2001 | 2502 | 2654 | 2549 | Inv7 | |
| 2/19/2001 | 3005 | 2944 | 2760 | Inv7 | |
| 3/12/2001 | 6720 | 6923 | 6789 | Inv7 | |
| 3/12/2001 | 6410 | 6399 | 6251 | Inv7 | |
| 3/12/2001 | 5598 | 5562 | 5452 | Inv7 | |
| 3/12/2001 | 5281 | 5260 | 5395 | Inv7 | |
| 3/12/2001 | 4981 | 5265 | 5281 | Inv7 | |
| 3/12/2001 | 4970 | 5006 | 5084 | Inv7 | |
| 3/12/2001 | 4883 | 4898 | 4934 | Inv7 | |
| 3/12/2001 | 5198 | 5198 | 5379 | Inv7 | |
| 3/12/2001 | 4783 | 4782 | 4706 | Inv7 | |
| 3/12/2001 | 4709 | 4669 | 4605 | Inv7 | |

| 4/6/2001 | 5609 | 5468 | 5562 | Inv7 | |
|---|---|---|---|---|---|
| 4/6/2001 | 6538 | 6596 | 6634 | Inv7 | |
| 4/6/2001 | 6586 | 6628 | 6538 | Inv7 | |
| 4/6/2001 | 6639 | 6918 | 6735 | Inv7 | |
| 4/6/2001 | 6245 | 6282 | 6161 | Inv7 | |
| 4/6/2001 | 6956 | 6709 | 6923 | Inv7 | |
| 4/6/2001 | 6367 | 6431 | 6532 | Inv7 | |
| 4/6/2001 | 4597 | 4525 | 4148 | Inv7 | |
| 4/6/2001 | 6452 | 6373 | 6309 | Inv7 | |
| 4/6/2001 | 6655 | 6810 | 6805 | Inv7 | |
| 5/1/2001 | 5379 | 5556 | 5437 | Inv7 | |
| 5/1/2001 | 4488 | 4359 | 4494 | Inv7 | |
| 5/1/2001 | 4429 | 4526 | 4466 | Inv7 | |
| 5/1/2001 | 4349 | 4142 | 4271 | Inv7 | |
| 5/1/2001 | 5306 | 5198 | 5084 | Inv7 | |
| 5/1/2001 | 5048 | 5761 | 5032 | Inv7 | |
| 5/1/2001 | 5509 | 5395 | 5390 | Inv7 | |
| 5/1/2001 | 5239 | 4778 | 5110 | Inv7 | |
| 5/1/2001 | 5110 | 5136 | 5012 | Inv7 | |
| 5/3/2001 | 6479 | 6522 | 6564 | Inv7 | |
| 5/3/2001 | 7117 | 6864 | 6859 | Inv7 | |
| 5/3/2001 | 6336 | 6410 | 6373 | Inv7 | |
| 5/3/2001 | 7404 | 7377 | 7432 | Inv7 | |
| 5/3/2001 | 7595 | 7546 | 7263 | Inv7 | |
| 5/3/2001 | 6399 | 6607 | 6325 | Inv7 | |
| 5/3/2001 | 5950 | 6039 | 5761 | Inv7 | |
| 5/3/2001 | 6655 | 6821 | 6607 | Inv7 | |
| 5/3/2001 | 5587 | 5416 | 5598 | Inv7 | |
| 5/3/2001 | 7052 | 7258 | 7144 | Inv7 | |
| 5/25/2001 | 6134 | 6219 | 6161 | Inv7 | |
| 5/25/2001 | 6789 | 6628 | 6961 | Inv7 | |
| 5/25/2001 | 6934 | 6950 | 6751 | Inv7 | |
| 5/25/2001 | 7188 | 7166 | 7101 | Inv7 | |
| 5/25/2001 | 7464 | 7568 | 7366 | Inv7 | |
| 5/25/2001 | 6720 | 6864 | 6800 | Inv7 | |
| 5/25/2001 | 6437 | 6516 | 6442 | Inv7 | |
| 5/25/2001 | 6548 | 6789 | 6661 | Inv7 | |
| 5/25/2001 | 6044 | 6415 | 6176 | Inv7 | |
| 5/25/2001 | 6515 | 6607 | 6469 | Inv7 | |
| 5/28/2001 | 8643 | 8671 | 8699 | Inv7 | |
| 5/28/2001 | 9465 | 9199 | 9250 | Inv7 | |
| 5/28/2001 | 7846 | 7945 | 7764 | Inv7 | |
| 5/28/2001 | 6119 | 6208 | 5965 | Inv7 | |
| 5/28/2001 | 9556 | 9329 | 9476 | Inv7 | |
| 5/28/2001 | 9301 | 9238 | 9357 | Inv7 | |
| 5/28/2001 | 8744 | 8582 | 8382 | Inv7 | |
| 5/28/2001 | 8121 | 8066 | 8227 | Inv7 | |
| 5/28/2001 | 7639 | 7846 | 7535 | Inv7 | |
| 5/31/2001 | 6426 | 6762 | 6720 | Inv7 | |
| 5/31/2001 | 7890 | 7715 | 7858 | Inv7 | |
| 5/31/2001 | 9414 | 9836 | 9784 | Inv7 | |
| 5/31/2001 | 6843 | 6773 | 6778 | Inv7 | |
| 5/31/2001 | | 7052 | 6875 | Inv7 | |
| 5/31/2001 | 9266 | 9352 | 9420 | Inv7 | |
| 5/31/2001 | 8028 | 7731 | 7731 | Inv7 | |
| 5/31/2001 | 10083 | 10308 | 10302 | Inv7 | |
| 5/31/2001 | 9767 | 9590 | 9710 | Inv7 | |
| 6/3/2001 | 9864 | 9779 | 10066 | Inv7 | |
| 6/3/2001 | 11462 | 11509 | 11161 | Inv7 | |
| 6/3/2001 | 10164 | 9996 | 10152 | Inv7 | |
| 6/3/2001 | 9967 | 10586 | 10756 | Inv7 | |
| 6/3/2001 | 9956 | 9819 | 9979 | Inv7 | |
| 6/3/2001 | 9471 | 9533 | 9522 | Inv7 | |
| 6/3/2001 | 10066 | 10401 | 10343 | Inv7 | |
| 6/3/2001 | 9493 | 9510 | 9830 | Inv7 | |
| 6/3/2001 | 11261 | 11208 | 11261 | Inv7 | |
| 6/5/2001 | 1668 | 1072 | 1642 | Inv7 | |
| 6/5/2001 | 1076 | 1131 | 1098 | Inv7 | |
| 6/5/2001 | 4557 | 4716 | 4518 | Inv7 | |
| 6/5/2001 | 3971 | 4078 | 3996 | Inv7 | |
| 6/5/2001 | 3919 | 3886 | 3878 | Inv7 | |
| 6/5/2001 | 1985 | 1949 | 1962 | Inv7 | |
| 6/5/2001 | 3083 | 3019 | 3061 | Inv7 | |
| 6/5/2001 | 3789 | 3614 | 3772 | Inv7 | |
| 6/5/2001 | 4934 | 4986 | 4924 | Inv7 | |
| 6/25/2001 | 6003 | 5803 | | Inv7 | |
| 6/25/2001 | 5067 | 5096 | 4980 | Inv7 | |
| 6/25/2001 | 6270 | 6373 | 6309 | Inv7 | |
| 6/25/2001 | 5934 | 5676 | 5619 | Inv7 | |
| 6/25/2001 | 5904 | 5828 | 5686 | Inv7 | |
| 6/25/2001 | 5903 | 5821 | 5843 | Inv7 | |
| 6/25/2001 | 5438 | 5627 | 5667 | Inv7 | |
| 6/25/2001 | 5456 | 5370 | 5454 | Inv7 | |
| 6/25/2001 | 5410 | | | Inv7 | |
| 6/25/2001 | 5637 | 5539 | 5497 | Inv7 | |
| 5/13/1996 | 765 | 818 | 768 | Inv5 | |
| 5/13/1996 | 516 | 496 | 545 | Inv5 | |
| 5/13/1996 | 476 | 450 | 496 | Inv5 | |
| 5/13/1996 | 484 | 476 | 500 | Inv5 | |
| 5/13/1996 | 449 | 457 | 468 | Inv5 | |
| 5/13/1996 | 484 | 505 | 521 | Inv5 | |
| 5/13/1996 | 417 | 411 | 399 | Inv5 | |
| 5/13/1996 | 464 | 476 | 429 | Inv5 | |
| 5/13/1996 | 312 | 375 | 366 | Inv5 | |
| 5/13/1996 | 353 | 345 | 333 | Inv5 | |
| 6/6/1996 | 469 | 479 | 455 | Inv5 | |
| 6/6/1996 | 577 | 567 | 586 | Inv5 | |
| 6/6/1996 | 292 | 281 | 271 | Inv5 | |
| 6/6/1996 | 514 | 510 | 485 | Inv5 | |
| 6/6/1996 | 583 | 592 | 625 | Inv5 | |
| 6/6/1996 | 541 | 507 | 537 | Inv5 | |
| 6/6/1996 | 564 | 558 | 523 | Inv5 | |
| 6/6/1996 | 505 | 519 | 522 | Inv5 | |
| 6/6/1996 | 427 | 442 | 452 | Inv5 | |
| 6/6/1996 | 240 | 222 | 229 | Inv5 | |
| 6/24/1996 | 518 | 543 | 513 | Inv5 | |
| 6/24/1996 | 584 | 624 | 628 | Inv5 | |
| 6/24/1996 | 608 | 578 | 582 | Inv5 | |
| 6/24/1996 | 504 | 521 | 530 | Inv5 | |
| 6/24/1996 | 485 | 497 | 523 | Inv5 | |
| 6/24/1996 | 538 | 555 | 485 | Inv5 | |
| 6/24/1996 | 487 | 467 | 474 | Inv5 | |
| 6/24/1996 | 604 | 619 | 615 | Inv5 | |
| 6/24/1996 | 580 | 550 | 535 | Inv5 | |
| 6/24/1996 | 457 | 411 | 418 | Inv5 | |
| 7/30/1996 | 788 | 735 | 684 | Inv5 | |
| 7/30/1996 | 650 | 738 | 715 | Inv5 | |
| 7/30/1996 | 707 | 678 | 743 | Inv5 | |
| 7/30/1996 | 683 | 654 | 631 | Inv5 | |
| 7/30/1996 | 703 | 721 | 648 | Inv5 | |
| 7/30/1996 | 503 | 499 | 452 | Inv5 | |
| 7/30/1996 | 795 | 824 | 791 | Inv5 | |
| 7/30/1996 | 551 | 479 | 493 | Inv5 | |

| Date | | | | |
|---|---|---|---|---|
| 7/30/1996 | 473 | 474 | 503 | Inv5 |
| 7/30/1996 | 510 | 561 | 579 | Inv5 |
| 8/20/1996 | 1234 | 1187 | 1101 | Inv5 |
| 8/20/1996 | 1434 | 1341 | 1298 | Inv5 |
| 8/20/1996 | 1061 | 1054 | 1080 | Inv5 |
| 8/20/1996 | 1219 | 1239 | 1192 | Inv5 |
| 8/20/1996 | 1237 | 1245 | 1236 | Inv5 |
| 8/20/1996 | 1179 | 1132 | 1128 | Inv5 |
| 8/20/1996 | 1303 | 1239 | 1303 | Inv5 |
| 8/20/1996 | 1304 | 1192 | 1211 | Inv5 |
| 8/20/1996 | 1166 | 1182 | 1121 | Inv5 |
| 8/20/1996 | 1226 | 1159 | 1227 | Inv5 |
| 4/30/1996 | 1068 | 1098 | 1052 | Inv5 |
| 4/30/1996 | 849 | 861 | 831 | Inv5 |
| 4/30/1996 | 801 | 725 | 772 | Inv5 |
| 4/30/1996 | 827 | 832 | 911 | Inv5 |
| 4/30/1996 | 703 | 687 | 667 | Inv5 |
| 4/30/1996 | 805 | 810 | 838 | Inv5 |
| 4/30/1996 | 839 | 831 | 819 | Inv5 |
| 4/30/1996 | 882 | 905 | 889 | Inv5 |
| 4/30/1996 | 776 | 750 | 751 | Inv5 |
| 4/30/1996 | 767 | 797 | 760 | Inv5 |
| 11/11/2002 | 2239 | 2226 | 2222 | Inv3 |
| 11/11/2002 | 1432 | 1365 | 1269 | Inv3 |
| 11/11/2002 | 1622 | 1703 | 1783 | Inv3 |
| 11/11/2002 | 1015 | 954 | 1057 | Inv3 |
| 11/11/2002 | 1278 | 1195 | 1313 | Inv3 |
| 11/11/2002 | 1172 | 1136 | 1118 | Inv3 |
| 11/11/2002 | 1143 | 1064 | 1130 | Inv3 |
| 11/11/2002 | 1585 | 1584 | 1430 | Inv3 |
| 11/11/2002 | 1509 | 1508 | 1521 | Inv3 |
| 11/11/2002 | 1705 | 1740 | 1663 | Inv3 |
| 11/12/2002 | 2225 | 2118 | 2104 | Inv3 |
| 11/12/2002 | 2118 | 2050 | 2027 | Inv3 |
| 11/12/2002 | 1973 | 2006 | 1981 | Inv3 |
| 11/12/2002 | 2350 | 2327 | 2285 | Inv3 |
| 11/12/2002 | 1952 | 1962 | 1968 | Inv3 |
| 11/12/2002 | 1831 | 1878 | 1856 | Inv3 |
| 11/12/2002 | 1894 | 1867 | 1888 | Inv3 |
| 11/12/2002 | 1916 | 1952 | 1966 | Inv3 |
| 11/12/2002 | 2120 | 2056 | 2091 | Inv3 |

| 11/12/2002 | 1956 | 1973 | 2002 | Inv3 |
|---|---|---|---|---|
| 5/26/2000 | 680 | 708 | 665 | Inv6 |
| 5/26/2000 | 690 | 672 | 693 | Inv6 |
| 5/26/2000 | 668 | 715 | 666 | Inv6 |
| 5/26/2000 | 670 | 701 | 679 | Inv6 |
| 5/26/2000 | 789 | 765 | 761 | Inv6 |
| 5/26/2000 | 904 | 843 | 851 | Inv6 |
| 5/26/2000 | 671 | 663 | 719 | Inv6 |
| 5/26/2000 | 774 | 736 | 758 | Inv6 |
| 5/26/2000 | 732 | 748 | 711 | Inv6 |
| 5/26/2000 | 804 | 759 | 758 | Inv6 |
| 7/28/2000 | 589 | 558 | 582 | Inv6 |
| 7/28/2000 | 537 | 500 | 541 | Inv6 |
| 7/28/2000 | 602 | 612 | 617 | Inv6 |
| 7/28/2000 | 533 | 568 | 512 | Inv6 |
| 7/28/2000 | 545 | 536 | 544 | Inv6 |
| 7/28/2000 | 615 | 595 | 585 | Inv6 |
| 7/28/2000 | 505 | 521 | 510 | Inv6 |
| 7/31/2000 | 107 | 91 | 82 | Inv6 |
| 7/31/2000 | 79 | 75 | 71 | Inv6 |
| 7/31/2000 | 36 | 36 | 55 | Inv6 |
| 7/31/2000 | 37 | 48 | 39 | Inv6 |
| 7/31/2000 | 64 | 49 | 64 | Inv6 |
| 7/31/2000 | 97 | 113 | 84 | Inv6 |
| 7/31/2000 | 71 | 77 | 75 | Inv6 |
| 7/31/2000 | 47 | 47 | 54 | Inv6 |
| 7/31/2000 | 81 | 72 | 65 | Inv6 |
| 7/31/2000 | 55 | 42 | 50 | Inv6 |
| 7/31/2000 | 272 | 300 | 268 | Inv6 |
| 7/31/2000 | 231 | 320 | 214 | Inv6 |
| 7/31/2000 | 251 | 258 | 297 | Inv6 |
| 7/31/2000 | 305 | 362 | 321 | Inv6 |
| 7/31/2000 | 484 | 499 | 473 | Inv6 |
| 7/31/2000 | 637 | 618 | 619 | Inv6 |
| 7/31/2000 | 650 | 641 | 626 | Inv6 |
| 7/31/2000 | 640 | 632 | 638 | Inv6 |
| 7/31/2000 | 608 | 690 | 636 | Inv6 |
| 7/31/2000 | 377 | 373 | 380 | Inv6 |
| 7/31/2000 | 441 | 414 | 455 | Inv6 |
| 7/31/2000 | 571 | 505 | 514 | Inv6 |
| 7/31/2000 | 475 | 439 | 432 | Inv6 |

| Date | | | | |
|---|---|---|---|---|
| 7/31/2000 | 405 | 416 | 441 | Inv6 |
| 8/11/2000 | 89 | 97 | 86 | Inv6 |
| 8/11/2000 | 331 | 316 | 329 | Inv6 |
| 8/11/2000 | 378 | 330 | 375 | Inv6 |
| 8/11/2000 | 333 | 404 | 367 | Inv6 |
| 8/11/2000 | 396 | 382 | 408 | Inv6 |
| 8/11/2000 | 342 | 331 | 344 | Inv6 |
| 8/11/2000 | 340 | 349 | 344 | Inv6 |
| 8/11/2000 | 325 | 347 | 304 | Inv6 |
| 8/11/2000 | 315 | 291 | 283 | Inv6 |
| 8/11/2000 | 307 | 339 | 323 | Inv6 |
| 8/11/2000 | 285 | 314 | 323 | Inv6 |
| 8/11/2000 | 260 | 262 | 284 | Inv6 |
| 8/11/2000 | 361 | 315 | 298 | Inv6 |
| 8/11/2000 | 355 | 324 | 356 | Inv6 |
| 10/14/1992 | 1257 | 1291 | 1224 | Inv9 |
| 10/14/1992 | 1032 | 987 | 1053 | Inv9 |
| 10/14/1992 | 1126 | 1081 | 1074 | Inv9 |
| 10/14/1992 | 1225 | 1248 | 1178 | Inv9 |
| 10/14/1992 | 1034 | 986 | 988 | Inv9 |
| 10/14/1992 | 994 | 988 | 1027 | Inv9 |
| 10/14/1992 | 932 | 900 | 917 | Inv9 |
| 10/14/1992 | 878 | 866 | 850 | Inv9 |
| 10/14/1992 | 927 | 874 | 885 | Inv9 |
| 10/14/1992 | 1947 | 1847 | 1815 | Inv9 |
| 4/15/1992 | 1547 | 1574 | 1523 | Inv9 |
| 4/15/1992 | 1617 | 1552 | 1570 | Inv9 |
| 4/15/1992 | 1258 | 1279 | 1284 | Inv9 |
| 4/15/1992 | 1273 | 1313 | 1286 | Inv9 |
| 4/15/1992 | 1071 | 1044 | 1044 | Inv9 |
| 4/15/1992 | 1014 | 1014 | 965 | Inv9 |
| 4/15/1992 | 1051 | 1012 | 990 | Inv9 |
| 4/15/1992 | 948 | 954 | 918 | Inv9 |
| 4/15/1992 | 1039 | 977 | 1060 | Inv9 |
| 4/15/1992 | 2023 | 1851 | 1830 | Inv9 |
| 4/29/1992 | 1427 | 1401 | 1447 | Inv9 |
| 4/29/1992 | 1181 | 1234 | 1109 | Inv9 |
| 4/29/1992 | 1147 | 1131 | 1195 | Inv9 |
| 4/29/1992 | 1252 | 1212 | 1267 | Inv9 |
| 4/29/1992 | 1224 | 1248 | 1211 | Inv9 |
| 4/29/1992 | 1297 | 1215 | 1194 | Inv9 |

| 4/29/1992 | 1161 | 1212 | 1112 | Inv9 | |
|---|---|---|---|---|---|
| 4/29/1992 | 1065 | 1058 | 1014 | Inv9 | |
| 4/29/1992 | 1049 | 1035 | 1038 | Inv9 | |
| 4/29/1992 | 1657 | 1696 | 1649 | Inv9 | |
| 6/30/1992 | 1898 | 1814 | 1849 | Inv9 | |
| 6/30/1992 | 1594 | 1508 | 1596 | Inv9 | |
| 6/30/1992 | 1460 | 1460 | 1519 | Inv9 | |
| 6/30/1992 | 1430 | 1385 | 1406 | Inv9 | |
| 6/30/1992 | 1279 | 1297 | 1234 | Inv9 | |
| 6/30/1992 | 1165 | 1185 | 1162 | Inv9 | |
| 6/30/1992 | 1128 | 1077 | 1099 | Inv9 | |
| 6/30/1992 | 1125 | 1095 | 1078 | Inv9 | |
| 6/30/1992 | 1297 | 1213 | 1283 | Inv9 | |
| 6/30/1992 | 1972 | 1903 | 1865 | Inv9 | |
| 8/3/1992 | 4523 | 4509 | 4568 | Inv9 | |
| 8/4/1992 | 1028 | 986 | 975 | Inv9 | |
| 8/4/1992 | 1020 | 940 | 927 | Inv9 | |
| 8/4/1992 | 1012 | 976 | 962 | Inv9 | |
| 8/4/1992 | 816 | 868 | 872 | Inv9 | |
| 8/4/1992 | 757 | 693 | 678 | Inv9 | |
| 8/4/1992 | 828 | 825 | 843 | Inv9 | |
| 8/4/1992 | 719 | 710 | 766 | Inv9 | |
| 8/4/1992 | 805 | 809 | 785 | Inv9 | |
| 8/4/1992 | 881 | 851 | 831 | Inv9 | |
| 8/4/1992 | 913 | 981 | 998 | Inv9 | |

| Outside Lab 2 Coulters | | | |
|---|---|---|---|
| Date | cou1 | cou2 | cou3 |
| 11/9/1998 | 914 | 1107 | 1146 |
| | 1867 | 1948 | 1974 |
| | 1678 | 1635 | 1976 |
| | 1535 | 1531 | 1516 |
| | 729 | 740 | 847 |
| | 381 | 462 | 393 |
| | 4679 | 1546 | 836 |
| | 1246 | 1295 | 1161 |
| | 1591 | 1690 | 1844 |
| | 605 | 809 | 1103 |
| | 405 | 511 | 650 |
| | 268 | 563 | 421 |
| | 2098 | 1748 | 1952 |
| 11/13/1998 | 183 | 438 | 402 |
| | 7915 | 8584 | 7969 |
| | 8755 | 9501 | 8694 |
| | 8311 | 8379 | 8748 |
| | 8017 | 8113 | 8079 |
| | 3155 | 3240 | 3254 |
| | 218 | 201 | 216 |
| | 5985 | 5078 | 7710 |
| | 5004 | 5099 | 4475 |
| | 1958 | 3446 | 3833 |
| | 481 | 561 | 1011 |
| | 287 | 368 | 322 |
| | 221 | 210 | 191 |
| 1/25/1999 | 128 | 111 | 123 |
| | 2251 | 2127 | 1897 |
| | 1625 | 1439 | |
| | 1639 | 1777 | |
| | 350 | 123 | |
| | 264 | 244 | |
| | 1823 | 1359 | |
| | 232 | 257 | |
| | 245 | 137 | |
| | 1742 | 1788 | |
| | 2196 | 1719 | |
| | 1426 | 1642 | |
| | 1782 | 1677 | |
| | 1459 | 1685 | |
| 1/29/1999 | 70 | 79 | 76 |
| | 533 | 512 | 921 |
| | 438 | 919 | 701 |
| | 1317 | 1295 | 1343 |
| | 202 | 146 | 310 |
| | 240 | 201 | 118 |
| | 839 | 823 | 1365 |
| | 146 | 189 | 93 |
| | 73 | 58 | 105 |
| | 1450 | 1305 | 1181 |
| | 1365 | 1648 | 1330 |
| | 1717 | 1677 | 1559 |
| | 1332 | 1073 | 1663 |
| | 1372 | 1669 | 1266 |
| 2/1/1999 | 1000 | 679 | 946 |
| | 2907 | 2040 | 1944 |
| | 61 | 73 | 37 |
| | 3739 | 3701 | 3355 |
| | 349 | 176 | 252 |
| | 215 | 371 | 218 |
| | 629 | 1105 | 2096 |
| | 276 | 181 | 387 |
| | 209 | 119 | 223 |
| | 502 | 614 | 477 |
| | 2155 | 2779 | 2425 |
| | 2379 | 2189 | 2836 |
| | 2195 | 2206 | 1792 |
| | 1226 | 987 | 1272 |
| 2/22/1999 | 1332 | 1105 | 1236 |
| | 2556 | 2597 | 2586 |
| | 1185 | 1357 | 1490 |
| | 1934 | 1496 | 1710 |
| | 1732 | 1970 | 1627 |
| | 2213 | 2014 | 1923 |
| 2/26/1999 | 615 | 1698 | 1393 |
| | 6884 | 9036 | 8120 |
| | 1394 | 2021 | 982 |
| | 8516 | 8612 | 8571 |
| | 747 | 1197 | 1309 |
| | 8796 | 7696 | 8352 |
| 3/1/1999 | 924 | 2528 | 3286 |
| | 6534 | 8495 | 8352 |
| | 2367 | 2406 | 1555 |
| | 6575 | 8250 | 8407 |
| | 1765 | 1359 | 476 |
| | 6850 | 8584 | 7963 |
| 4/5/1999 | 1426 | 2812 | 2638 |

|  | 3235 | 3232 | 2805 |
|---|---|---|---|
|  | 2444 | 2233 | 2318 |
|  | 2358 | 2556 | 2782 |
|  | 1815 | 1550 | 2284 |
|  | 1824 | 1958 | 2094 |
| 4/9/1999 | 3179 | 4576 | 3764 |
|  | 8414 | 8680 | 8536 |
|  | 4914 | 5677 | 4196 |
|  | 11434 | 9966 | 8831 |
|  | 4206 | 3589 | 2747 |
|  | 6081 | 5660 | 7331 |
| 6/7/1991 | 2266 | 2321 | 2192 |
|  | 676 | 554 | 478 |
|  | 601 | 723 | 520 |
|  | 582 | 516 | 881 |
|  | 3200 | 2719 | 3747 |
|  | 425 | 626 | 785 |
|  | 2200 | 1042 | 1847 |
|  | 1561 | 987 | 919 |
|  | 1141 | 1788 | 1957 |
|  | 1210 | 747 | 2233 |
|  | 1132 | 1286 | 2210 |

PAGE 195: Statistical Detection of Potentially Fabricated Data: A Case Study| Outside Lab 3 Coulters | | | |
|---|---|---|---|
| Date | cou1 | cou2 | cou3 |
| 6.6.2008 | 5868 | 5838 | 5691 |
| | 3451 | 3343 | 3315 |
| | 4844 | 4854 | 4695 |
| 6.10.08 | 4851 | 4549 | 4532 |
| | 3010 | 3018 | 2982 |
| | 4009 | 3989 | 3785 |
| 6.11.08 | 531 | 502 | 527 |
| | 558 | 550 | 511 |
| | 4417 | 4239 | 4381 |
| | 2076 | 2017 | 2039 |
| 6.14.08 | 4476 | 4710 | 4501 |
| | 4124 | 3985 | 3893 |
| | 9561 | 9164 | 9370 |
| | 3072 | 3007 | 3017 |
| | 2679 | 2622 | 2652 |
| 6.19.08 | 3274 | 3020 | 3008 |
| | 1590 | 1558 | 1538 |
| | 3184 | 3123 | 3221 |
| | 2911 | 2833 | 2739 |
| | 1309 | 1174 | 1077 |
| | 1374 | 1316 | 1312 |
| 6.20.08 | 941 | 870 | 828 |
| | 1694 | 1630 | 1637 |
| | 1320 | 1395 | 1373 |
| | 1549 | 1465 | 1492 |
| | 4066 | 4078 | 4024 |
| 6.24.08 | 4908 | 4716 | 4657 |
| | 4014 | 3915 | 3823 |
| | 4673 | 4621 | 4624 |
| | 4816 | 4501 | 4622 |
| | 7215 | 7032 | 6900 |
| | 5836 | 5923 | 5858 |
| | 380 | 334 | 371 |
| | 595 | 577 | 540 |
| | 6797 | 6650 | 6625 |
| | 3481 | 3435 | 3534 |
| | 2353 | 2335 | 2160 |
| 6.27.08 | 1691 | 1668 | 1597 |
| | 1126 | 1062 | 1076 |
| | 1202 | 1222 | 1162 |
| | 1883 | 1776 | 1730 |
| | 1448 | 1454 | 1368 |

PAGE 195: Statistical Detection of Potentially Fabricated Data: A Case Study

|         | 1583 | 1606 | 1631 |
|---------|------|------|------|
| 6.30.08 | 325  | 287  | 276  |
|         | 2259 | 2156 | 2136 |
|         | 1747 | 1695 | 1640 |
|         | 2029 | 1971 | 1928 |
|         | 8729 | 8692 | 8759 |
|         | 2535 | 2425 | 2470 |
|         | 4221 | 4170 | 4224 |
|         | 5151 | 5037 | 4997 |
|         | 6723 | 6621 | 6553 |
| 7.1.08  | 2733 | 2605 | 2547 |
|         | 3018 | 2966 | 3019 |
|         | 2779 | 2763 | 2820 |
|         | 1748 | 1857 | 1711 |
| 7.3.08  | 2752 | 2694 | 2672 |
|         | 5701 | 5805 | 5681 |
|         | 3940 | 3989 | 4048 |
|         | 4403 | 4463 | 4294 |
|         | 1477 | 1468 | 1529 |
| 7.6.08  | 1651 | 1673 | 1643 |
|         | 1224 | 1229 | 1236 |
|         | 1649 | 1497 | 1632 |
| 7.7.08  | 4505 | 4471 | 4459 |
|         | 5042 | 4850 | 4635 |
|         | 5316 | 5326 | 5137 |
|         | 5323 | 5259 | 5351 |
|         | 3155 | 2878 | 3104 |
|         | 336  | 370  | 358  |
| 6.26.08 | 8986 | 8960 | 8811 |
|         | 4056 | 4273 | 4132 |
|         | 1538 | 1523 | 1545 |
|         | 1538 | 1523 | 1545 |
|         | 3588 | 3582 | 3490 |
|         | 4589 | 4568 | 4950 |
|         | 3903 | 3691 | 3817 |
|         | 4859 | 4891 | 4912 |
|         | 215  | 185  | 200  |
|         | 3134 | 3073 | 3101 |
|         | 2464 | 2454 | 2420 |
| 6.30.08 | 3767 | 3759 | 3765 |
|         | 8665 | 8756 | 8956 |
|         | 4502 | 4271 | 4256 |
|         | 4502 | 4419 | 4572 |
|         | 5327 | 5239 | 5369 |
|         | 2435 | 2334 | 2460 |
|        | 3544 | 3543 | 3658 |
|        | 702  | 654  | 657  |
|        | 4262 | 4273 | 4232 |
|        | 4136 | 4132 | 3963 |
|        | 3646 | 3735 | 3708 |
|        | 4025 | 3949 | 4082 |
| 7.1.08 | 3912 | 3852 | 3735 |
|        | 1068 | 1055 | 1089 |
|        | 617  | 616  | 679  |
|        | 1135 | 1217 | 1251 |
|        | 1853 | 1775 | 1790 |
|        | 1451 | 1475 | 1495 |
|        | 579  | 616  | 594  |
|        | 124  | 122  | 122  |
|        | 1835 | 1825 | 1744 |
|        | 3062 | 3033 | 2947 |
|        | 7673 | 7464 | 7462 |
|        | 864  | 807  | 807  |
| 7.2.08 | 6048 | 6000 | 5917 |
|        | 6892 | 7016 | 6827 |
|        | 6572 | 6438 | 6405 |
|        | 6192 | 6158 | 5762 |
|        | 5222 | 5320 | 4953 |
|        | 6289 | 6202 | 6069 |
|        | 3205 | 3132 | 3221 |
|        | 3568 | 3488 | 3564 |
|        | 8170 | 8020 | 7961 |
|        | 3215 | 3286 | 3172 |
|        | 1743 | 1622 | 1597 |
|        | 253  | 221  | 238  |
|        | 6267 | 6296 | 6197 |
|        | 6051 | 6073 | 6155 |
|        | 5707 | 5618 | 5650 |